\documentclass{SCGE}
\usepackage[colorlinks,linkcolor=blue,citecolor=blue,urlcolor=blue]{hyperref}

\usepackage[latin1]{inputenc}
\usepackage{rotating}
\usepackage{footnote}
\newcolumntype{C}[1]{>{\centering\let\newline\\\arraybackslash\hspace{0pt}}m{#1}}

\usepackage{times}
\usepackage{newtxtext,newtxmath}
\usepackage{graphicx}	
\usepackage{amsmath}	
\usepackage{amssymb}	
\usepackage{xspace}
\usepackage{bigints}
\usepackage{siunitx}
\usepackage{amsthm}
\usepackage{mathtools}
\usepackage[normalem]{ulem}
\usepackage{float}
\usepackage{color}
\usepackage{tikz}
\usetikzlibrary{shapes,arrows}

\usepackage{CJK}

\let\citedash\relax
\makeatletter \providecommand{\citedash}{\hbox{-}\penalty\@m}
\makeatother

\begin{document}

\begin{picture}(0,0){\rm
\put(0,-20){\makebox[160truemm][l]{\bf {\sanhao\raisebox{2pt}{.}}
Article  {\sanhao\raisebox{1.5pt}{.}}}}}
\put(0,-34){\jiuwuhao {\textcolor[rgb]{0.5,0.5,0.5}{\sf
}}}
\end{picture}

\def\bm{\boldsymbol}

\def\dl{\displaystyle}
\def\du{\end{document}}
\def\d{{\rm d}}
\def\e{{\rm e}}
\def\i{{\rm i}}

\Year{2022} %
\Month{September} %
\Vol{65} %
\No{9} %
\BeginPage{299511} %
\AuthorMark{{\rm JIAO}, et al.}  
\DOI{https://doi.org/10.1007/s11433-021-1902-3} 
\ArtNo{299511}

\title[The J-comb image combination algorithm]{J-comb: An Image Fusion Algorithm to Combine Observations Covering Different Spatial Frequency Ranges}

\author[1,2]{Sihan Jiao}{}
\author[3]{Yuxin Lin}{}
\author[4]{Xiangyu Shui}{}
\author[1,2]{Jingwen Wu}{}
\author[1,2]{Zhiyuan Ren}{}
\author[1,2,5]{Di Li*}{}
\footnote{*Corresponding author: dili@nao.cas.cn}

\address[{\rm1}]{National Astronomical Observatories, Chinese Academy of Sciences, Beijing 100101, China}
\address[{\rm2}]{University of Chinese Academy of Sciences, Beijing 100049, China}
\address[{\rm3}]{Centre for Astrochemical Studies, Max-Planck-Institut f\"{u}r Extraterrestrische Physik, 85748 Garching, Germany}
\address[{\rm4}]{Astrophysics Department, Munich University, 81679, Munich, Germany}
\address[{\rm5}]{NAOC-UKZN Computational Astrophysics Centre, University of KwaZulu-Natal, Durban 4000, South Africa}

\maketitle\vspace{-3.5mm}{\footnotesize
\begin{center} Received November 23, 2021; accepted April 2, 2022; published online July 27, 2022
\end{center}}
\vspace*{-5mm}

\begin{center}
\rule{16.5cm}{0.4pt}
\parbox{16.5cm}
{\begin{abstract}
Ground-based, high-resolution bolometric (sub)millimeter continuum mapping observations on spatially extended target sources are often subject to significant missing fluxes. 
This hampers accurate quantitative analyses. 
Missing flux can be recovered by fusing high-resolution images with observations that preserve extended structures.
However, the commonly adopted image fusion approaches do not maintain the simplicity of the beam response function and do not try to elaborate the details of the yielded beam response functions. 
These make the comparison of the observations at multiple wavelengths not straightforward. We present a new algorithm, J-comb, which combines the high and low-resolution images linearly. 
By applying a taper function to the low-pass filtered image and combining it with a high-pass filtered image using proper weights, the beam response functions of our combined images are guaranteed to have near-Gaussian shapes.
This makes it easy to convolve the observations at multiple wavelengths to share the same beam response functions.
Moreover, we introduce a strategy to tackle the specific problem that the imaging at 850 $\mu$m from the present-date ground-based bolometric instrument and that taken with the {\it Planck} satellite do not overlap in the Fourier domain.
We benchmarked our method against two other widely-used image combination algorithms, \textsc{CASA}-feather and \textsc{MIRIAD}-immerge, with mock observations of star-forming molecular clouds. 
We demonstrate that the performance of the J-comb algorithm is superior to those of the other two algorithms.
We applied the J-comb algorithm to real observational data of the Orion A star-forming region.
We successfully produced dust temperature and column density maps with $\sim$10$''$ angular resolution, unveiling much greater details than the previous results.
A \textsc{python} code release of J-comb and implementation of the algorithm are available at \href{https://github.com/SihanJiao/J-comb}{https://github.com/SihanJiao/J-comb}.
\end{abstract}}
\end{center}\vspace*{-0.6cm}

\begin{center}
\parbox{16.5cm}
{\bf\jiuhao Star formation, molecular clouds, Image processing}
\end{center}

\begin{center}
{}
\end{center}

\textwidth=178truemm \textheight=236truemm

\wuhao\vspace*{1.5mm}

\begin{multicols}{2}

\renewcommand{\baselinestretch}{1.08} \baselineskip 12.2pt\parindent=10.8pt

\section{Introduction}\label{sec:intro}


Bolometric (sub)millimeter continuum mapping observations of extended sources from ground-based observatories are largely subject to artifacts induced by the subtraction of bright and time-varying atmospheric emissions.
The subtraction procedure often leads to considerable missing fluxes on extended angular scales.
In addition, it can produce significant negative-intensity bowls around bright sources. 
Such artifacts hamper the accuracy of quantitative analyses of (sub)millimeter continuum images, an essential observational avenue for the studies of star-forming molecular clouds.
The sky condition during the observations determines the details of the missing extended structures in the ground-based bolometric images.
And we can not easily predict the missing flux accurately.
Therefore, it is non-trivial to correctly fit the spectral energy distribution (SED) solely with the ground-based bolometric images. 
It may prevent the identification and quantitative analyses of faint structures.

\begin{figure}[H]
\hspace{-0.3cm}
\vspace{-0.1cm}
\includegraphics[width=9.cm]{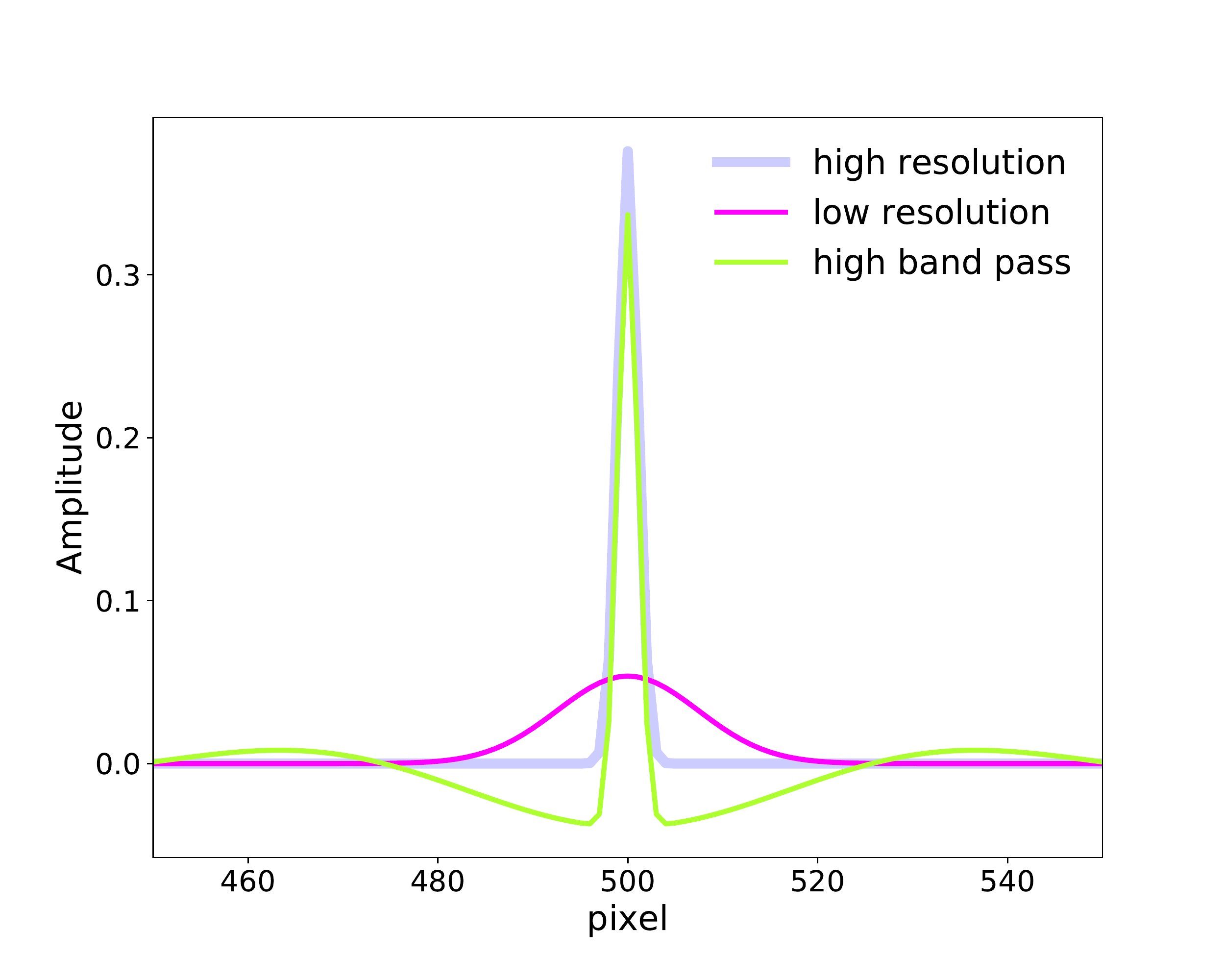}
\caption{
Appearance of a one-dimensional point source under different observational conditions.
The observations with FWHM = 3 and 25 pixels of Gaussian point-spread-function are shown in gray and magenta lines, respectively. Green line shows the high-pass filtered observations with a FWHM = 3 pixels Gaussian beam response function.}
\label{fig:1d_model}
\end{figure}

\begin{figure*}
\begin{tabular}{p{0.3\linewidth}p{0.3\linewidth}p{0.3\linewidth} }
\hspace{-0.6cm}\includegraphics[scale=0.263]{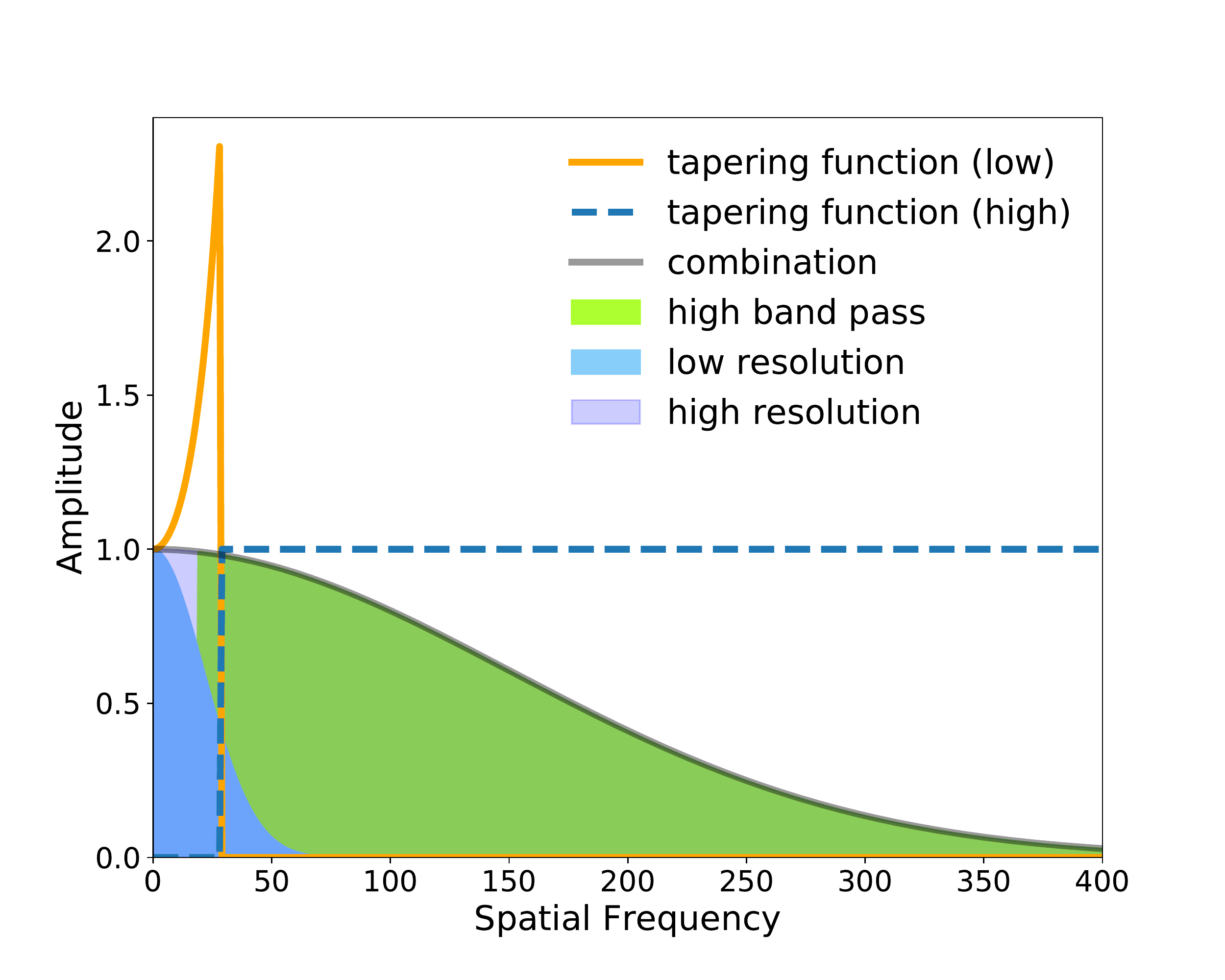} &
\hspace{-0.2cm}\includegraphics[scale=0.263]{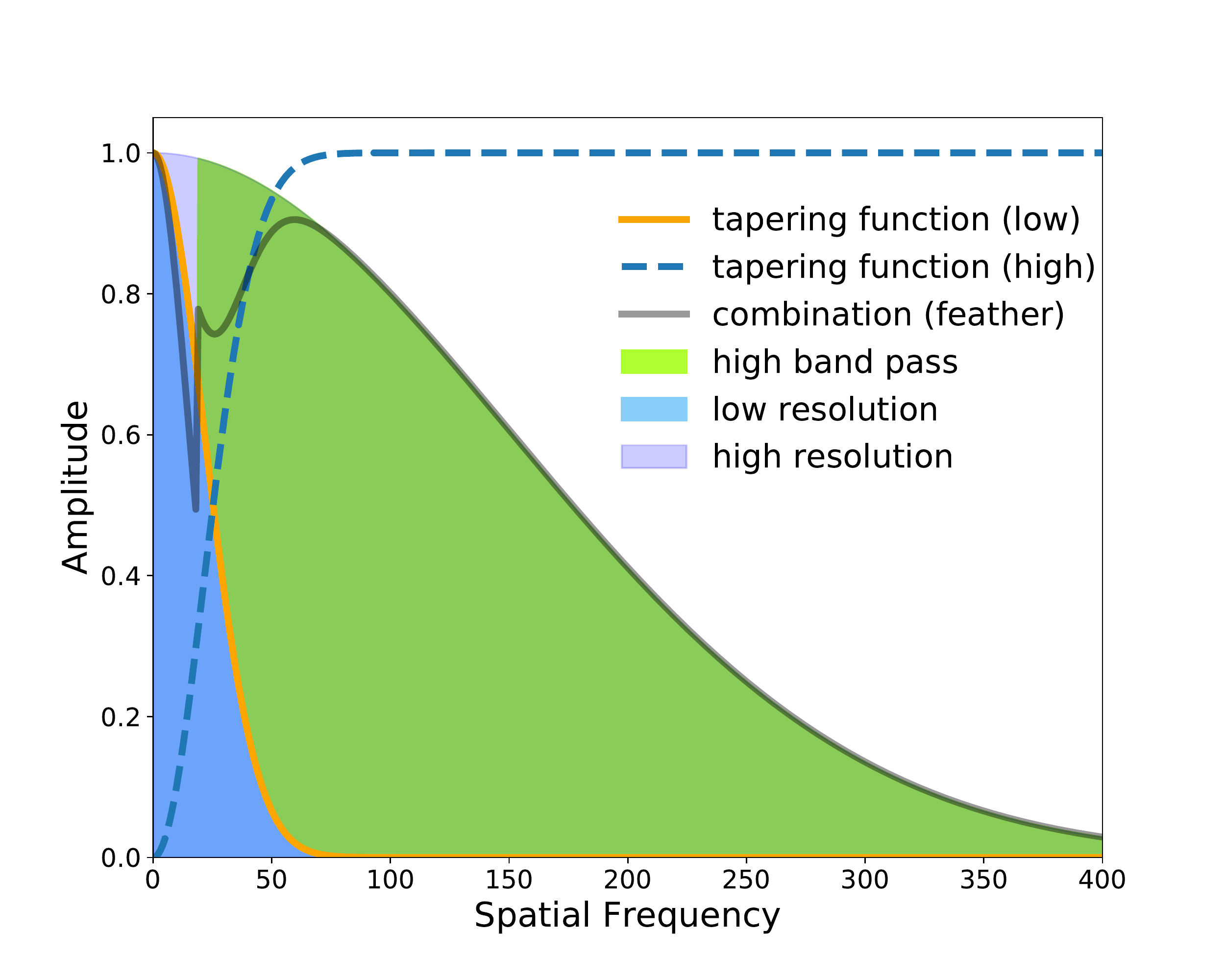}&
\hspace{0.13cm}\includegraphics[scale=0.263]{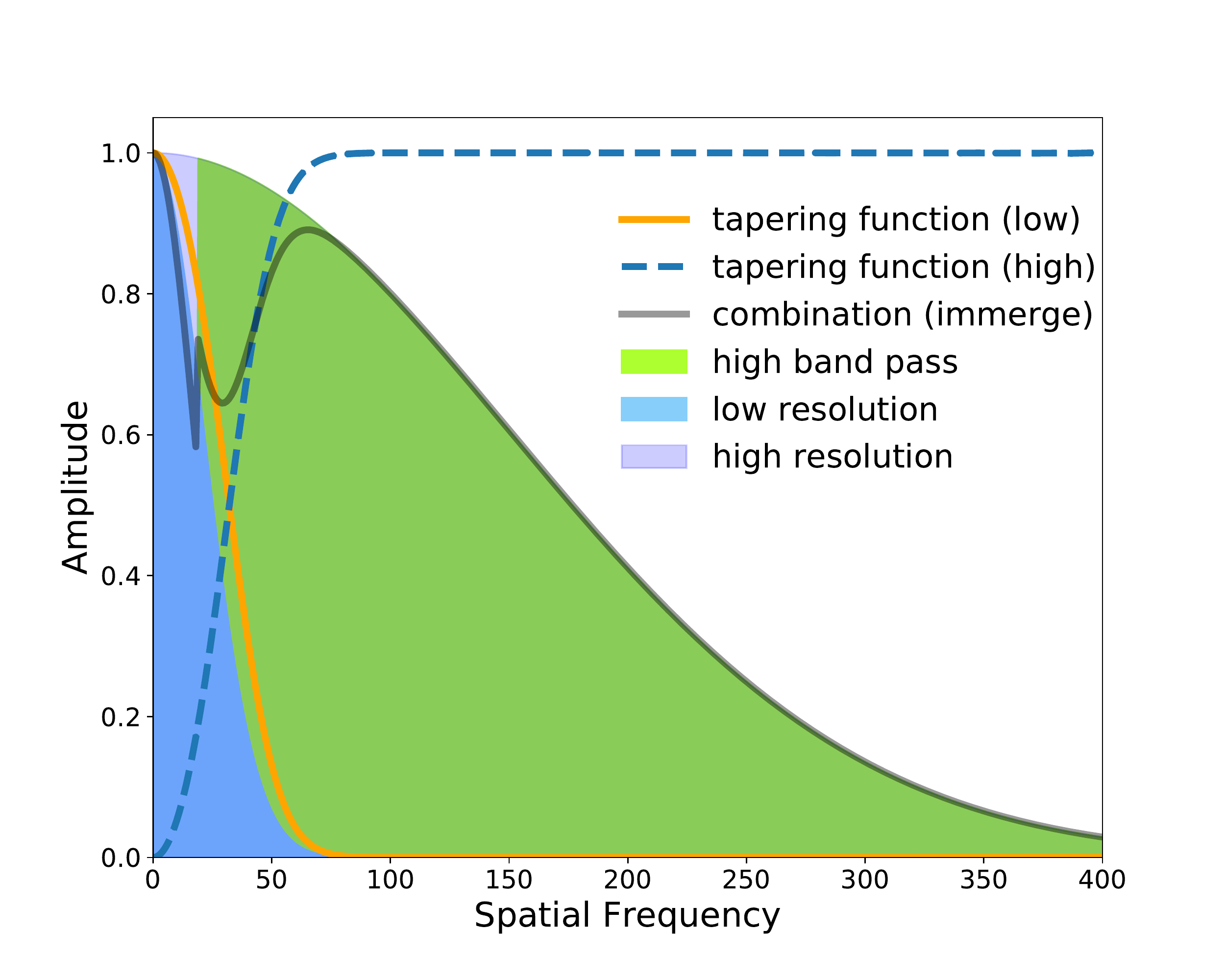}  \\
\end{tabular}
\begin{tabular}{p{0.3\linewidth}p{0.3\linewidth}p{0.3\linewidth} }
\hspace{-0.6cm}\includegraphics[scale=0.38]{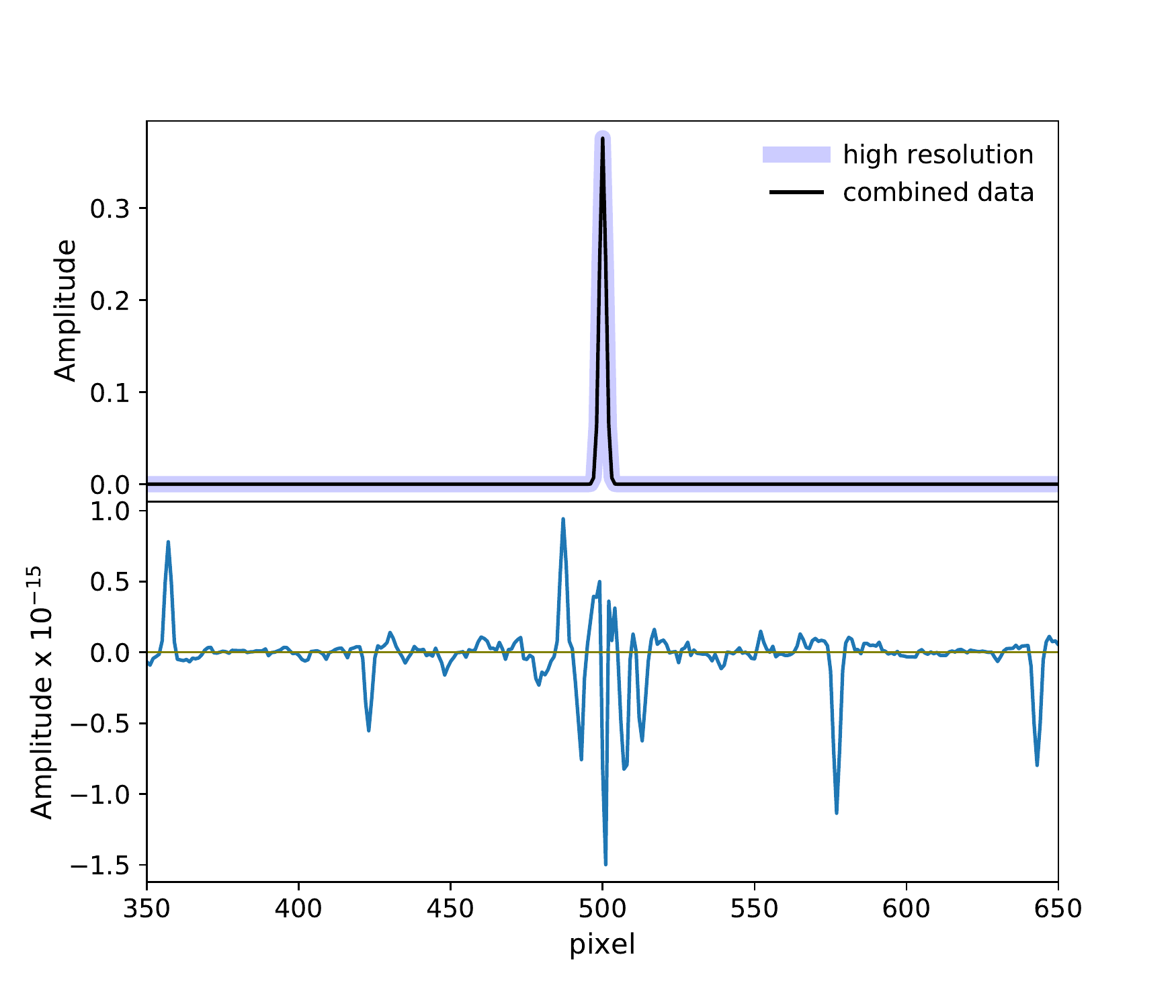} &
\hspace{-0.2cm}\includegraphics[scale=0.38]{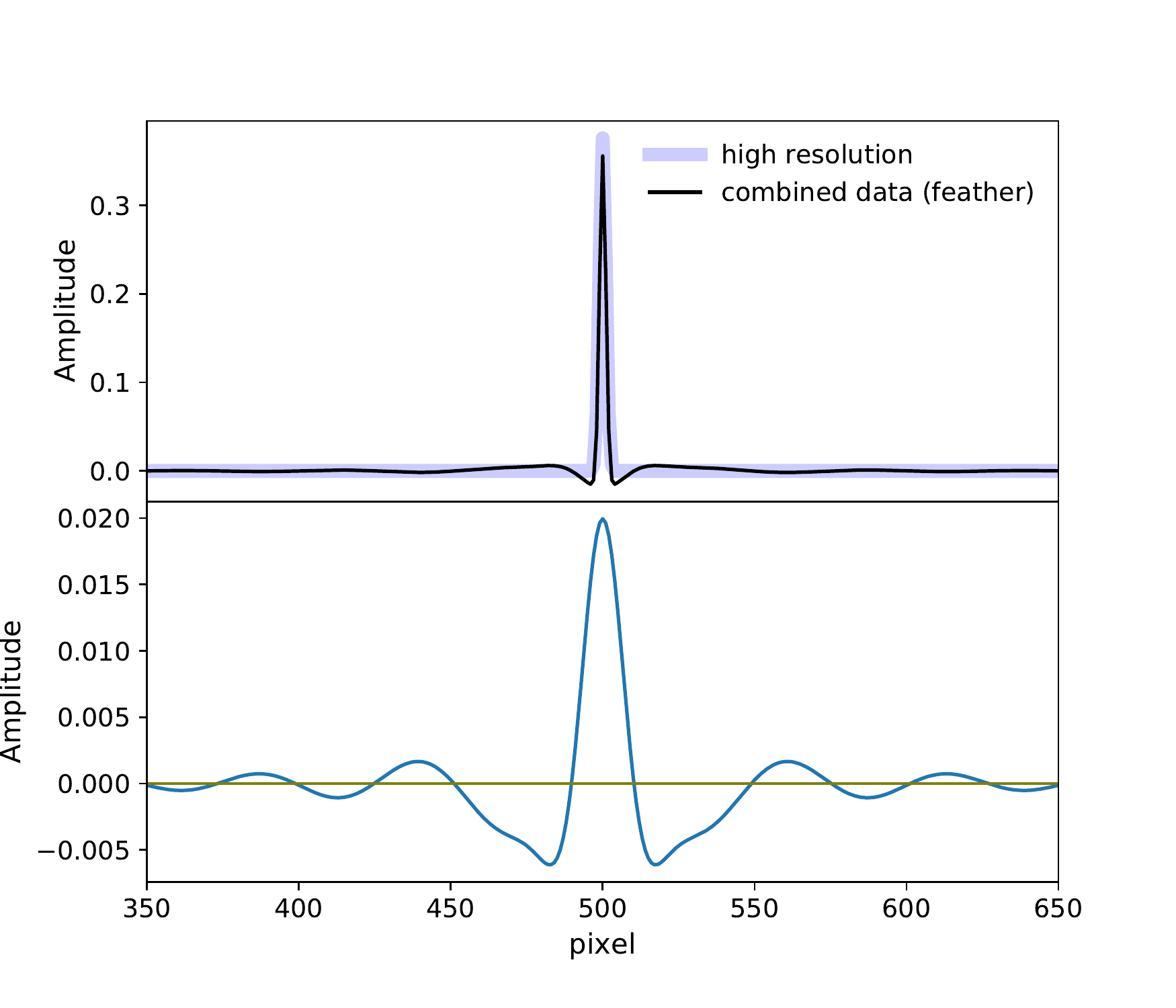}&
\hspace{0.13cm}\includegraphics[scale=0.38]{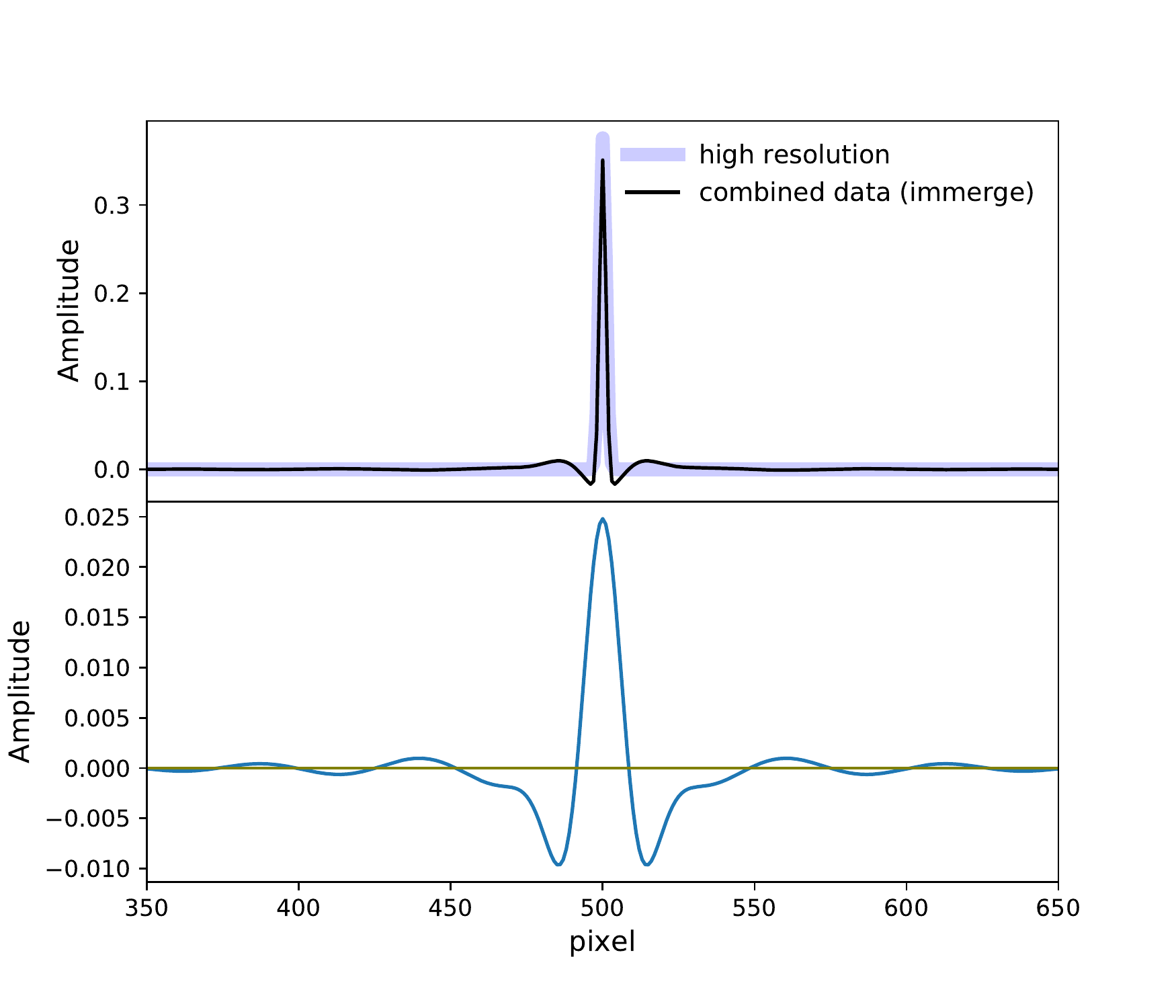}  \\
\end{tabular}
\caption{
The performances of J-comb, feather, and immerge algorithms in the one-dimensional point source simulation.
(Top panels) The tapering function of J-comb, feather, and immerge algorithms. 
The blue and green shaded areas show the Fourier transformed Gaussian beam response function of the low-resolution data and high-resolution data, respectively.
The orange line and blue dashed line show the tapering functions for low and high-resolution data.
(Bottom panels) The one-dimensional combination results (upper panel) and residuals (lower panel) of J-comb, feather, and immerge algorithms.
In the upper panel, combined data are shown in black line. In the lower panel, the residual estimated from subtracting the high-resolution data (gray line in the top panel) from the combined data, are shown in blue line.
}
\label{fig:1d_combine}
\end{figure*}


Observing from space (e.g., using the {\it Herschel} \cite{Pilbratt2010} or {\it Planck} Space Observatory \cite{Planck2011}) can avoid the artifacts induced by atmospheric subtraction and can better preserve extended structures.
However, the (sub)millimeter space telescopes typically have much smaller dish sizes than the ground-based ones. The diffraction limit implies that the celestial brightness distributions are convolved with the large beam response functions before being recorded. 
In other words, the observations taken with the (sub)millimeter space telescopes have worse angular resolutions.
The beam response function is the relation between the input and the output images for an observational sense of beam pattern.
Before conducting quantitative analyses, such as SED fittings, we often need to adjust all utilized images to have the same beam response function (or angular resolution, simply). This compels us to adopt the coarsest angular resolution, which is often given by the space telescope observations at the longest wavelength band (e.g., 5$'$ for {\it Planck} observations at 850 $\mu m$ wavelength \cite{Planck2011}).
In this case, the details of high-density gas structures in the star-forming molecular clouds, such as the dense gas filaments, clumps, cores, or other irregularly shaped gas streams, become confused with the diffuse cloud structures.
Molecular clouds ubiquitously exhibit a complex and filamentary morphology \cite{Andre2016}.
The cold molecular clumps represent the initial state of star-formation in giant molecular clouds (GMCs), which typically have spatial scales of 0.5 pc \cite{Liu2012}.
These cold molecular clump structures can be easily missed in the smeared, lower angular resolution temperature and column density maps derived by traditional methods.

To avoid such drawbacks, initiated by \cite{Liu2015}, we have been working on developing an iterative procedure \cite{Lin2016,Lin2017,Lin2019} by linearly combining the images taken from ground-based and space telescope observations.
In essence, the emission of extended structures recovered from space telescope observations can be used to complement the images taken by ground-based telescopes to obtain high angular resolution combined images without missing flux, which helps to better constrain the global properties of the target sources \cite{Csengeri2016}.
Linear image combination in the Fourier domain has been widely applied and implemented in all commonly used radio interferometric data reduction packages, including task immerge in \textsc{MIRIAD} software package \cite{Sault1995}, task feather in \textsc{CASA} software package \cite{McMullin2007}, task {\tt IMERG} in \textsc{AIPS} \cite{Greisen1990} software package and task feather in the \textsc{Obit} software package \cite{Cotton2008}.
These algorithms combine one image of a narrow beam response function, which is incompletely sampled over the extended angular scales, with an image of a broad beam response function. 
These algorithms have successfully yielded combined images of high fidelity for single wavelength observations.
However, the resultant beam response functions of the combined images are not elaborated in most cases. 
This sometimes makes the quantitative analyses uncertain: it is not straightforward to compare the observations taken at multiple wavelengths with different observational facilities.
Specifically, the beam response functions of the combined images derived from the previous linear data combination algorithm are not Gaussian and different from that of the high-resolution images.
The non-Gaussianity of beam response functions makes the angular resolution ill-defined.
To maintain the same Gaussian beam as the high-resolution data and recover the information from different spatial frequencies so that the residuals are maximally suppressed, we developed the tool of image combination, J-comb. 
In this context, we also introduce a strategy to tackle the problem that the imaging at 850 $\mu m$ from present-date ground-based observations and that taken with the {\it Planck} satellite do not overlap in the Fourier domain.
J-comb is a new image-fusion algorithm that can combine two images describing the same region but covering different spatial frequencies.
In this work, the ultimate goal of the J-comb algorithm is to
derive dust temperature and column density distributions of star-forming molecular clouds precisely by combining bolometric observations taken from ground-based instruments and space telescopes.
To benchmark our development, the J-comb algorithm is tested on mock observations, together with \textsc{CASA}-feather and \textsc{MIRIAD}-immerge.
By quantitatively comparing the emission levels and the power spectrum, we demonstrate that J-comb performs better than the other algorithms for the specific purpose of combining bolometric observations from ground-based and space telescopes.

To further test the performance of the J-comb algorithm, we apply it to the real observational data of the Orion A star-forming region.
We successfully produced dust temperature and column density maps of $\sim$10$''$ angular resolution, which unveil much greater details than the previous results and precisely constrain physical properties on all angular scales.

The paper is laid out as follows: we outline the background of the linear image combination and the J-comb algorithm in Section \ref{section:background}. 
We then present benchmarks of different algorithms on mock data in Section \ref{sub:benchmark}.
In Section \ref{subsub:orion}, we apply the J-comb algorithm to observational data of the Orion A region.
Finally, we summarise our developments of the J-comb algorithm in Section \ref{section:summary}.

\section{Linear image combination and the J-comb algorithm}\label{section:background}

\subsection{Linear approaches for image fusion}

For the specific purpose of combining bolometric observations taken from ground-based and space telescopes, we assume that the low-resolution space telescopes observations ($I_{\mbox{\scriptsize low}}$) can better probe the large-scale structures (corresponding to high spatial frequencies) and the high-resolution ground-based observations ($I_{\mbox{\scriptsize high}}$) can better reflect details of the small-scale structures (corresponding to low spatial frequencies).
Before performing the combination, flux calibration is needed to guarantee that the input low-resolution and high-resolution images are at the same flux level. 
The extended structures seen by the space telescope observations can therefore be used to complement the ground-based images with missing flux.
Conceptually, the combination of $I_{\mbox{\scriptsize low}}$ and $I_{\mbox{\scriptsize high}}$ in the Fourier domain can be expressed by the following generic forms:
\begin{equation}\label{eq:linearcombine}
\begin{aligned}
C(\vec{k}) = A^{\mbox{\scriptsize low}} W^{\mbox{\scriptsize low}} FT(I_{\mbox{\scriptsize low}}) + A^{\mbox{\scriptsize high}} W^{\mbox{\scriptsize high}} FT(I_{\mbox{\scriptsize high}}),
\end{aligned}
\end{equation}
\begin{equation}\label{eq:linearcombine2}
\begin{aligned}
W^{\mbox{\scriptsize low}} + W^{\mbox{\scriptsize high}} = 1.0,
\end{aligned}
\end{equation}
where FT denotes Fourier transform, A$^{\mbox{\scriptsize low}}$ and A$^{\mbox{\scriptsize high}}$ are the amplitude modulation factors for low and high-resolution images, W$^{\mbox{\scriptsize low}}$ and W$^{\mbox{\scriptsize high}}$ are the relative weighting functions for low and high-resolution image, $I_{\mbox{\scriptsize low}}$ and $I_{\mbox{\scriptsize high}}$ are the low and high-resolution images, and $\vec{k}$ is the spatial frequency vector in the Fourier domain.
We refer to the product of the amplitude modulation factor and the weighting function as tapering function.
The combined image is then:
\begin{equation}
I = FT^{-1}(C).
\end{equation}
Ideally, this technique requires the input low and high-resolution images to have overlapping spatial frequencies.

\subsection{Canonical methods: feather and immerge }

The feather and the immerge algorithms have been widely used for linear image combination. 
Here we present the case of combining the one-dimensional, low and high-resolution data of a point source, to introduce concepts of these two methods and that of linear data combination algorithms.
For simplicity, we only present the cases in which Gaussian functions can approximate the observational beam response functions.
We simulated a one-dimensional point source and then convolved it with different Gaussian beams of FWHM = 3 pixel and FWHM = 25 pixel to generate the high-resolution image (gray line in Figure \ref{fig:1d_model}) and the low-resolution image (magenta line in Figure \ref{fig:1d_model}).
To mimic the effect of missing flux at short spacings, we manually extracted the large scale structures from the high-resolution image, by applying a filter threshold (spatial frequency \textless 20 pixel$^{-1}$) in the Fourier domain (green line in Figure \ref{fig:1d_model}).
Finally, we adopted the above algorithms to combine the low-resolution image and the high-pass filtered high-resolution image to obtain the combined images.

Some variants of the feather algorithm differ in the forms of the weighting functions and the amplitude modulation factors.
Here we equivalently express parameters of the feather algorithm as the forms in Equation \ref{eq:linearcombine} and \ref{eq:linearcombine2} \cite{Stanimirovic2002,Rau2019} :
\begin{equation}\label{eq:feather1}
A^{\mbox{\scriptsize low}} = A^{\mbox{\scriptsize high}} = 1.0,
\end{equation}
and
\begin{equation}\label{eq:feather2}
        W^{\mbox{\scriptsize low}} = G_{\mbox{\scriptsize low}},
\end{equation}
where $G_{\mbox{\scriptsize low}}$ is the Fourier transform of the low-resolution beam ($B_{\mbox{\scriptsize low}}$), i.e., $G_{\mbox{\scriptsize low}}$ = FT($B_{\mbox{\scriptsize low}}$).
The tapering functions of feather method are presented in Figure \ref{fig:1d_combine}.
For the case of a one-dimensional point source,
\begin{equation}
FT(I_{\mbox{\scriptsize low}}) = G_{\mbox{\scriptsize low}}, \,\,\,\,\,\,
FT(I_{\mbox{\scriptsize high}}) = G_{\mbox{\scriptsize high}}.
\end{equation}
Substituting feather tapering functions of Equation \ref{eq:feather1} and \ref{eq:feather2} into Equation \ref{eq:linearcombine}, we have
\begin{equation}
C(\vec{k}) = G_{\mbox{\scriptsize high}}-G_{\mbox{\scriptsize low}}(G_{\mbox{\scriptsize high}}-G_{\mbox{\scriptsize low}})\neq G_{\mbox{\scriptsize high}},
\end{equation}
which means that the resultant beam response function of the combined image differs from that of the high-resolution image.

Similarly, the tapering functions of immerge algorithm introduced in the official \textsc{MIRIAD} document from Australia Telescope National Facility (ANTF) \footnote{An official documentation of immerge method can be found at https://www.cfa.harvard.edu/sma/miriad/manuals/ATNFuserguide\_US.pdf} can be expressed as,
\begin{equation}\label{eq:immerge1}
A^{\mbox{\scriptsize low}} = A^{\mbox{\scriptsize high}} = 1.0,
\end{equation}
and
\begin{equation}\label{eq:immerge2}
        W^{\mbox{\scriptsize low}} = G_{\mbox{\scriptsize high}}/(G_{\mbox{\scriptsize low}}+G_{\mbox{\scriptsize high}}),
\end{equation}
where $G_{\mbox{\scriptsize high}}$ is the Fourier transform of the high-resolution beam ($B_{\mbox{\scriptsize high}}$), i.e., $G_{\mbox{\scriptsize high}}$ = FT($B_{\mbox{\scriptsize high}}$).
Substituting the documented immerge tapering functions of Equation \ref{eq:immerge1} and \ref{eq:immerge2} into Equation \ref{eq:linearcombine}, we have
\begin{equation}
C(\vec{k}) = G_{\mbox{\scriptsize high}}\frac{2G_{\mbox{\scriptsize low}}}{(G_{\mbox{\scriptsize low}}+G_{\mbox{\scriptsize high}})}\neq G_{\mbox{\scriptsize high}}
\end{equation}

However, looking into the source code of \textsc{MIRIAD} software package, we found that the actual implemented forms are different from the above equations from the official document.
In fact, a new Fourier transformed beam ($G_{\mbox{\scriptsize new}}$) is introduced based on the standard deviation of the low and high-resolution Gaussian beam, $\sigma_{\mbox{\scriptsize low}}$ and $\sigma_{\mbox{\scriptsize high}}$, and has the form following 
\begin{equation}\label{eq:immerge_sec1}
G_{\mbox{\scriptsize new}}=e^{-\frac {k^{2}}{2(\sigma^{2}_{\mbox{\tiny low}}-\sigma^{2}_{\mbox{\tiny high}})}}.
\end{equation}
In the implementation, it was elaborated that when $G_{\mbox{\scriptsize new}}<2\times10^{-9}$,
\begin{equation}\label{eq:immerge_sec2}
A^{\mbox{\scriptsize low}}W^{\mbox{\scriptsize low}} = 0.0 ,
\end{equation}
and
\begin{equation}\label{eq:immerge_sec3}
A^{\mbox{\scriptsize high}}W^{\mbox{\scriptsize high}} = 1.0.
\end{equation}
And when $G_{\mbox{\scriptsize new}}>2\times10^{-9}$,
\begin{equation}\label{eq:immerge_sec4}
A^{\mbox{\scriptsize low}} = G_{\mbox{\scriptsize new}},
A^{\mbox{\scriptsize high}} = 1.0,
\end{equation}
and
\begin{equation}\label{eq:immerge_sec5}
        W^{\mbox{\scriptsize low}} = 
        1/G_{\mbox{\scriptsize new} }= 
        e^{\frac {k^{2}}{2(\sigma^{2}_{\mbox{\tiny low}}-\sigma^{2}_{\mbox{\tiny high}})}}
\end{equation}
Substituting the actual immerge tapering functions of Equation \ref{eq:immerge_sec1} and \ref{eq:immerge_sec5} in Equation \ref{eq:linearcombine}, when $G_{\mbox{\scriptsize new}}>2\times10^{-9}$,
\begin{equation}
C(\vec{k}) = G_{\mbox{\scriptsize low}}+(1-1/G_{\mbox{\scriptsize new}})G_{\mbox{\scriptsize high}}\neq G_{\mbox{\scriptsize high}}.
\end{equation}

Again, the discrepancy between $C(\vec{k})$ and $G_{\mbox{\scriptsize high}}$ implies that the beam response functions of the combined images do not preserve that of the high-resolution data, which is usually Gaussian.
The non-Gaussianity of the yielded beam response functions can be immediately seen from the combined data of a point source, as presented in the bottom panels of Figure \ref{fig:1d_combine}.
Using the feather method, the ratio of the residual and the flux level of the high-resolution data is $\sim$10 \% around the peak.
And there are ripples of residuals all over the image domain.
With the immerge method, there is similar behavior of residuals.
These plots showcase the effect of the non-Gaussian beam in the combined image.
Moreover, for the purpose of, e.g., performing SED fittings, typically one needs to convert multi-wavelength images to have the same beam.
This process can be complicated when the beams of the combined images are not Gaussian.

\subsection{The J-comb algorithm}\label{sub:jcomb}

\begin{figure*}
\begin{tabular}{ p{0.5\linewidth}p{0.5\linewidth} }
\includegraphics[scale=0.5]{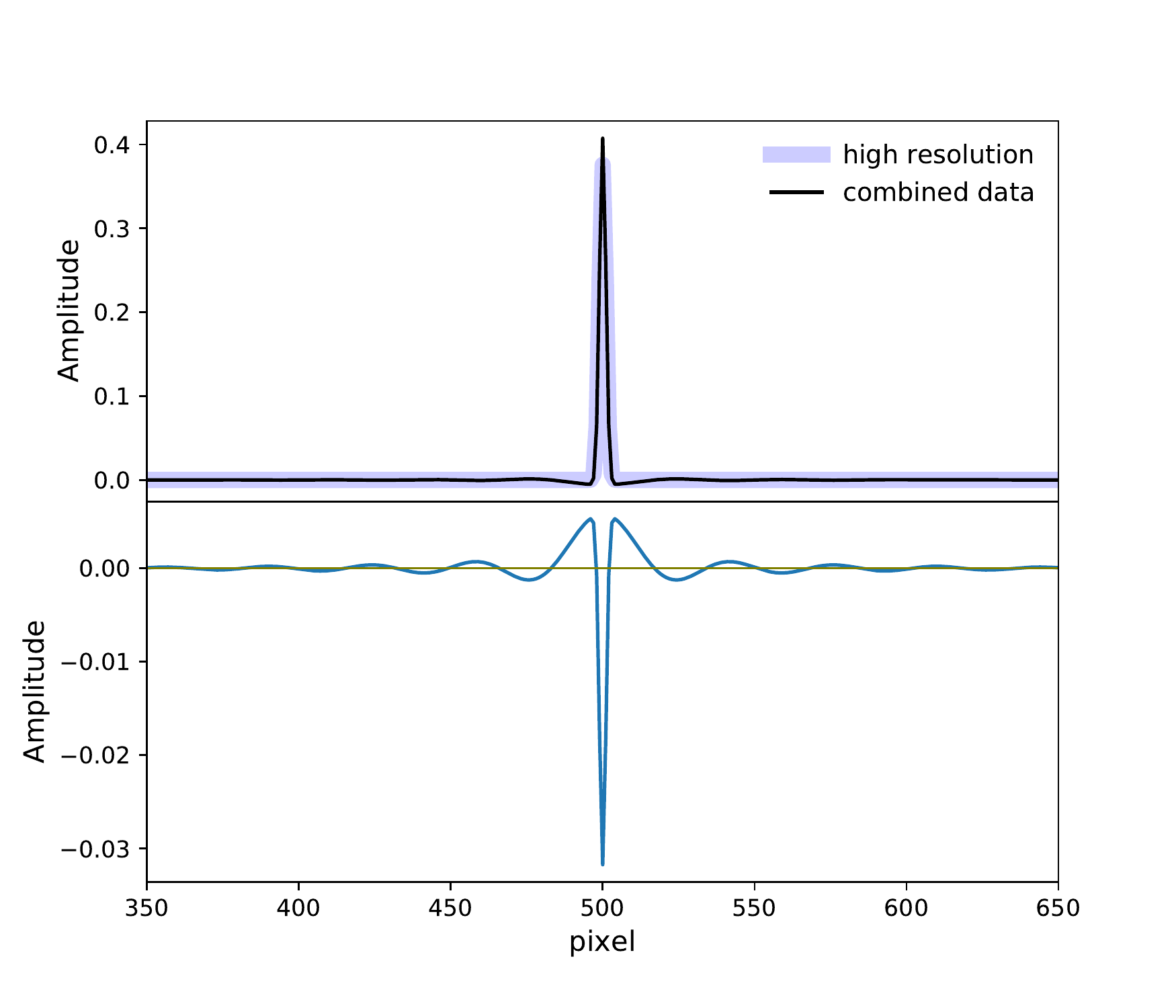} & \includegraphics[scale=0.5]{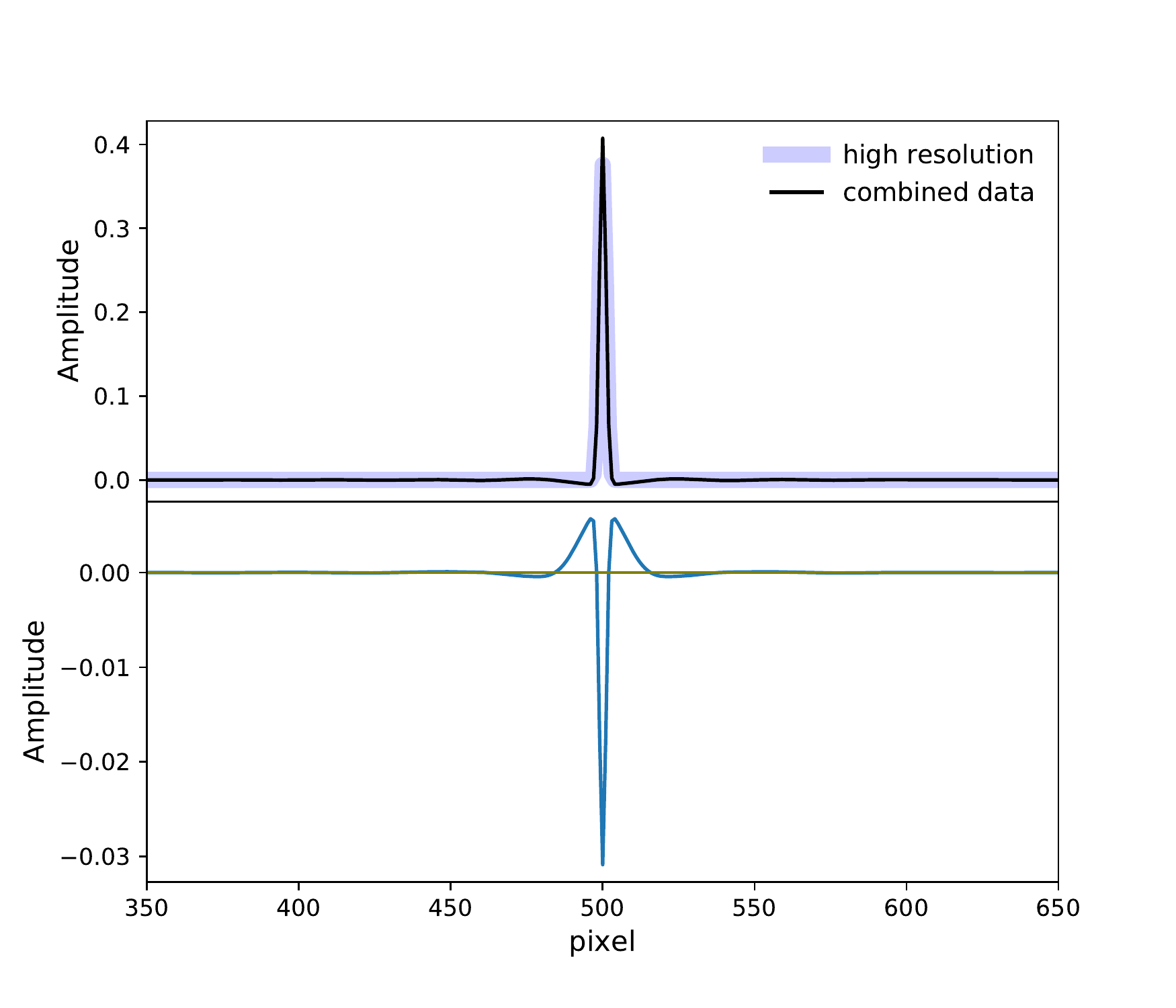}  \\
\end{tabular}
\vspace{-0.3cm}
\caption{
The performances of the J-comb algorithm adopting the step filter and the Butterworth filter.
Left panel shows results of the step filter.
The combined data (black line) and high-resolution data (gray line) are shown in the top panels and the residual (blue line) is shown in the bottom panels.
Right panel shows results from the Butterworth filter.
}
\label{fig:sub_err}
\end{figure*}

To retain the same Gaussian beam as the high-resolution data and preserve the information of all spatial frequencies in a way that residuals are maximally suppressed, we introduce a new combination scheme, namely the J-comb algorithm.

In this scheme, 
\begin{equation}\label{eq:j-comb1}
A^{\mbox{\scriptsize low}} = G_{\mbox{\scriptsize high}} / G_{\mbox{\scriptsize low}}, \,\,\,\,\,\,
A^{\mbox{\scriptsize high}} = 1.0,
\end{equation}
and
\begin{equation}\label{eq:j-comb2}
        W^{\mbox{\scriptsize low}}=
        \begin{cases}
            1.0 &(k < k_{\mbox{\scriptsize threshold}}^{\mbox{\scriptsize low}}) \\
            \frac{f(k)-f(k_{\mbox{\scriptsize threshold}}^{\mbox{\scriptsize high}})}{f(k_{\mbox{\scriptsize threshold}}^{\mbox{\scriptsize low}})-f(k_{\mbox{\scriptsize threshold}}^{\mbox{\scriptsize high}})} &(k_{\mbox{\scriptsize threshold}}^{\mbox{\scriptsize low}} < k < k_{\mbox{\scriptsize threshold}}^{\mbox{\scriptsize high}}) \\
            0.0 &(k > k_{\mbox{\scriptsize threshold}}^{\mbox{\scriptsize high}})
        \end{cases}
\end{equation}
where $G_{\mbox{\scriptsize low}}$ is the Fourier transform of the low-resolution beam ($B_{\mbox{\scriptsize low}}$), i.e., $G_{\mbox{\scriptsize low}}$ = FT($B_{\mbox{\scriptsize low}}$), $G_{\mbox{\scriptsize high}}$ is the Fourier transform of the high-resolution beam ($B_{\mbox{\scriptsize high}}$), i.e., $G_{\mbox{\scriptsize high}}$ = FT($B_{\mbox{\scriptsize high}}$), $k_{\mbox{\scriptsize threshold}}^{\mbox{\scriptsize low}}$ and $k_{\mbox{\scriptsize threshold}}^{\mbox{\scriptsize high}}$ are the spatial frequency thresholds, and $f(k)$ is a monotonic function with values in between 1.0 to 0.0 for spatial frequency $k$ ranging from $k_{\mbox{\scriptsize threshold}}^{\mbox{\scriptsize high}}$ to $k_{\mbox{\scriptsize threshold}}^{\mbox{\scriptsize low}}$.


\tikzstyle{decision} = [diamond, draw,  
text width=7em, text badly centered, node distance=3cm, inner sep=0pt]
\tikzstyle{block} = [rectangle, draw,  
text width=9em, text centered, rounded corners, minimum height=3em]
\tikzstyle{block_3} = [rectangle, draw,  
text width=18em, text centered, rounded corners, minimum height=3em]
\tikzstyle{block_sec} = [rectangle, draw,  
text width=11em, text centered, minimum height=3em]
\tikzstyle{block_4} = [rectangle, draw,  
text width=8em, text centered, minimum height=3em]
\tikzstyle{line} = [draw, -latex']
\tikzstyle{cloud} = [draw, ellipse, node distance=3cm,
minimum height=3em,text width=9em,text centered]
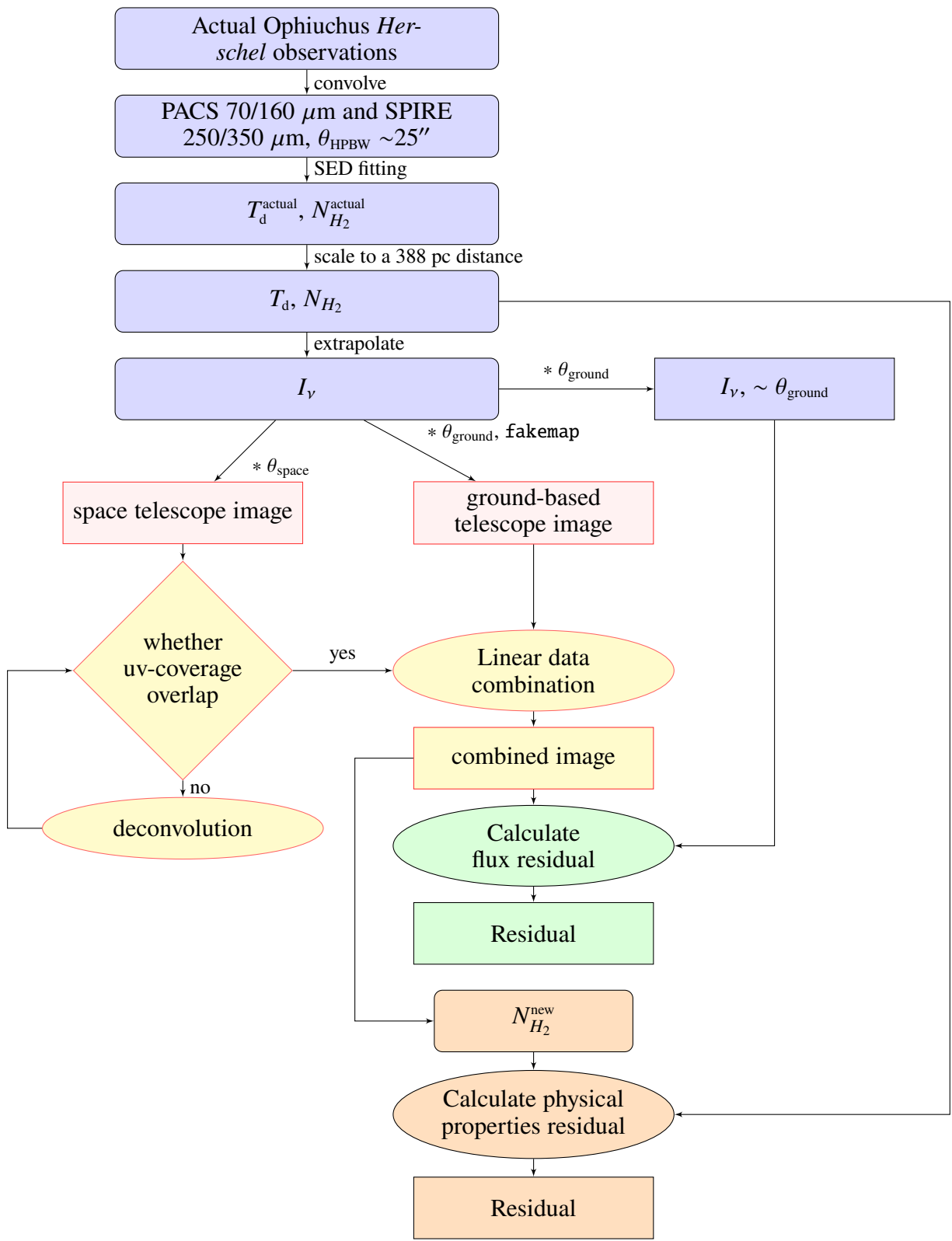
\begin{figure*}[!ht]
\centering
\hspace*{-100pt}
\begin{tikzpicture}[node distance = 1.5cm, auto]
    \hspace{2.5cm}
    \node [block_3,fill=blue!15] (model) {\large Actual Ophiuchus {\it Herschel} observations};
    \node [block_3, below of = model, node distance = 1.5 cm,fill=blue!15] (sm_data) {\large PACS 70/160 $\mu$m and SPIRE 250/350 $\mu$m, $\theta_{\mbox{\scriptsize HPBW}}$ $\sim$25$''$};
     \node [block_3, below of = sm_data, node distance = 1.5cm,fill=blue!15] (td_N) {\large $T_{\mbox{\scriptsize d}}^{\mbox{\scriptsize actual}}$, $N_{H_{2}}^{\mbox{\scriptsize actual}}$};
     \node [block_3, below of = td_N, node distance = 1.5cm,fill=blue!15] (td_N_sec) {\large $T_{\mbox{\scriptsize d}}$, $N_{H_{2}}$};
     \node [block_3, below of = td_N_sec, node distance = 1.5cm,fill=blue!15] (I) {\large $I_{\nu}$};
     \node [block_sec, below left of = I, node distance = 3.cm, draw=red!80, fill=red!5] (herschel) {\large space telescope image};
    \node [block_sec, right of = I, node distance = 8.cm, fill=blue!15] (I_high) {\large $I_{\nu}$, $\sim$ $\theta_{\mbox{\scriptsize ground}}$};
    \node [block_sec, right of = herschel, node distance = 6.cm, draw=red!80, fill=red!5] (high_850) {\large ground-based telescope image};
    \node [decision, below of = herschel,node distance = 2.7cm, draw=red!60, fill=yellow!25](decide){\large whether uv-coverage overlap};
    \node [cloud, below of = decide, node distance = 2.7cm, draw=red!60, fill=yellow!25] (lr) {\large deconvolution};
    \node [cloud, below of = high_850,node distance = 2.7cm, draw=red!60, fill=yellow!25] (combine) {\large Linear data combination};
    \node [block_sec, below of = combine, node distance = 1.5cm, draw=red!80, fill=yellow!25] (combine_data) {\large combined image};
    \node [cloud, below of = combine_data,node distance = 1.5cm,fill=green!15] (residual) {\large Calculate flux residual};
    \node [block_sec, below of = residual,node distance = 1.5cm,,fill=green!15] (residual_data) {\large Residual};
    \node [block, below of = residual_data, node distance = 1.5cm,fill=orange!25] (sed_sec) {\large $N_{H_{2}}^{\mbox{\scriptsize new}}$};
    \node [cloud, below of = sed_sec,node distance = 1.6cm,fill=orange!25] (residual_SED) {\large Calculate physical properties residual};
     \node [block_sec, below of = residual_SED,node distance = 1.6cm,fill=orange!25] (residual_SED_data) {\large Residual};

    \path [line] (model) -- node {convolve}(sm_data);
    \path [line] (sm_data) -- node {SED fitting}(td_N);
    \path [line] (td_N) -- node {scale to a 388 pc distance}(td_N_sec);
    \path [line] (td_N_sec) -- node {extrapolate}(I);
    \path [line] (I) -- node {$\ast$ $\theta_{\mbox{\scriptsize space}}$}(herschel);
    \path [line] (I) -- node {$\ast$ $\theta_{\mbox{\scriptsize ground}}$, {\tt fakemap}}(high_850);
    \path [line] (herschel) -- (decide);
    \path [line] (decide) -- node {no}(lr);
    \path [line] (decide) -- node {yes}(combine);
    \path [line] (lr) -- ++(-30mm,0mm) |- (decide);
    \path [line] (high_850) -- (combine);
    \path [line] (I) -- node {$\ast$ $\theta_{\mbox{\scriptsize ground}}$}(I_high);
    \path [line] (I_high) |- (residual);
    \path [line] (combine) -- (combine_data);
    \path [line] (combine_data.west) -- ++(-10mm,0mm) |- (sed_sec);
    \path [line] (sed_sec) -- (residual_SED);
    \path [line] (residual_SED) -- (residual_SED_data);
    \path [line] (combine_data) -- (residual);
    \path [line] (residual) -- (residual_data);
    \path [line] (td_N_sec) -- ++(110mm,0mm) |- (residual_SED);
\end{tikzpicture}
\caption{Flow chart for benchmarking the J-comb algorithm.}
\label{fig:flowchart_benchmarking}
\end{figure*}

\begin{figure*}[!ht]
\hspace{1.7cm}
\vspace{-0.1cm}
\begin{tabular}{ p{0.5\linewidth} p{0.5\linewidth}}
\hspace{-0.79cm}
\includegraphics[scale=0.54]{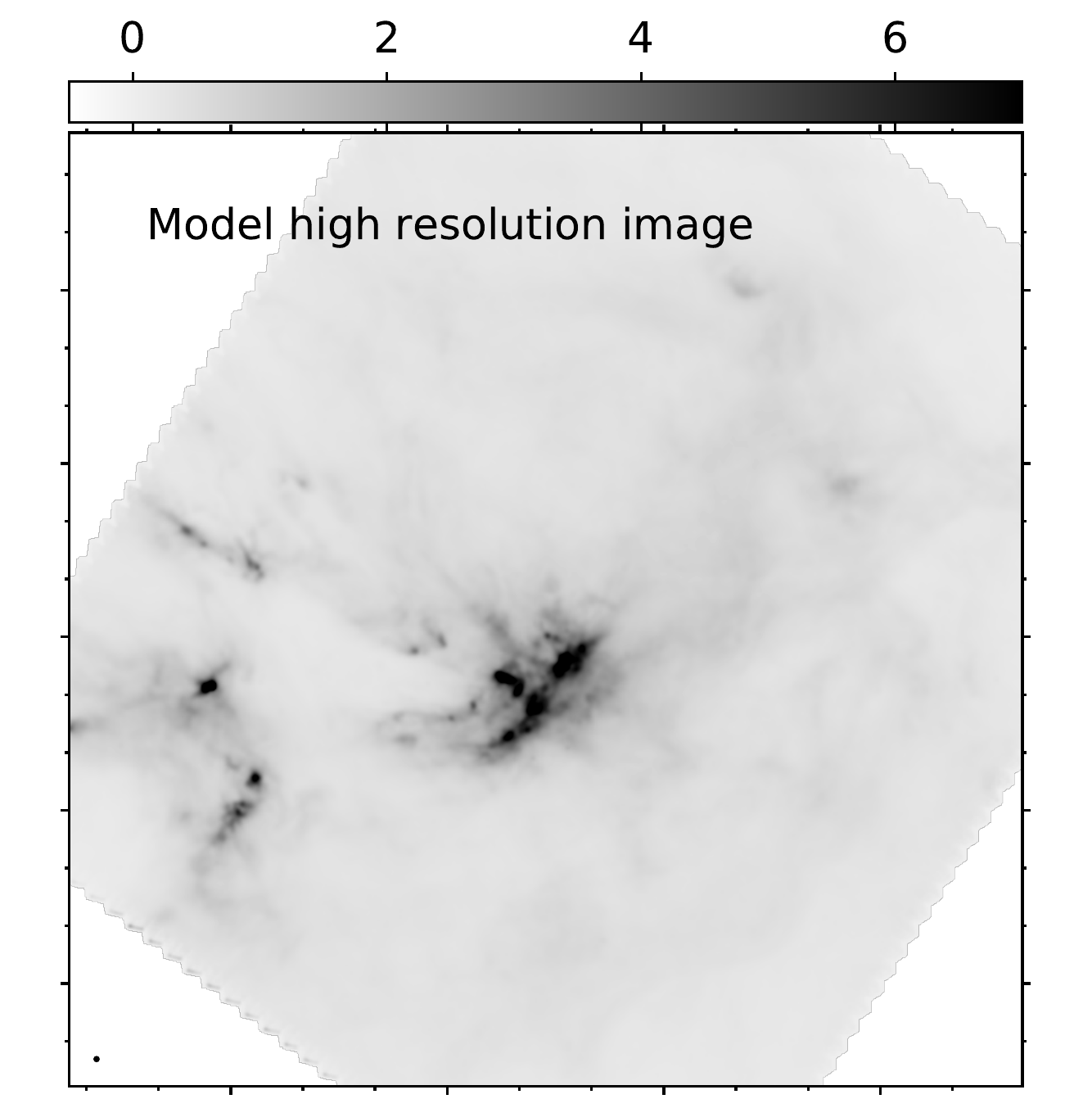} & 
\hspace{-2.2cm}
\includegraphics[scale=0.54]{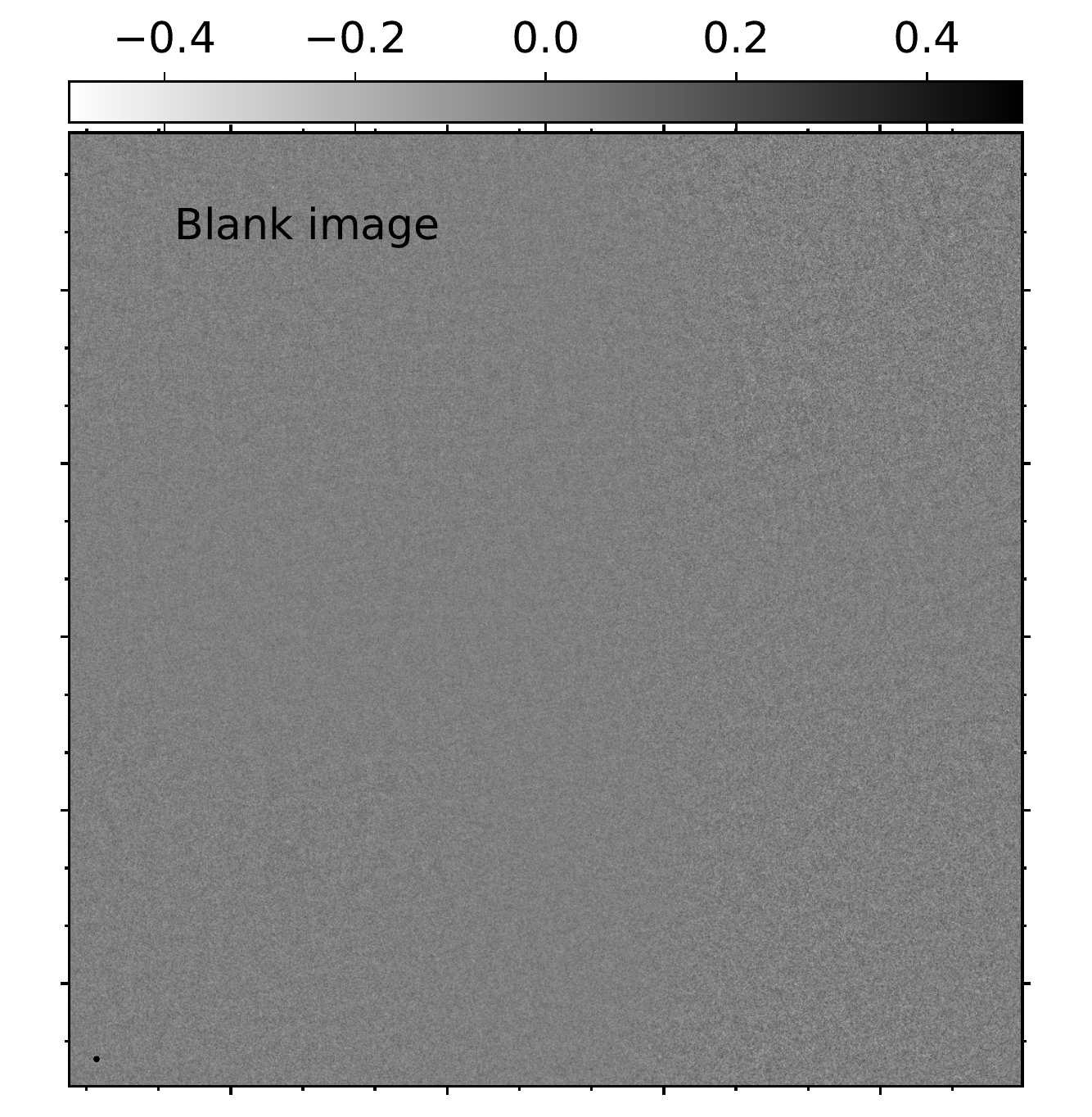} \\
\end{tabular}
\hspace{-1.7cm}
\vspace{-0.1 cm}
\begin{tabular}{ p{0.5\linewidth} p{0.5\linewidth}}
\hspace{-0.59cm}
\vspace{0.2 cm}
\includegraphics[scale=0.54]{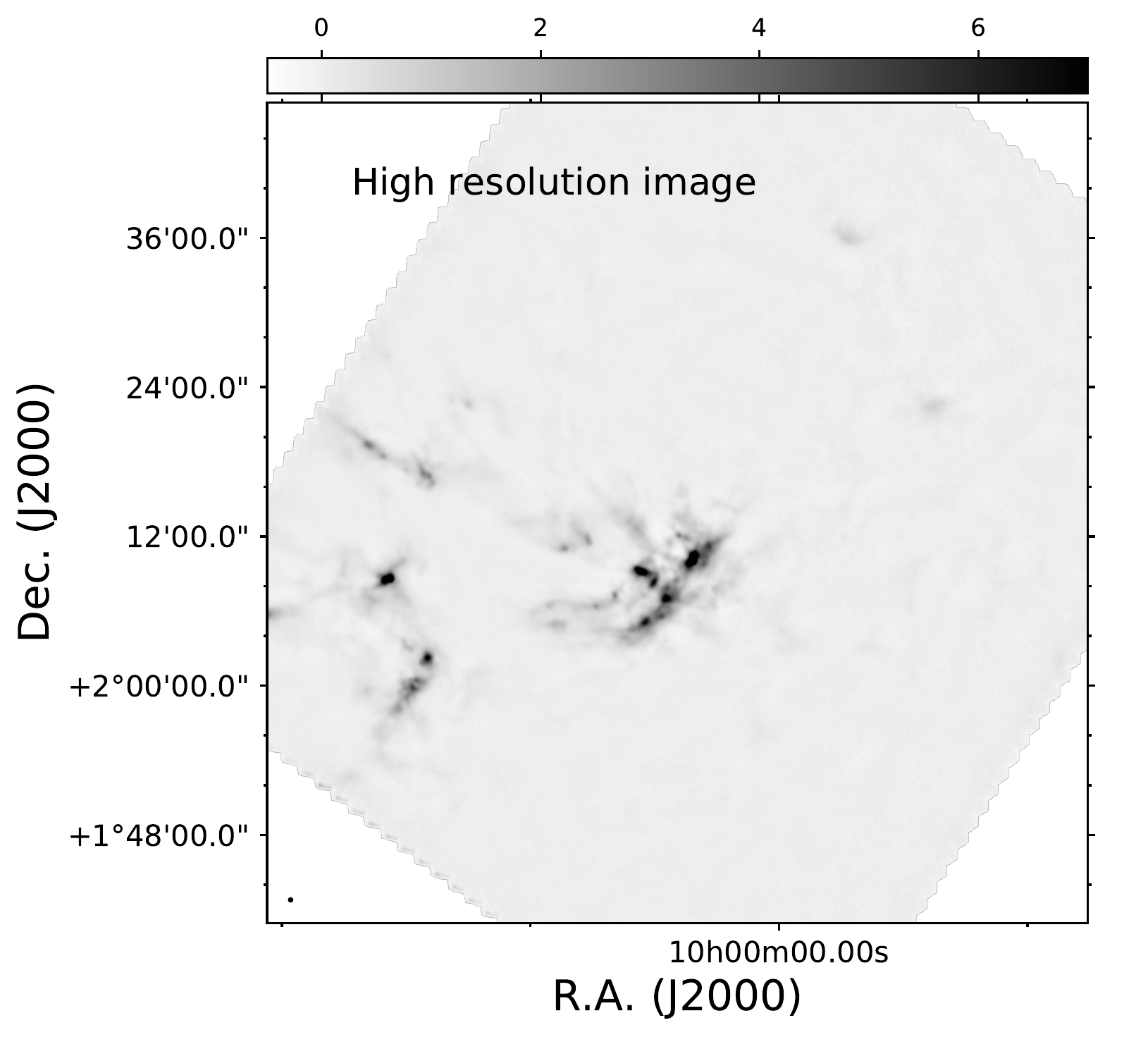} & 
\vspace{-8.15cm}
\hspace{-0.37cm}
\includegraphics[scale=0.54]{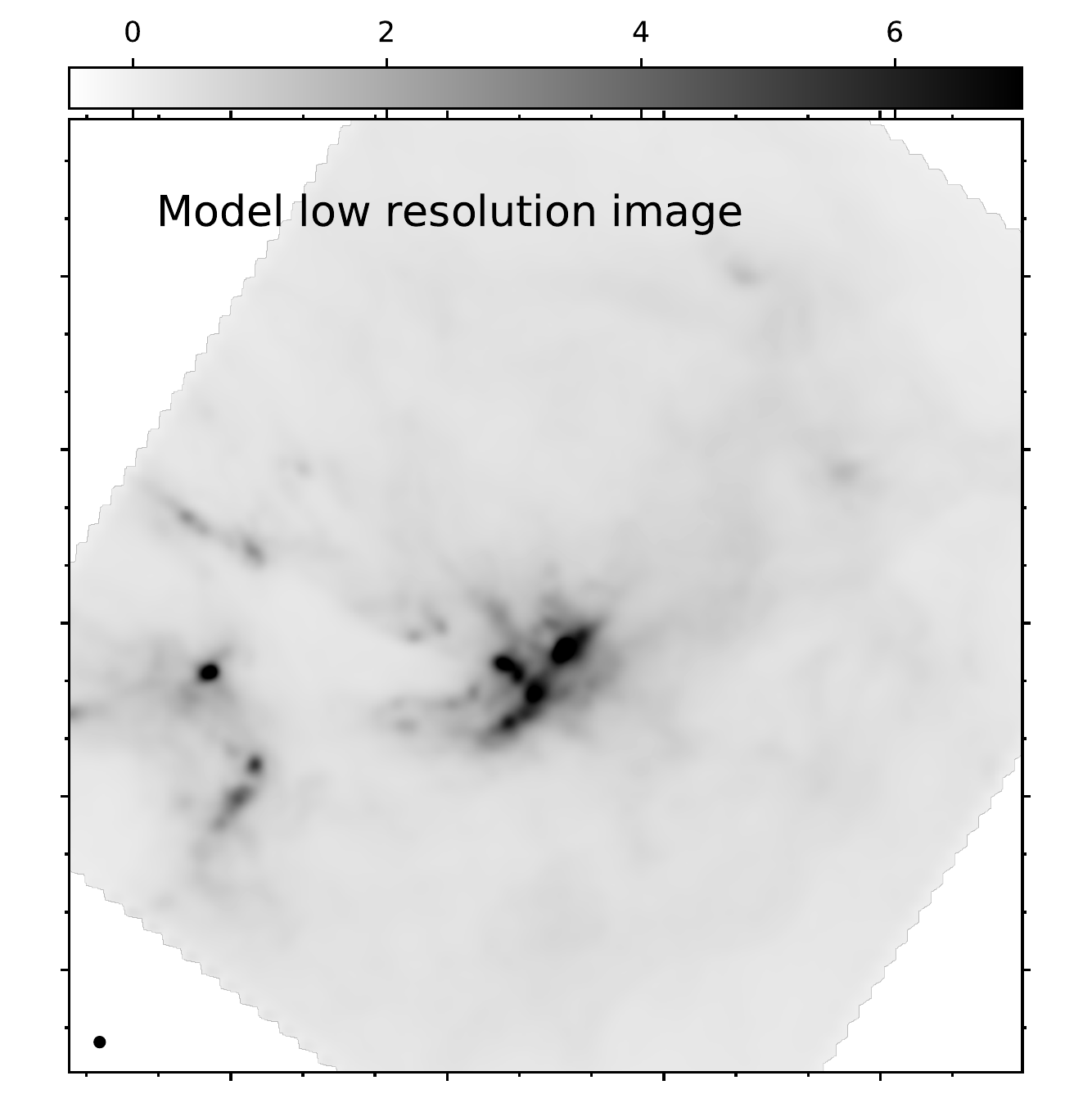} \\
\end{tabular}
\caption{
The input simulated 850 $\mu$m data for the following linear data combination benchmark.
Besides the blank image, all images are presented by using the same color scale with unit pW$\times$10$^{-3}$.
Upper left panel shows the high angular resolution, full spatial coverage image model, and the angular resolution is 14$''$. 
Upper right panel shows a JCMT-SCUBA2 observations towards COSMOS field \cite{Casey2013}, which can be regarded essentially as a blank field for our purpose, and the angular resolution is 14$''$.
Lower left panel shows the {\tt fakemap} simulation results based on the image model presented in the upper left panel, and the JCMT-SCUBA2 observations presented in the upper right panel.
This image is used as high-resolution image in the combination, and the angular resolution is 14$''$.
Lower right panel shows the simulated, {\it Planck}-850 $\mu$m image, which was deconvolved using Lucy-Richardson algorithm. This image is used as low-resolution image in the combination, and the angular resolution is 37$''$.
}
\label{fig:input}
\end{figure*}

\begin{figure*}
\hspace{-1.cm}
\begin{tabular}{p{0.5\linewidth}p{0.5\linewidth}}
\includegraphics[scale=0.4]{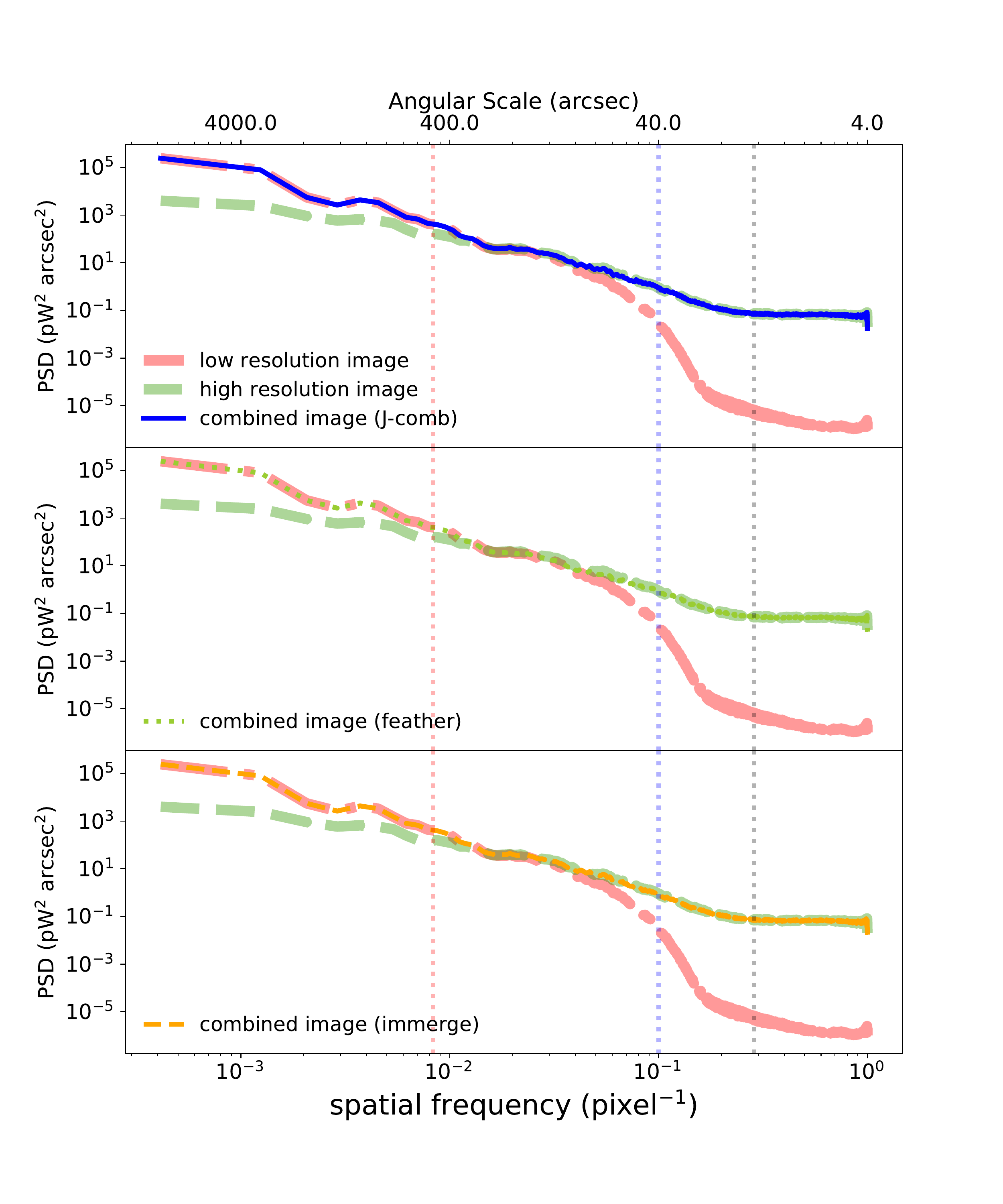}& \includegraphics[scale=0.4]{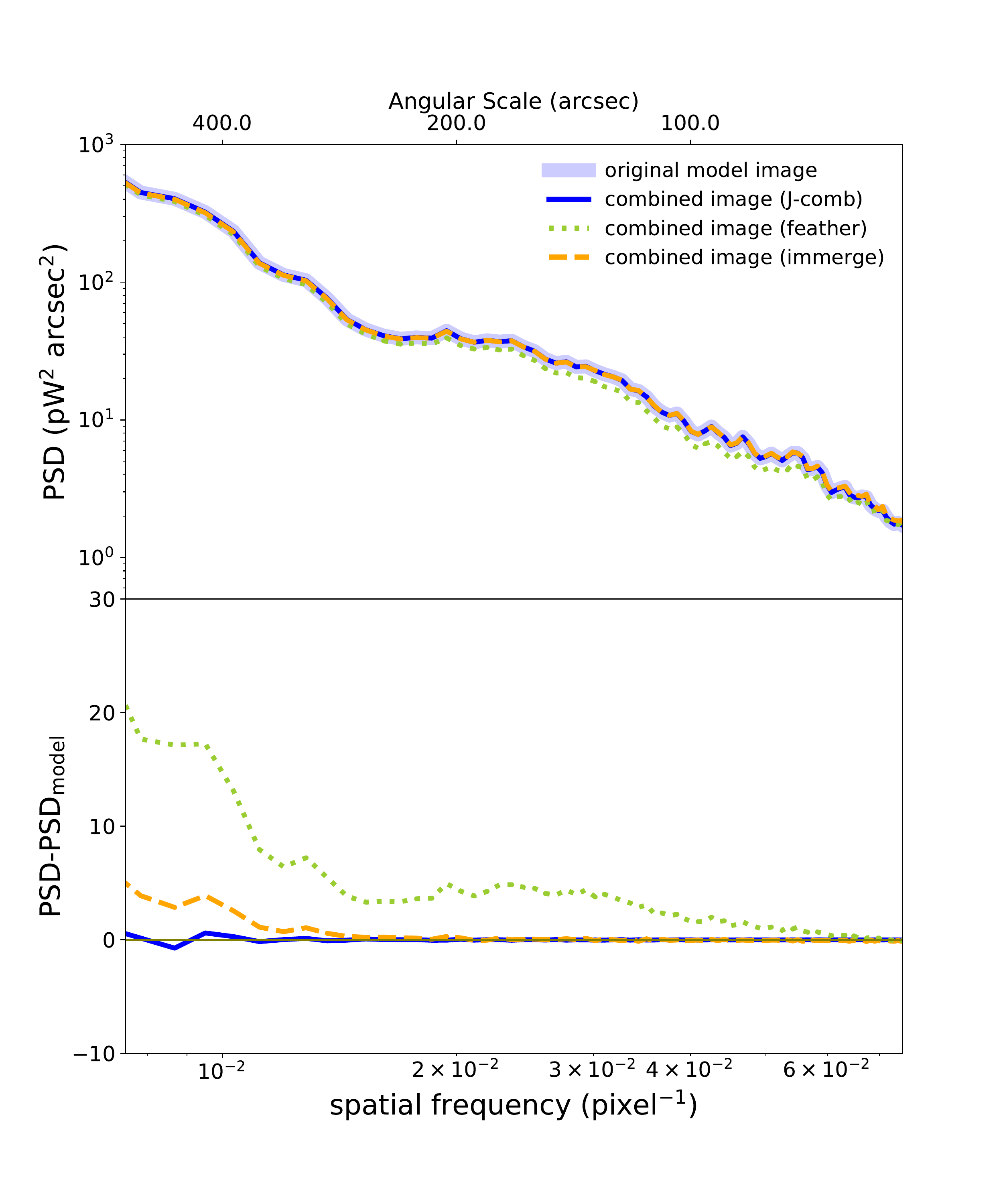} 
\end{tabular}
\vspace{-1.cm}
\caption{ 
Power spectra for the combined images using different linear data combination methods at 850 $\mu$m band.
Left panel shows the power spectrum for all spatial frequencies. 
At larger (\textgreater 800 $''$) and smaller angular scales ( \textless 40 $''$), the combined images generated by the J-comb algorithm, task immerge and feather preserve the high-resolution model image well. 
The vertical dashed black, blue and red lines mark the the angular scale of 14$''$ (the angular resolution of JCMT 850 $\mu$m observations), 37$''$ (the angular resolution of {\it Herschel} 500 $\mu$m observations), and 480$''$ (the default filter size for JCMT 850 $\mu$m data reduction pipeline).
Right panel shows a zoom-up view of the angular scale from 50 $''$ to 400 $''$, the combined image generated by the J-comb algorithm preserves the high-resolution model image better than that generated by \textsc{MIRIAD}-immerge, and \textsc{CASA}-feather. 
}
\label{fig:2dpsd}
\end{figure*}

\begin{figure*}[!ht]
\hspace{2.7cm}
\vspace{-0.1cm}
\begin{tabular}{ p{0.26\linewidth} p{0.26\linewidth} p{0.26\linewidth} }
\hspace{-3.3cm}\includegraphics[scale=0.46]{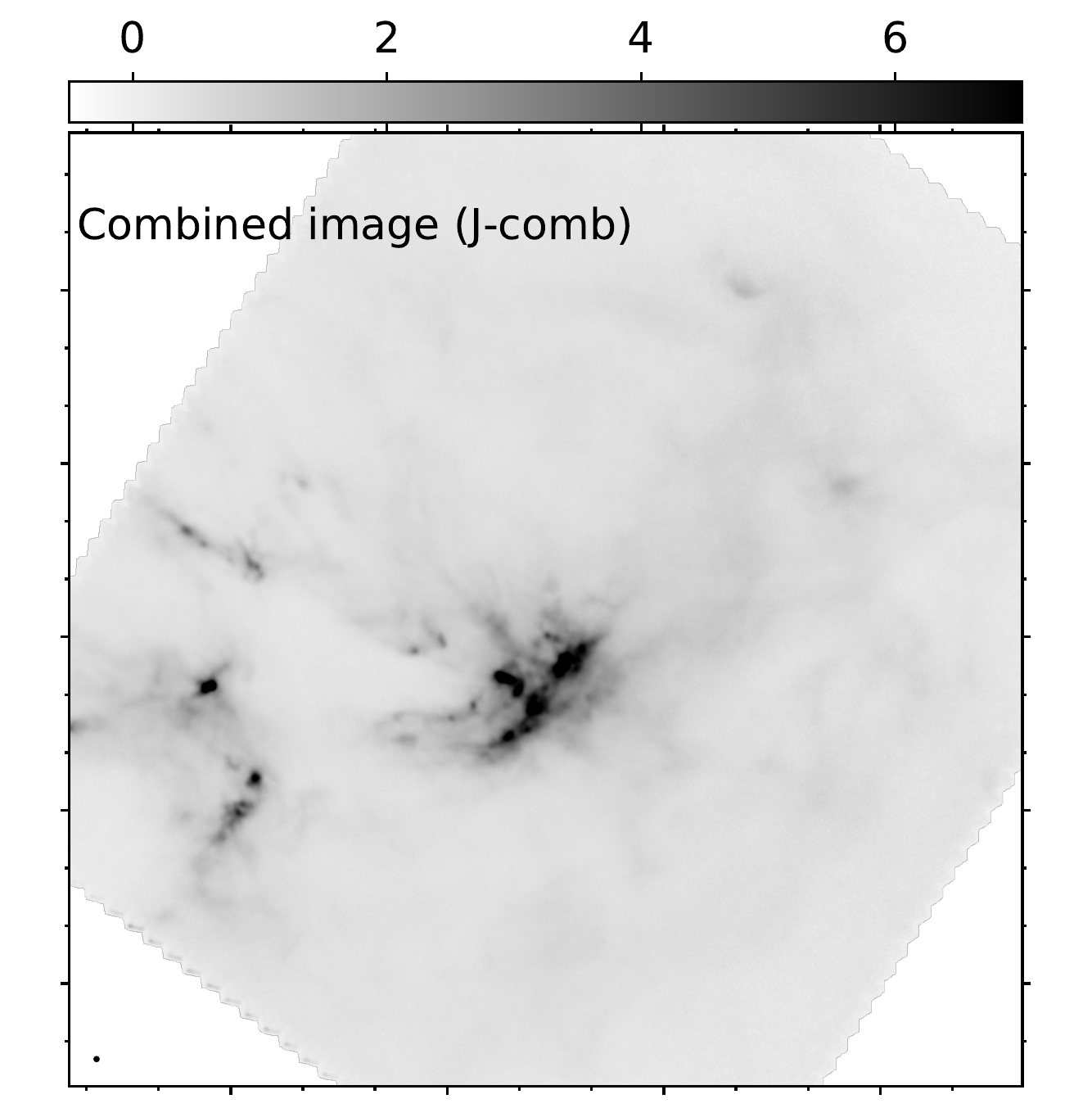} & 
\hspace{-2.2cm}\includegraphics[scale=0.46]{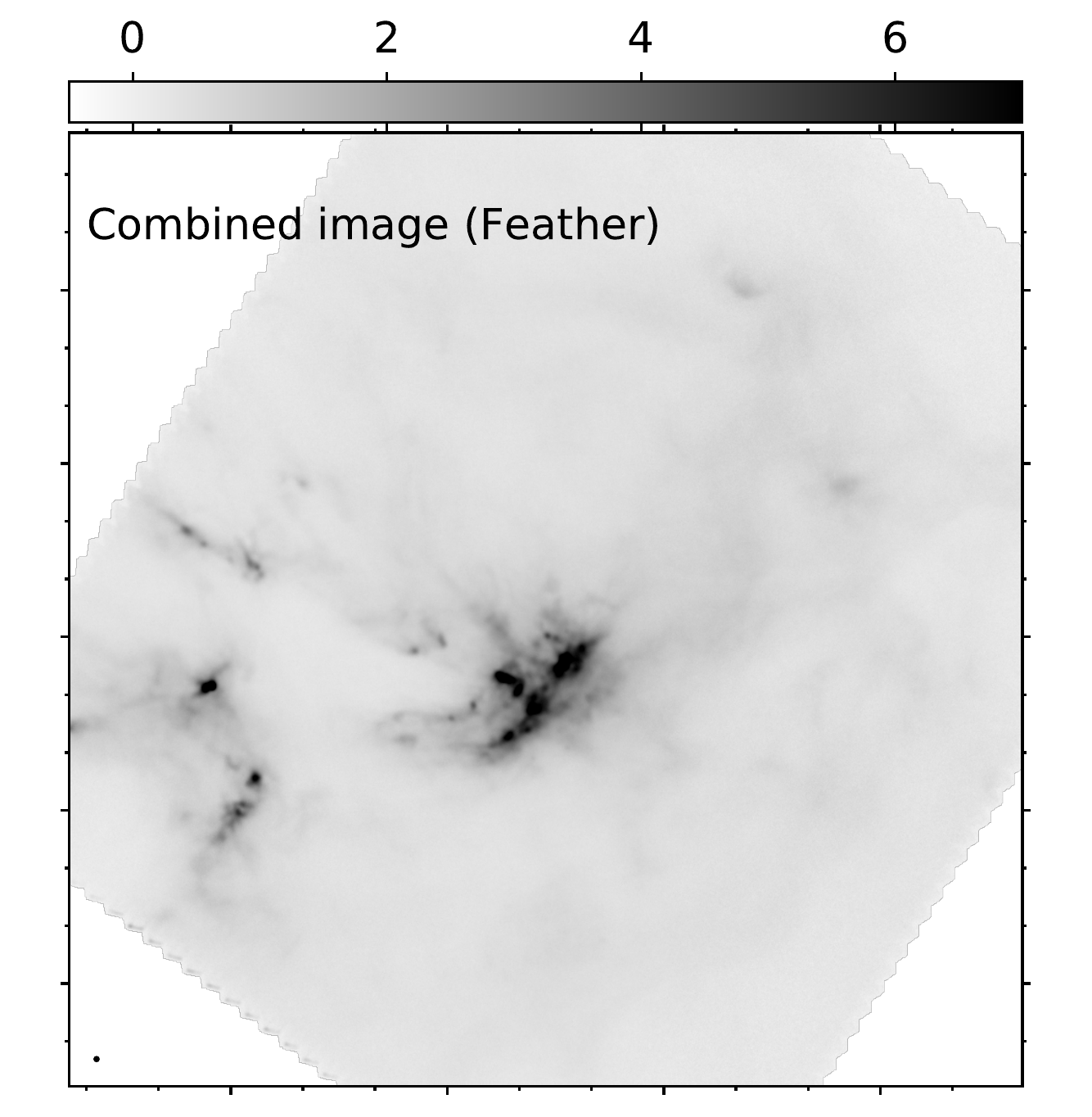} & 
\hspace{-1.1cm}\includegraphics[scale=0.46]{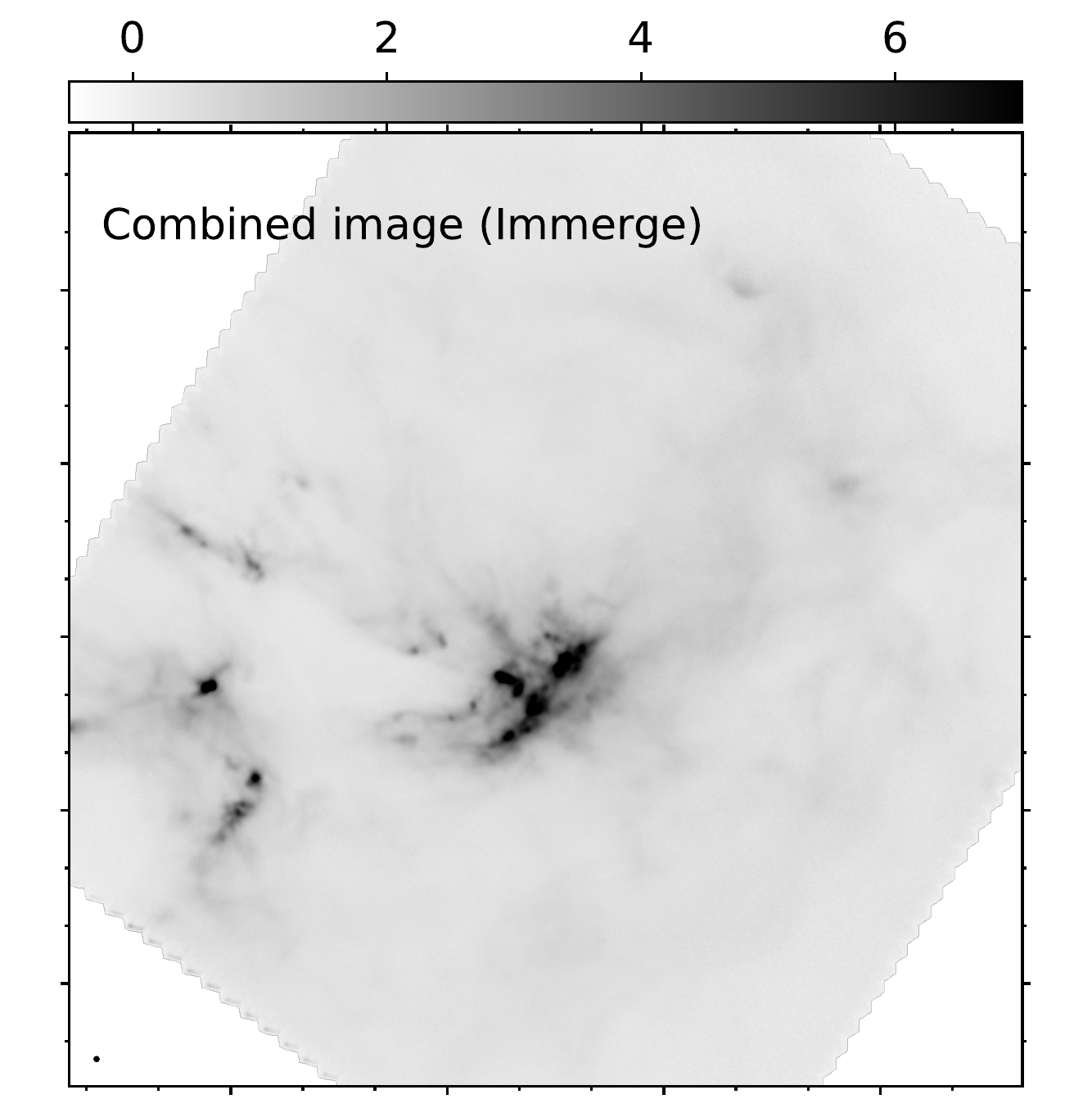} \\
\end{tabular}

\hspace{2.7cm}
\vspace{-0.1 cm}
\begin{tabular}{ p{0.26\linewidth} p{0.26\linewidth} p{0.26\linewidth} }
\hspace{-3.25cm}\includegraphics[scale=0.46]{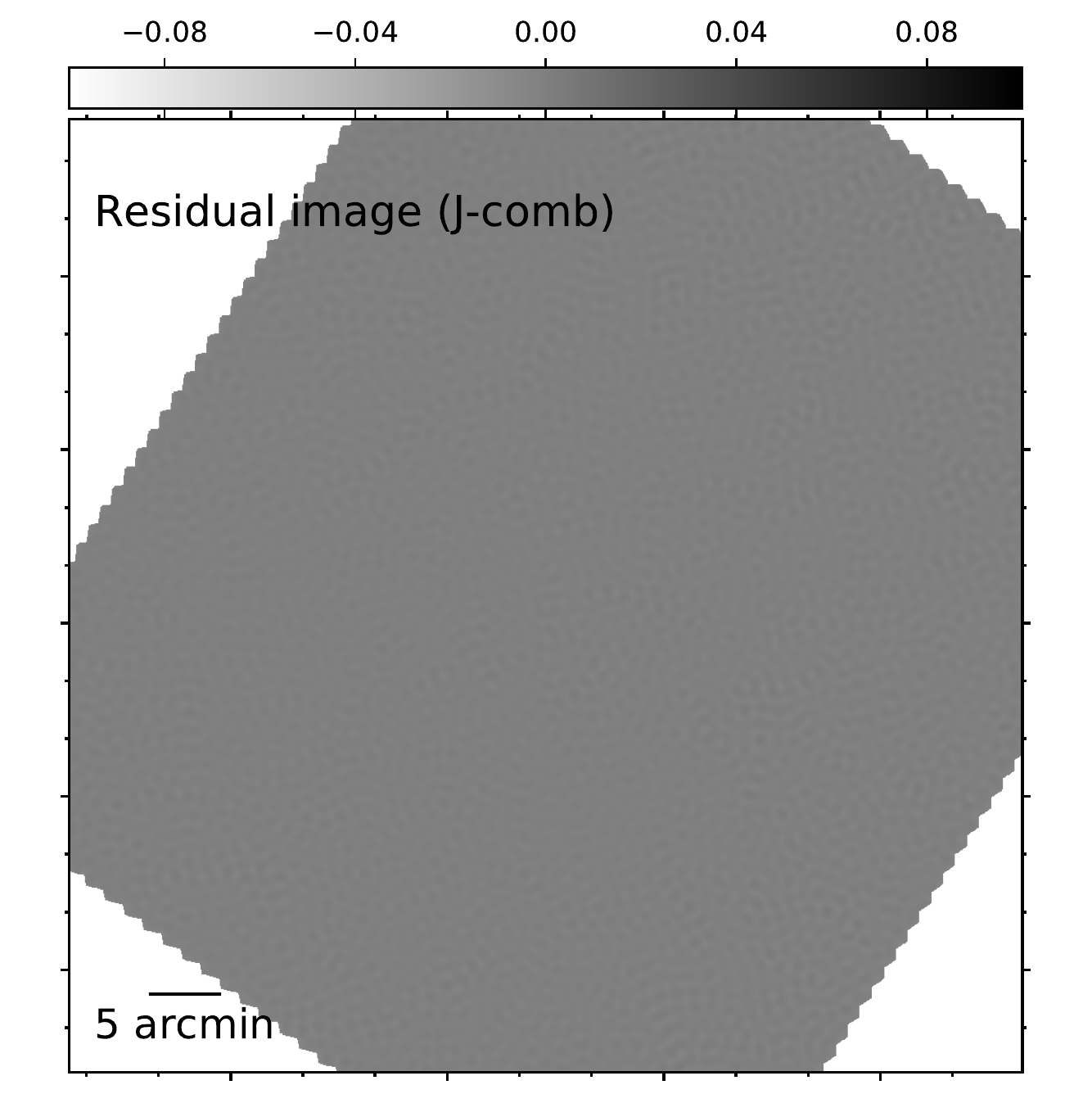} & 
\hspace{-2.2cm}\includegraphics[scale=0.46]{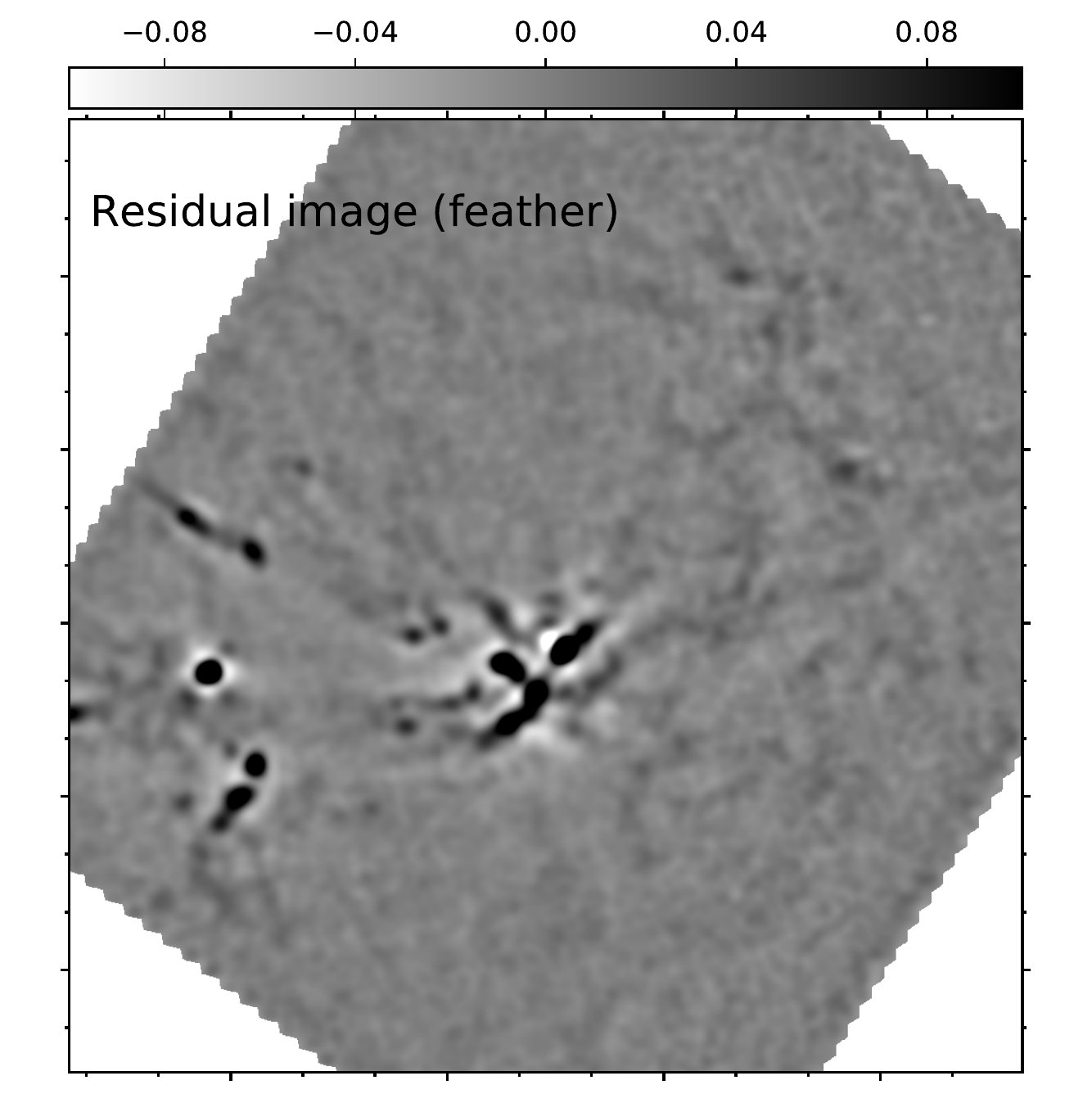} & 
\hspace{-1.1cm}\includegraphics[scale=0.46]{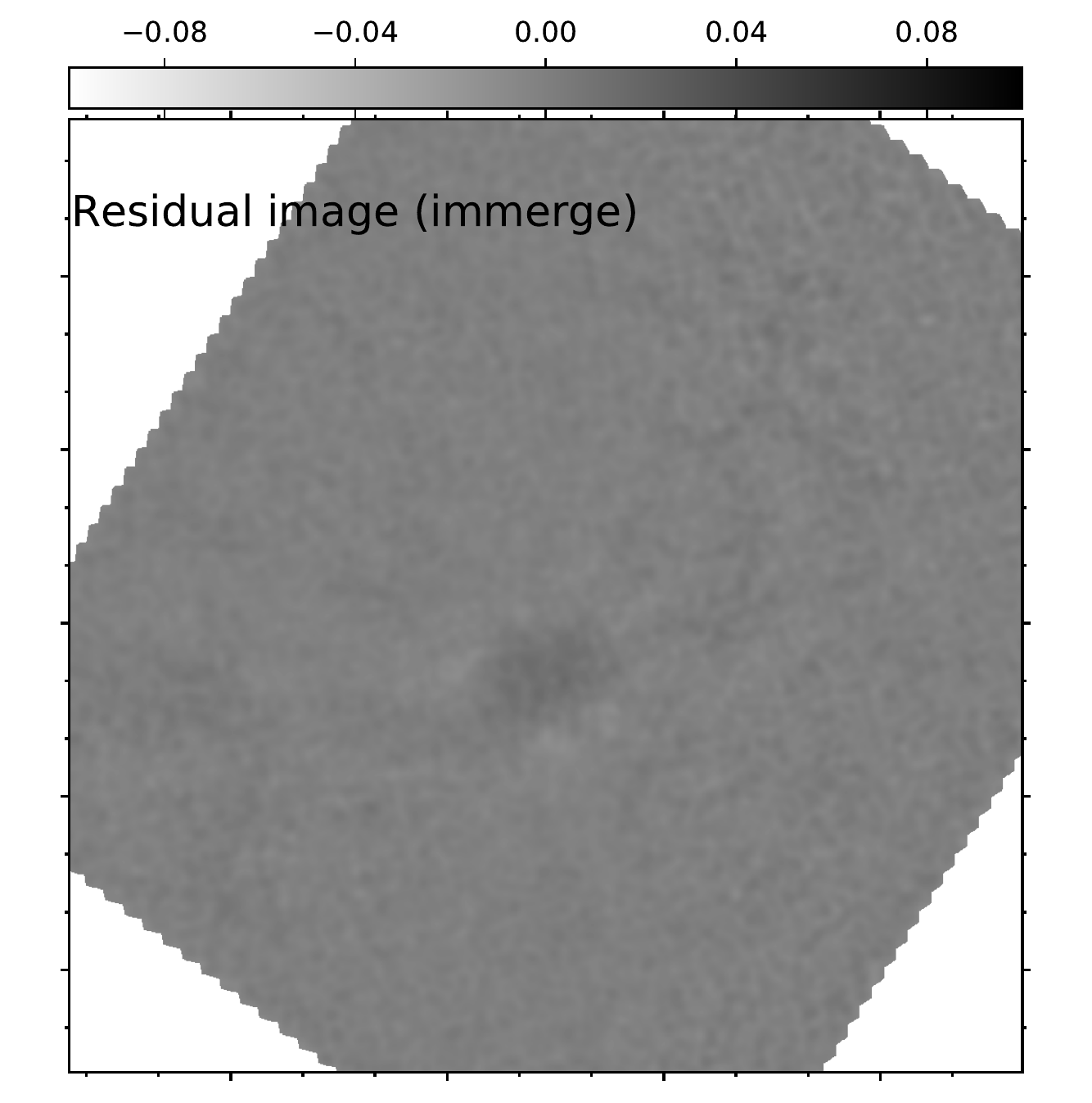} \\
\end{tabular}
\caption{
The combined images and the residual of the 50$''$ images. 
(Top panel) The combined images based on the J-comb algorithm, task immerge and task feather.
All images are presented using the same color scale with the unit pW$\times$10$^{-3}$ and the angular resolution is 14$''$.
(Bottom panel) The residual generated by subtracting the 50$''$ combined images from the 50$''$ model high resolution image.
}
\label{fig:2dcombinedresidual}
\end{figure*}

\begin{figure*}
\includegraphics[scale=0.6]{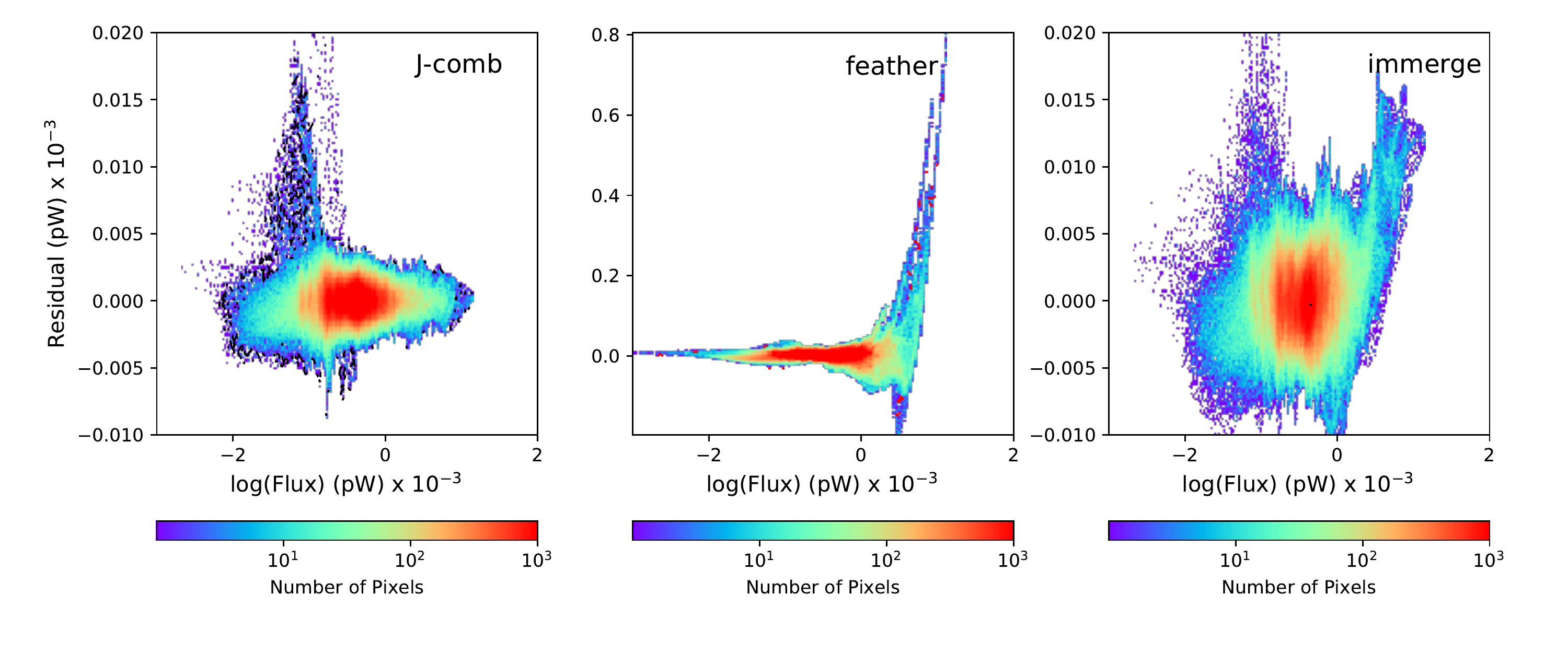}
\vspace{-1.1cm}
\caption{
Distributions of the pixel-by-pixel flux and residuals of the 50$''$ images at 850 $\mu$m Band.
Left panel shows the results of image combination based on the J-comb algorithm; middle panel shows the results of image combination based on task feather in the \textsc{CASA} software package; right panel shows the results of image combination based on task immerge in the \textsc{MIRIAD} software package.
}
\label{fig:pixel}
\end{figure*}

\begin{figure*}[!ht]
\hspace{1.7cm}
\vspace{-0.1cm}
\begin{tabular}{ p{0.5\linewidth} p{0.5\linewidth}}
\hspace{-0.79cm}
\includegraphics[scale=0.54]{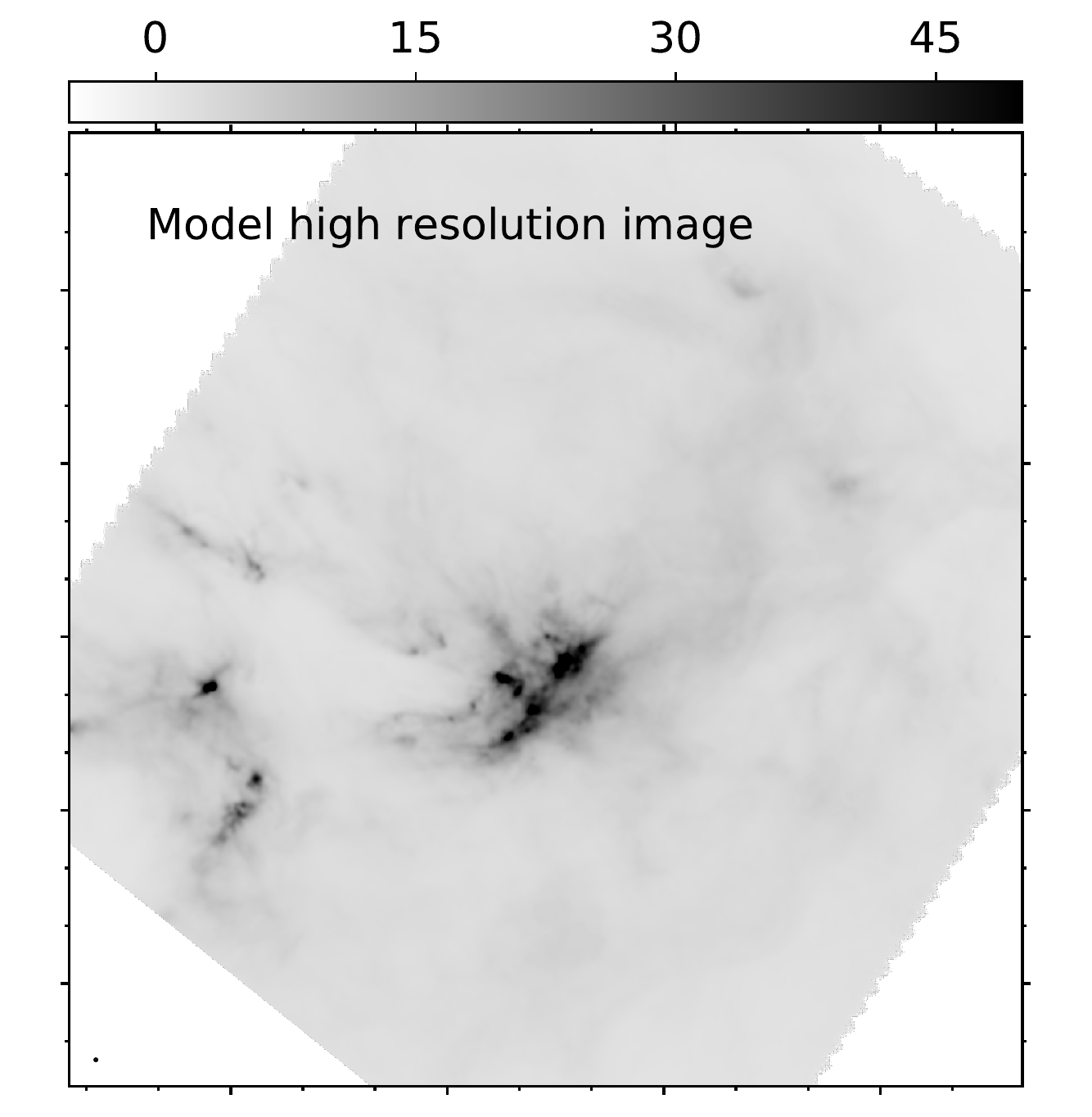} & 
\hspace{-2.2cm}
\includegraphics[scale=0.54]{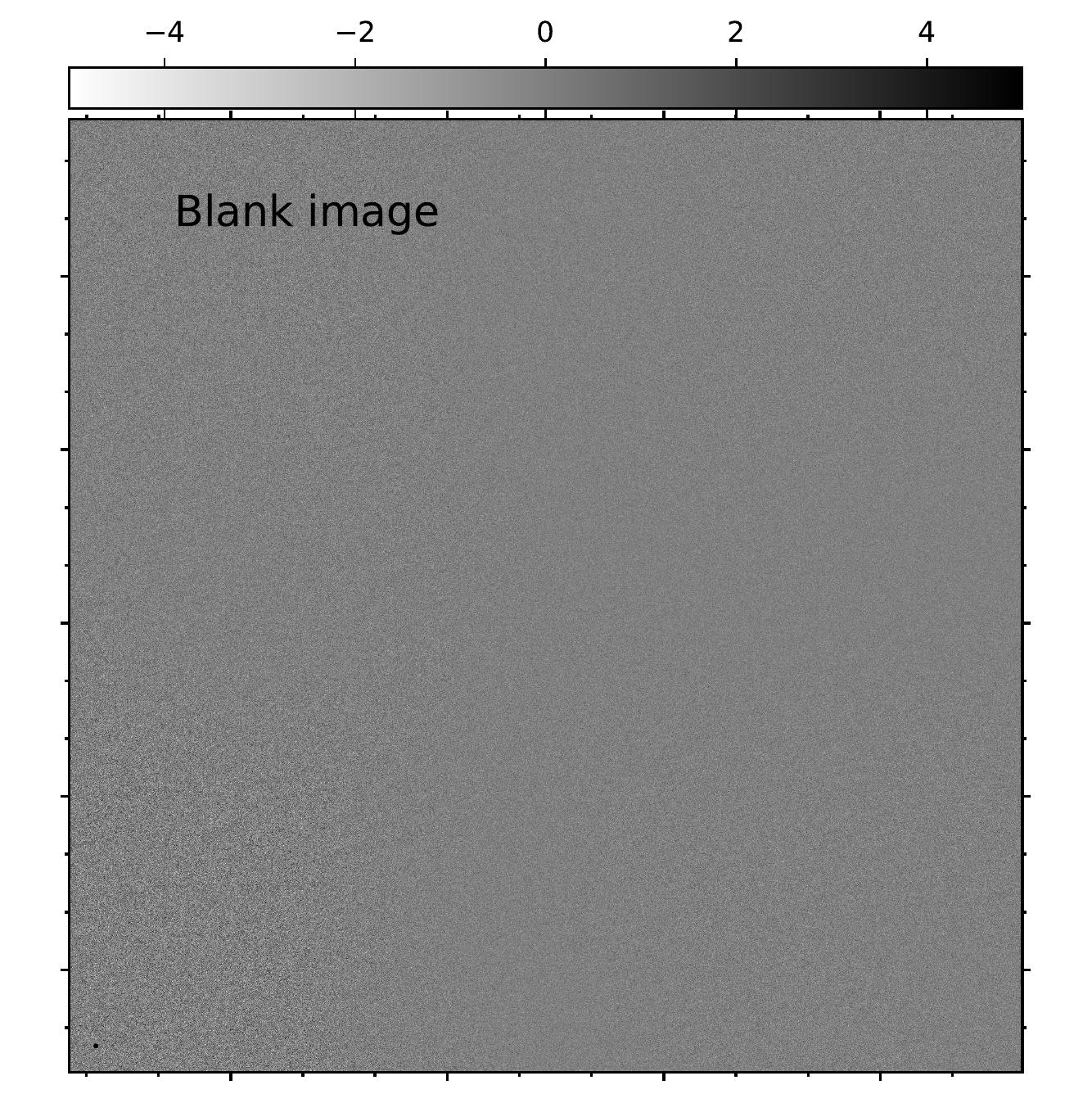} \\
\end{tabular}
\hspace{-1.7cm}
\vspace{-0.1 cm}
\begin{tabular}{ p{0.5\linewidth} p{0.5\linewidth}}
\hspace{-0.59cm}
\vspace{0.2 cm}
\includegraphics[scale=0.54]{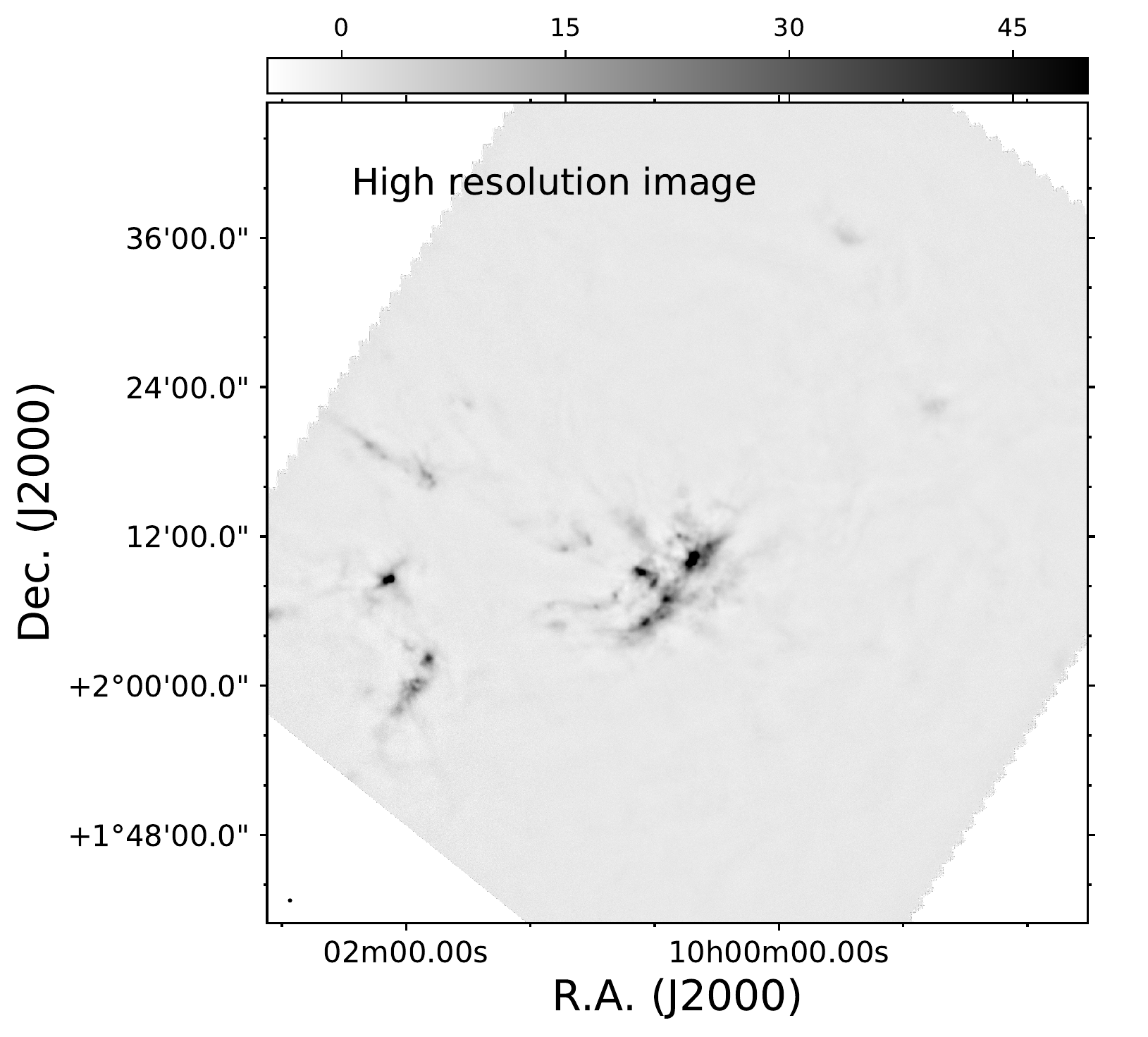} & 
\vspace{-8.15cm}
\hspace{-0.37cm}
\includegraphics[scale=0.54]{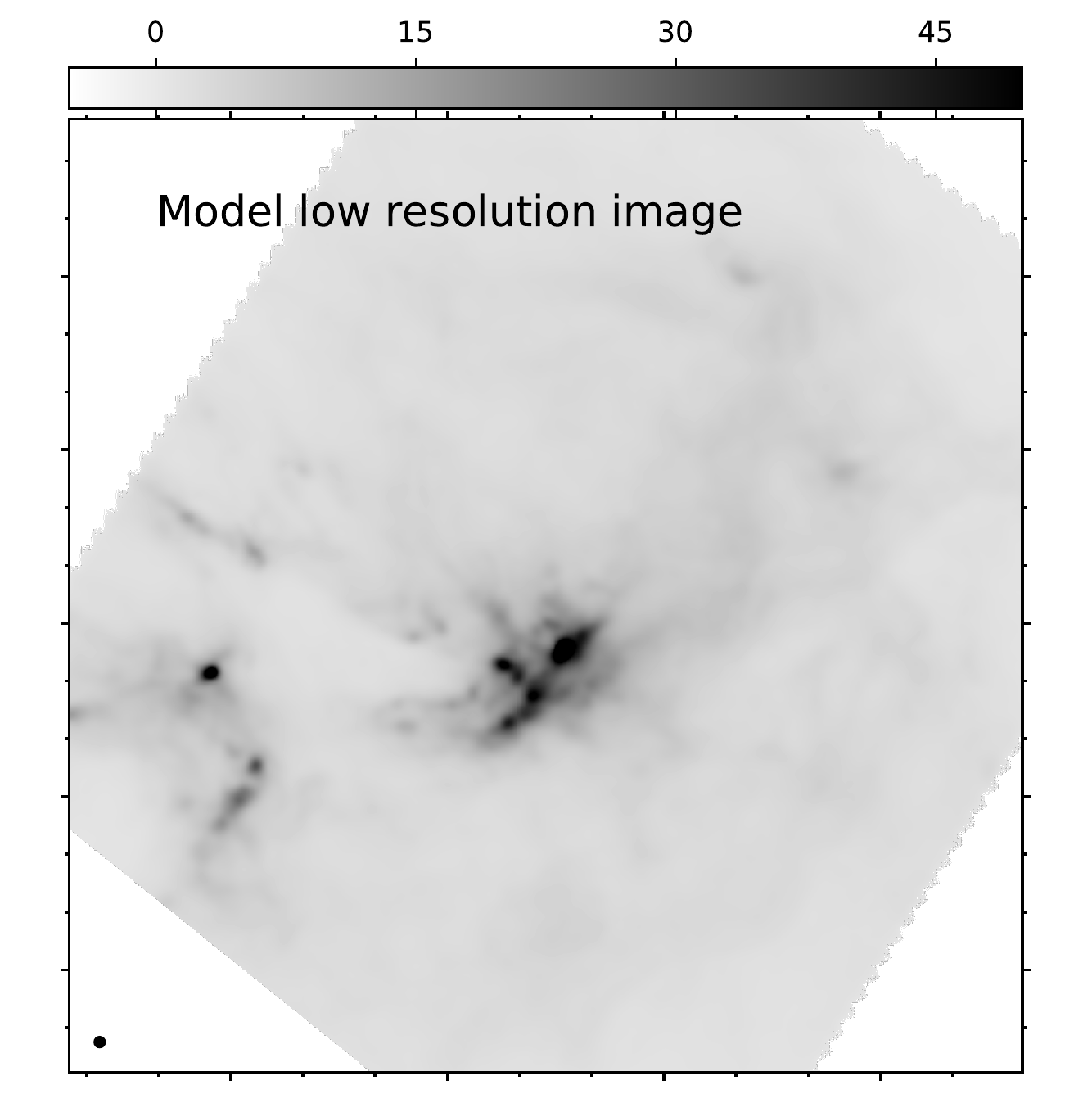}\\
\end{tabular}
\vspace{-0.25cm}
\caption{
Similar to Figure \ref{fig:input} but for 450 $\mu$m band. 
}
\label{fig:input_450}
\end{figure*}

With these functional forms, it can be seen that when $k$ > $k_{\mbox{\scriptsize threshold}}^{\mbox{\scriptsize high}}$, $C(\vec{k})$ = $G_{\mbox{\scriptsize high}}$.
In other words, the beam response function of the high-resolution image is preserved in the combined image at high spatial frequencies.
The values of $k_{\mbox{\scriptsize threshold}}^{\mbox{\scriptsize low}}$ and $k_{\mbox{\scriptsize threshold}}^{\mbox{\scriptsize high}}$ need to be chosen such that (1) only the low spatial frequency parts of the low-resolution images significantly contribute to the combined images, and (2) some artifacts due to the imperfect calibration or atmospheric subtraction procedures, scanning patterns, or any image defects caused by unidentified hardware/software issues are suppressed.
Because the artifacts in the high-resolution images (e.g., those taken with the ground-based telescopes) can have different characteristics, which usually also depend on the weather condition during the observations and the data reduction software.
For this moment, we optimized the choice of $k_{\mbox{\scriptsize threshold}}^{\mbox{\scriptsize low}}$ and $k_{\mbox{\scriptsize threshold}}^{\mbox{\scriptsize high}}$ in a quasi-empirical way following the procedure described below.
First, we based on cross-comparing the flux densities of the relatively isolated and spatially compact sources that were detected both in the low- and high-resolution images to match the flux scaling.
The flux re-scaling was done by multiplying the pixel values in the high-resolution image by a constant calibration factor.
Then, we identified the overlapping $k$ range from the power spectra of the low- and high-resolution images. 
If the input images are not overlapped in the $k$ space, the low-resolution images were deconvolved beforehand.
For the next step, in the ideal scenario, we can choose any $k_{\mbox{\scriptsize threshold}}$ in the range of $k_{\mbox{\scriptsize overlap}}^{\mbox{\scriptsize low}}$ and $k_{\mbox{\scriptsize overlap}}^{\mbox{\scriptsize high}}$. 
While in reality, it may require a series of trials to find the optimal $k_{\mbox{\scriptsize threshold}}$ such that the significant artifacts due to the imperfect calibration or atmospheric subtraction, scanning patterns, or any unidentified hardware/software issues are suppressed.
Finally, we can assess if the values of $k_{\mbox{\scriptsize threshold}}$ are optimal by visually inspecting the combined image and its power spectrum in comparison to the power spectra of the high angular resolution and low angular resolution images.
By adopting an optimal $k_{\mbox{\scriptsize threshold}}$, the combined image should not present any visually recognizable artifact defects.
And the combined image's power spectrum should not maintain the oscillations induced by these artifact patterns at the corresponding spatial frequencies.

Here, we use the combination of JCMT and deconvolved {\it Planck} data as an example.
The overlapped spatial scale of the input observations is $\sim$80$''$ to $\sim$200$''$. 
We should adopt the corresponding angular scales of  $k_{\mbox{\scriptsize threshold}}^{\mbox{\scriptsize low}}$ smaller than 200$''$ to satisfy $k_{\mbox{\scriptsize overlap}}^{\mbox{\scriptsize low}}$ < $k_{\mbox{\scriptsize threshold}}^{\mbox{\scriptsize low}}$.
The atmospheric subtractions can induce ripple or stripe patterns in the high-resolution maps. 
In our example, the scale of the ripple patterns is $\sim$ 120$''$.
These ripple patterns will produce oscillations at the corresponding spatial frequencies in the power spectrum.
We can then use a $k_{\mbox{\scriptsize threshold}}$ that ranges from 80$''$ to 100$''$ to remove the 120$''$ artifacts in the high-resolution image.

To suppress the artificial high-order oscillations in the combined image caused by the inconsistent absolute flux calibration of the high and low-resolution images, we use the normalized Butterworth filter \cite{Butterworth1930} to make the weighting function of low-resolution data drop from 1.0 to 0.0 smoothly.
It has a form of 
\begin{equation}
f(k) = \frac {1}{1-(\frac{k_{0}}{k})^{2n}},
\end{equation}
where n is the number of poles in the filter, ${k_{0}}$ is the mean value of $k_{\mbox{\scriptsize threshold}}^{\mbox{\scriptsize low}}$ and $k_{\mbox{\scriptsize threshold}}^{\mbox{\scriptsize high}}$.
The value of n is chosen based on the selected spatial frequency threshold and the input images' absolute flux level.
$f(k)$ turns into a step function when using large numbers of poles.
If the absolute flux calibrations for the high- and low-resolution images are consistent, it is sufficient to simply adopt $f(k)$ as a step function (e.g., the top left panel of Figure \ref{fig:1d_combine}).
The bottom left panel of Figure \ref{fig:1d_combine} shows the combination results and residual data of our new method.
The combined data recover the model data well. 
And the residual, generated by subtracting the high-resolution data from the combined data, is around 10$^{-15}$, which are merely numeric errors.

When the absolute flux of the high-resolution image matches that of the low-resolution image, with the J-comb algorithm, combining low and high-resolution data will obtain high-resolution data regardless of the adopted mathematical form of $f(k)$.
However, when the absolute flux of the high-resolution image is different from that of the low-resolution image, rapid changing of weighting function can result in considerable ripples in the combined image. 
For example, we scaled the flux of high-resolution data by a factor of 1.1, and then combined it with the low-resolution data utilizing a step function of $f(k)$ (see left panel of Figure \ref{fig:sub_err}) and Butterworth filter (see right panel of Figure \ref{fig:sub_err}).
The smoothly changing weighting function can suppress the ripples in the residuals, which affect extended spatial scales.

\section{Benchmark the J-comb algorithm }\label{sub:benchmark}

The ultimate goal of this development is to enable accurately deriving dust temperature and column density distributions in star-forming molecular clouds by combining bolometric observations taken from ground-based observatories and space telescopes and assessing the precision of these derivations.
To benchmark our development, we utilized the archival {\it Herschel}\footnote{{\it Herschel} is an ESA space observatory with science instruments provided by European-led Principal Investigator consortia and with important participation from NASA.} observations towards the Ophiuchus molecular cloud (d$\sim$135 pc \cite{Ortiz-Leon20017}) to construct a realistic model of dust temperature and column density distributions.
We scaled this model to represent more distant molecular clouds by scaling the pixel sizes.
We then carried out mock  observations mimicking the ground-based and space telescopes observations at (sub)millimeter and far-infrared wavelengths based on this model.
Finally, we based on these simulated observations at JCMT-SCUBA2 observing bands (450 $\mu$m and 850 $\mu$m) to test and compare the performance of the J-comb, \textsc{CASA}-feather, and \textsc{MIRIAD}-immerge algorithms.
An overall flowchart of our benchmark procedure is given in Figure \ref{fig:flowchart_benchmarking}.

In Section \ref{subsub:model} we introduce how the realistic model of temperature and dust column density distribution were generated.
In Section \ref{subsub:2d} we introduce how we simulate the ground-based and space telescopes observations.
In Section \ref{subsub:2dcombine} we benchmark the performance of the J-comb, \textsc{CASA}-feather, and \textsc{MIRIAD}-immerge algorithms.
We demonstrate that the performance of J-comb is improved over the other algorithms for our specific purpose of combining bolometric observations from ground-based and space telescopes, and therefore we adopt it in our combination algorithm for further analyses.
In Section \ref{subsub:sed} we quantify the bias of the dust column density map derived from the SED fittings.

\subsection{Constructing a model of dust temperature and column density distributions}\label{subsub:model}

The Ophiuchus Molecular Cloud is a well-studied star-forming cloud located at $\sim$135 pc from the Sun \cite{Ortiz-Leon20017}.
We retrieved the level 2.5 processed, archival {\it Herschel} images at 70/160 $\mu$m taken by the PACS instrument \cite{Poglitsch2010} and at 250/350 $\mu$m taken by the SPIRE instrument \cite{Griffin2010} (obsID: 1342227149, 1342205094).
Since we are interested in extended structures, we adopt the extended emission products, which have been absolute zero-point corrected based on the images taken by the {\it Planck} Space Observatory.

We first convolved the data to the same resolution of the 350 $\mu$m map (the largest FWHM, 25$''$). 
Based on the modified blackbody SED fitting to these 25$''$ resolution images, adopting a uniform dust opacity index ($\beta$) of 1.8, we obtained the dust temperature ($T_{\mbox{\scriptsize d}}^{\mbox{\scriptsize actual}}$) and dust column density ($N_{\mbox{\scriptsize d}}^{\mbox{\scriptsize actual}}$) maps of the Ophiuchus molecular cloud.
By scaling the column density map and temperature map to the distance of the Orion Nebula Cluster (d$\sim$388 pc \cite{Kounkel2017}), we obtained the modeled $H_{2}$ column density ($N_{\mbox{\scriptsize d}}$) and dust temperature ($T_{\mbox{\scriptsize d}}$) maps.

We assume that the model $T_{\mbox{\scriptsize d}}$, $N_{\mbox{\scriptsize d}}$ images are reflecting the properties of a molecular cloud unbiasedly.
From the $T_{\mbox{\scriptsize d}}$ and $N_{\mbox{\scriptsize d}}$ images, we can generate the intensity distribution ($I_{\nu}$) at various observing frequencies $\nu$ with different angular resolutions.
In the intensity distribution maps, we assume the value of each pixel reflects the real flux at this position precisely.
To mimic the actual observations, when we simulate a specific observation, we re-grid the intensity distribution maps to have the same pixel size as that of the real ground-based and space telescopes observations.
The procedure for obtaining the input simulated data for the following benchmark of linear data combination is summarized in the blue blocks of Figure \ref{fig:flowchart_benchmarking}.

\begin{tablehere}
\caption{Observational Parameters of the Multi-wavelength Data}\label{tab:obs_parameters}
\begin{center} \doublerulesep 0.2pt \tabcolsep 4.2 pt
\begin{tabular}{p{2.cm}p{2.cm}p{1.7cm}p{1.6cm}}
\hline
$\lambda$ ($\mu$m)   & Beam FWHM  & Pixel Size & Flux Unit \\
Camera               & (arcsec)   & (arcsec) & \\
\hline
70/PACS & 5.8 $\times$ 12.1    &  3.2 & Jy ${\mbox{\scriptsize pixel}}^{{\mbox{\scriptsize -1}}}$ \\
160/PACS & 11.4 $\times$ 13.4    &  3.2 & Jy ${\mbox{\scriptsize pixel}}^{{\mbox{\scriptsize -1}}}$ \\
250/SPIRE & 18.1    &  6.0 & MJy ${\mbox{\scriptsize Sr}}^{{\mbox{\scriptsize -1}}}$\\
350/SPIRE & 25.2    &  9.7 & MJy ${\mbox{\scriptsize Sr}}^{{\mbox{\scriptsize -1}}}$\\
500/SPIRE & 36.9    &  14.0 & MJy ${\mbox{\scriptsize Sr}}^{{\mbox{\scriptsize -1}}}$\\
450/SCUBA2 & 8.0    &  2.0 & mJy ${\mbox{\scriptsize arcsec}}^{{\mbox{\scriptsize -2}}}$\\
850/SCUBA2 & 14.0    &  4.0 & mJy ${\mbox{\scriptsize arcsec}}^{{\mbox{\scriptsize -2}}}$\\
850/{\it Planck} & 279.0    &  60.0 &  ${\mbox{\scriptsize K}}_{{\mbox{\scriptsize cmb}}}$\\
\hline
\end{tabular}
\end{center}
\end{tablehere}

\subsection{Simulating ground-based and space telescopes observations}\label{subsub:2d}

To simulate the space telescopes images, we smooth the intensity distribution image ($I_{\nu}$) to have the same angular resolution of the space telescope observations (i.e., 36.9$''$ for {\it Herschel}-SPIRE 500 $\mu$m observations and 279$''$ for {\it Planck} 850 $\mu$m observations). 
Observational parameters of the multi-wavelength data are listed in Table \ref{tab:obs_parameters}.

For the ground-based telescopes' observations, we smooth $I_{\nu}$ to have the same resolution of the ground-based telescopes' observations, which is 8$''$ for JCMT-SCUBA2 450 $\mu$m observations and 14$''$ for JCMT-SCUBA2 850 $\mu$m observations.
On extended angular scales, the missing flux of the ground-based telescopes' observations is complex, related to the instantaneous field-of-view of the instrument, the selected filtering threshold during the data reduction, and the condition of atmospheric emission when the observation is taken.
To see realistically how the extended scale structures are filtered and to gauge how far we can recover them by combining with the space telescope observations, we mimic the actual ground-based observations by adding $I_{\nu}$$\ast$$B_{\mbox{\scriptsize ground}}$ to the raw time stream of the real ground-based telescope observations.
We insert $I_{\nu}$$\ast$$B_{\mbox{\scriptsize ground}}$ as the model into a pure noise field via {\tt fakemap} parameter in {\tt makemap}  (see the lower-left panel in Figure \ref{fig:input})\footnote{A documentation of {\tt fakemap} can be found at http://pipelinesandarchives.blogspot.co.uk/search?q=fakemap}. 
In this manner, $I_{\nu}$$\ast$$B_{\mbox{\scriptsize ground}}$ is added to the raw time stream of the data and is subject to the common JCMT-SCUBA2 data reduction steps.
We retrieve the available nightly observations of JCMT-SCUBA2 at 450 and 850 $\mu$m from the online data archive (Program ID: M16AL002, JCMT large project S2COMSOS) as an approximate blank field image (see the upper right panel in Figure \ref{fig:input}).
S2COMSOS aims to create the deepest, wide-field image at 850 $\mu$m in the key extragalactic survey field, COSMOS \cite{Casey2013}.
The COSMOS field is a blank region but with submillimetre sources.
These submillimetre sources have a flux of $\sim$ 5 mJy at 850 $\mu$m band, which is much lower than the flux of nearby molecular cloud dense region (e.g., OMC-2, 1 Jy beam$^{-1}$ from SCUBA2 850 $\mu$m observation).
Hence, the S2COMSOS observations can serve as blank images of the pure noise fields for our simulation.

The procedure for simulating the input images for linear images combination is summarized in the red blocks of Figure \ref{fig:flowchart_benchmarking}.

\subsection{Linear images combination for two-dimensional case}\label{subsub:2dcombine}

\begin{table*} 
\begin{center}
\caption{Parameters of $W^{\mbox{\scriptsize low}}$ used on the J-comb combination. }\label{tab:parameters}
\begin{tabular}{p{1.8cm}p{1.8cm}p{1.8cm}p{1.8cm}p{1.8cm}p{1.8cm}}\hline\hline
$\lambda$ & $k_{\mbox{\scriptsize threshold}}^{\mbox{\scriptsize low}}$ & $\theta_{\mbox{\scriptsize threshold}}^{\mbox{\scriptsize low}}$ & $k_{\mbox{\scriptsize threshold}}^{\mbox{\scriptsize high}}$ &  $\theta_{\mbox{\scriptsize threshold}}^{\mbox{\scriptsize high}}$ & n\\
($\mu$m) & ($\mbox{\scriptsize pixel}^{\mbox{\scriptsize -1}}$)  & (arcsec) &($\mbox{\scriptsize pixel}^{\mbox{\scriptsize -1}}$) & (arcsec) & \\
\hline
450    & 0.017 & 120 & 0.020 & 100 & 5\\
850    & 0.033 & 120 & 0.040 & 100 & 6\\
\hline
\end{tabular}
\end{center}
\end{table*}

\begin{figure*}
\hspace{-1cm}
\begin{tabular}{p{0.5\linewidth}p{0.5\linewidth}}
\includegraphics[scale=0.38]{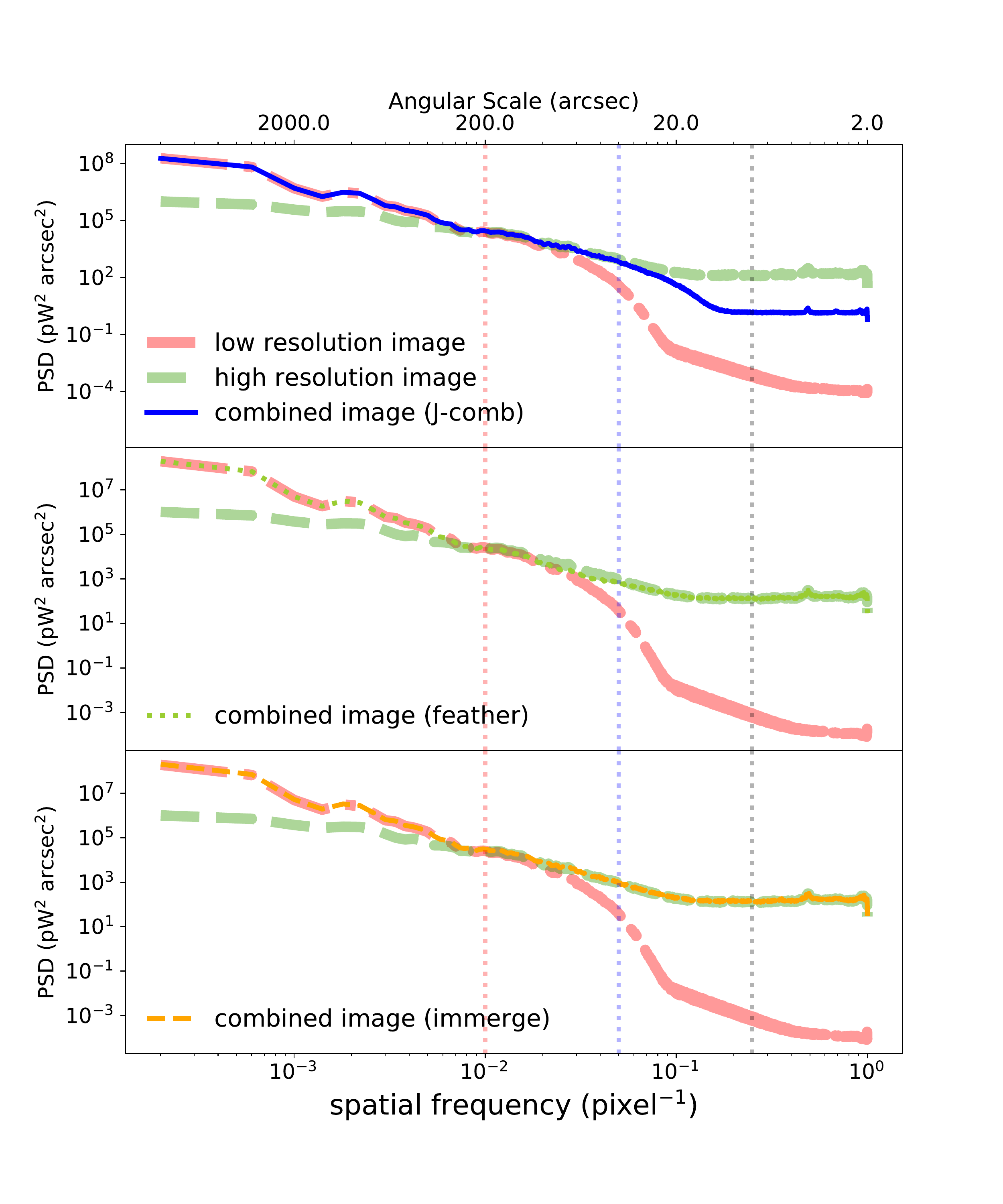}& \includegraphics[scale=0.38]{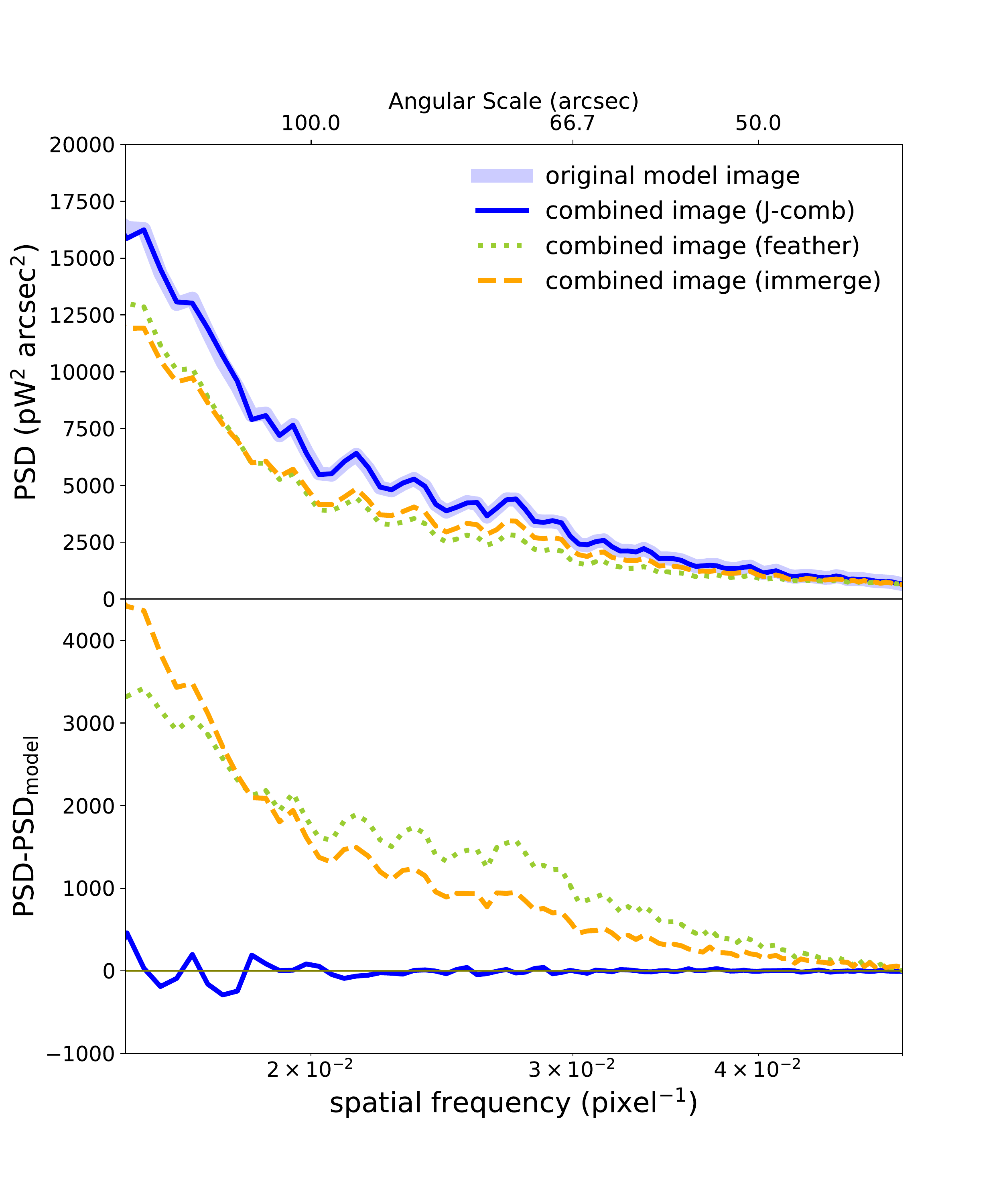} 
\end{tabular}
\vspace{-0.4cm}
\hspace{-1.55cm}
\caption{
Similar to Figure \ref{fig:2dpsd} but for 450 $\mu$m band.}
\label{fig:2dpsd_450}
\end{figure*}

\begin{figure*}
\begin{tabular}{ p{0.3\linewidth} p{0.3\linewidth} p{0.3\linewidth} }
\includegraphics[scale=0.43]{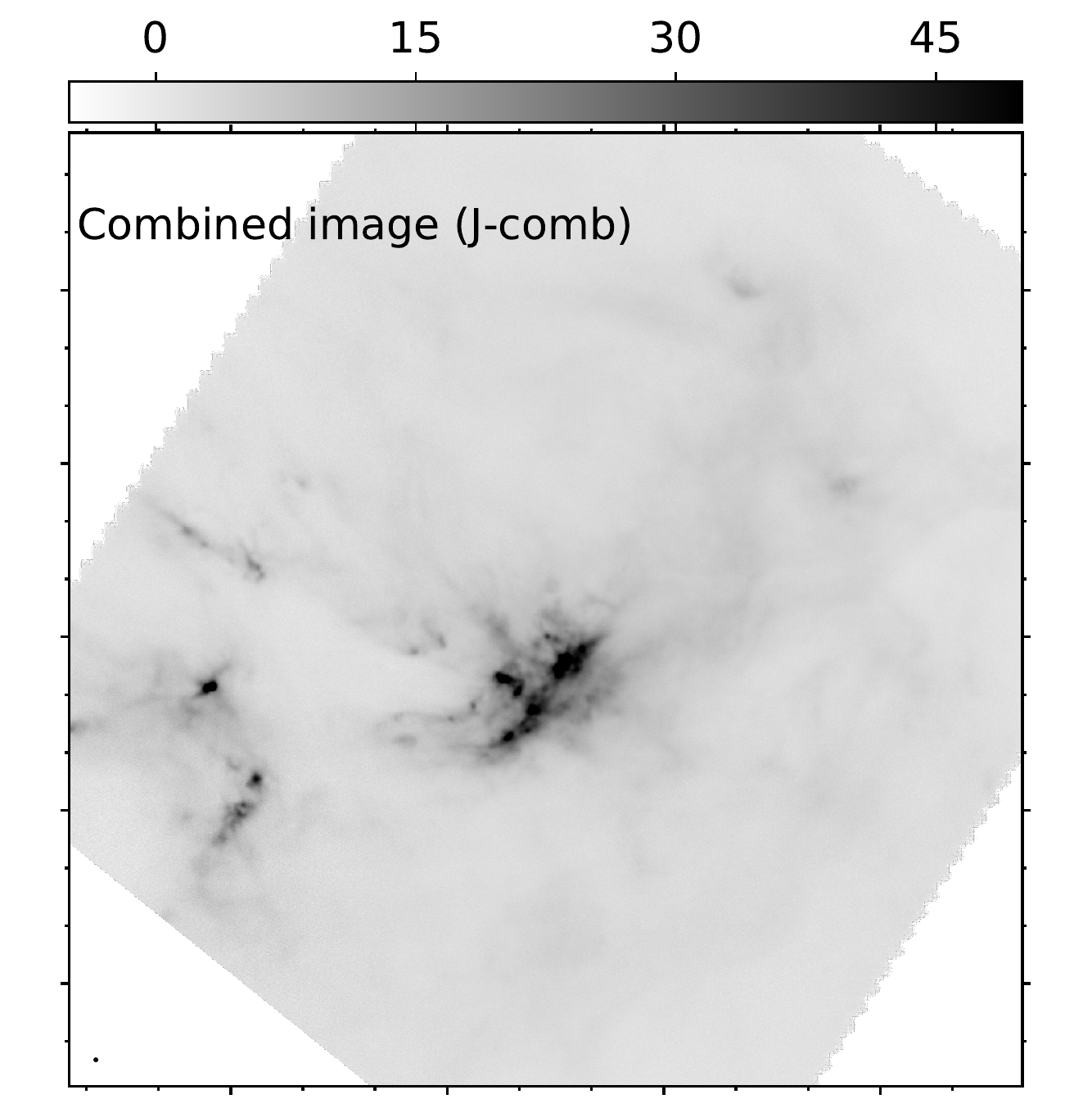} & 
\includegraphics[scale=0.43]{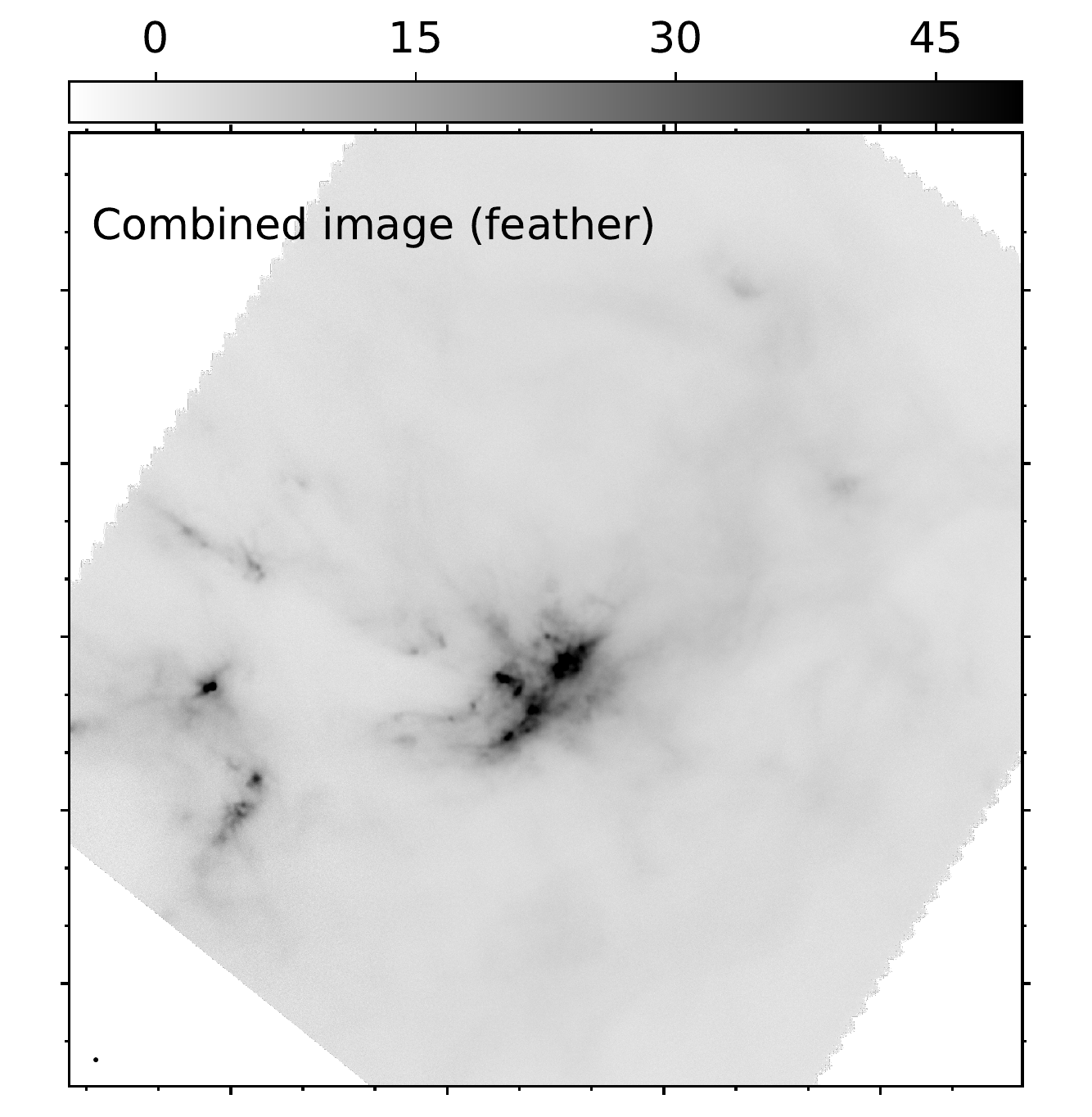} & 
\includegraphics[scale=0.43]{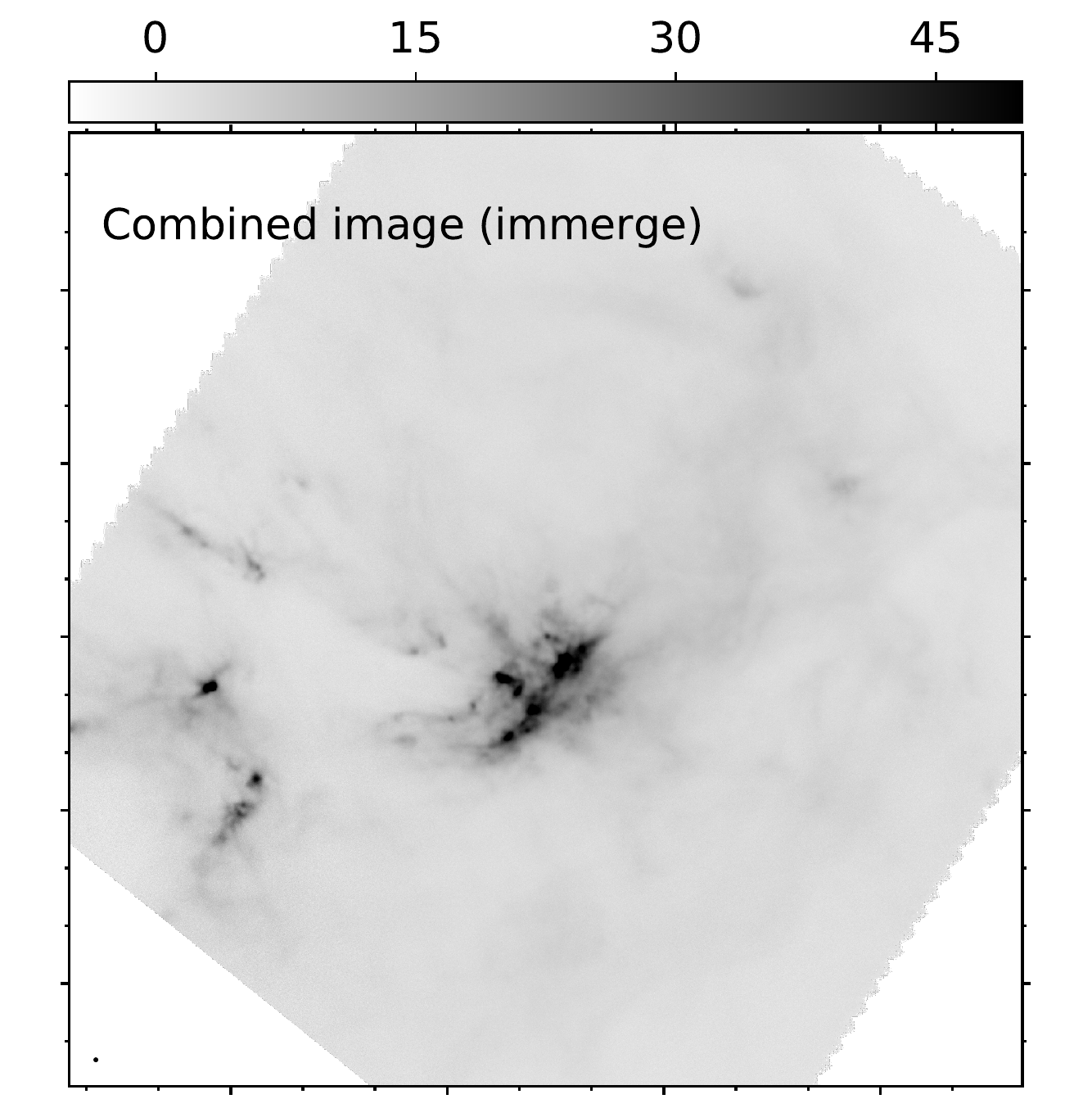} \\
\end{tabular}

\begin{tabular}{ p{0.3\linewidth} p{0.3\linewidth} p{0.3\linewidth} }
\includegraphics[scale=0.43]{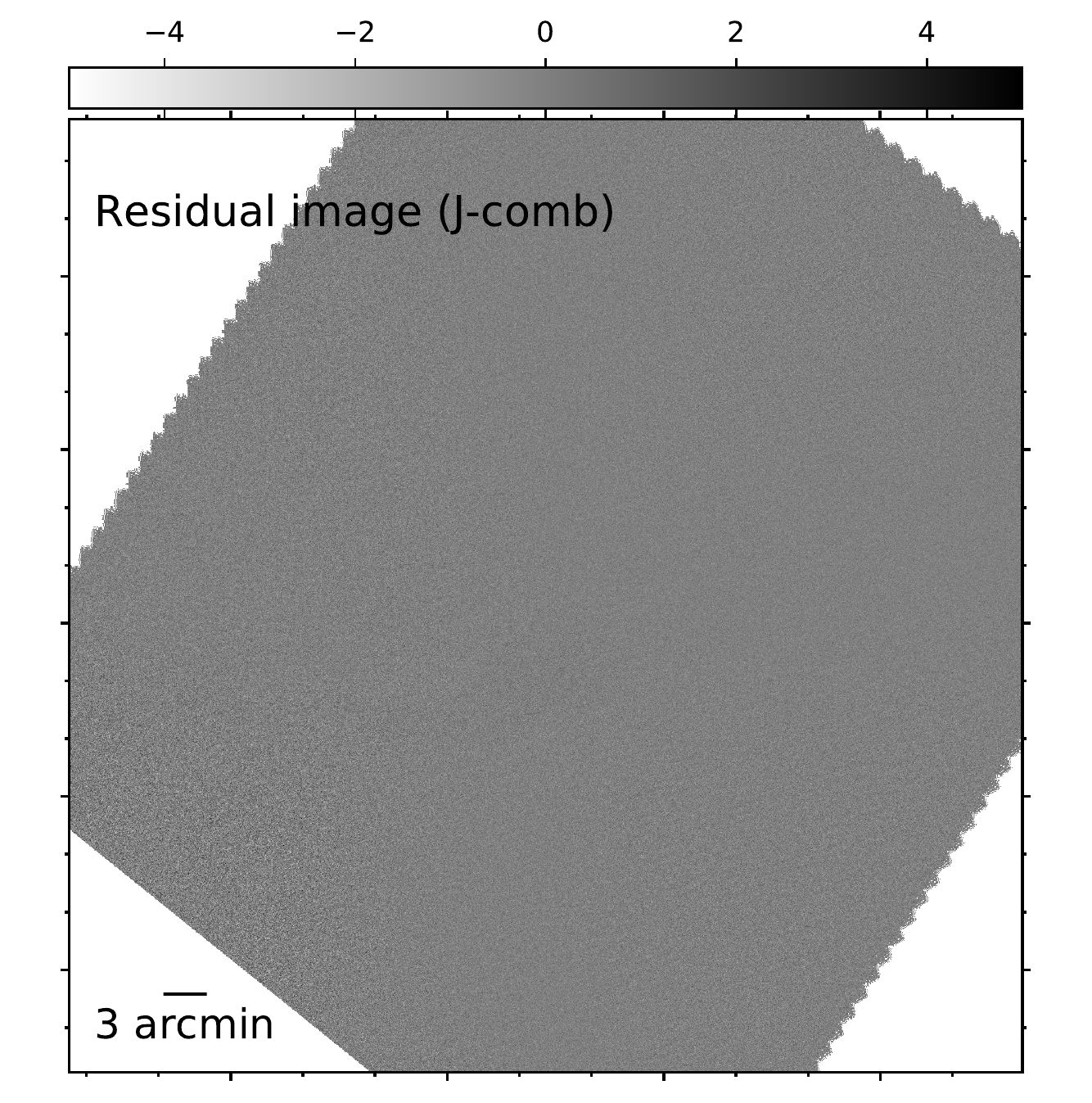} & 
\includegraphics[scale=0.43]{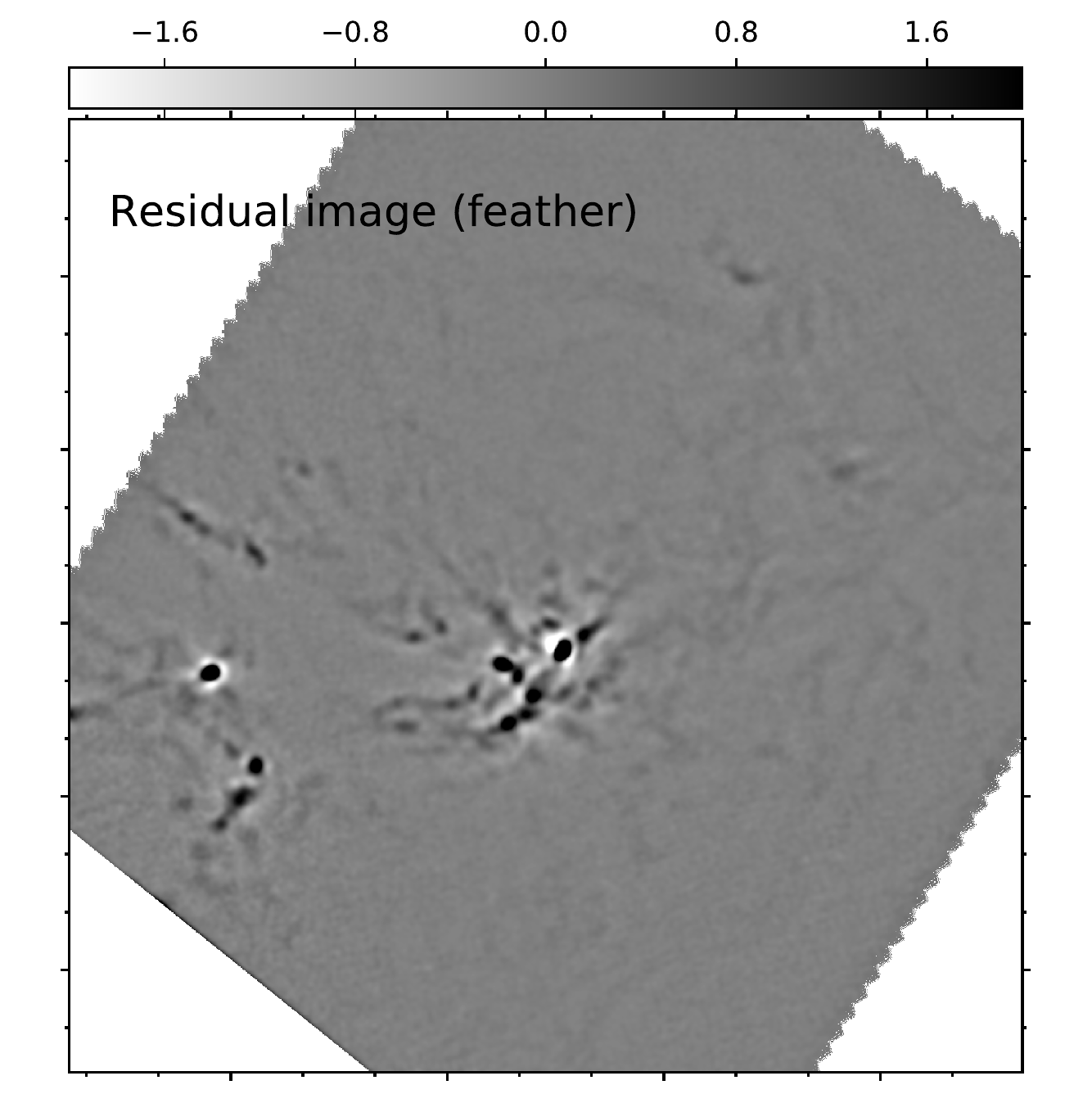} & 
\includegraphics[scale=0.43]{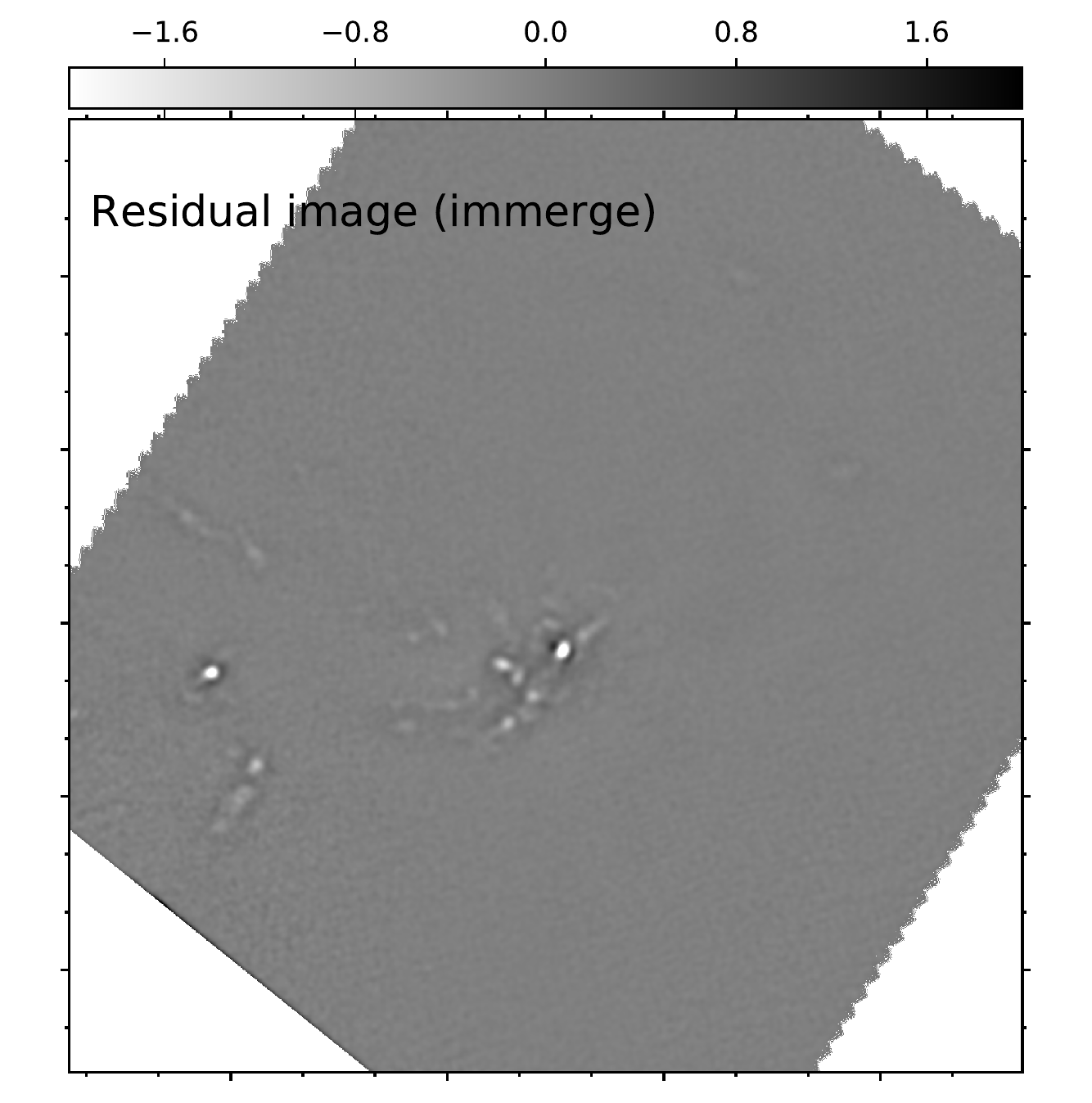}\\
\end{tabular}
\vspace{-0.2cm}
\caption{
Similar to Figure \ref{fig:2dcombinedresidual} but for 450 $\mu$m band. 
}
\label{fig:2dcombinedresidual_450}

\end{figure*}

\begin{figure*}
\vspace{-0.5cm}
\includegraphics[scale=0.55]{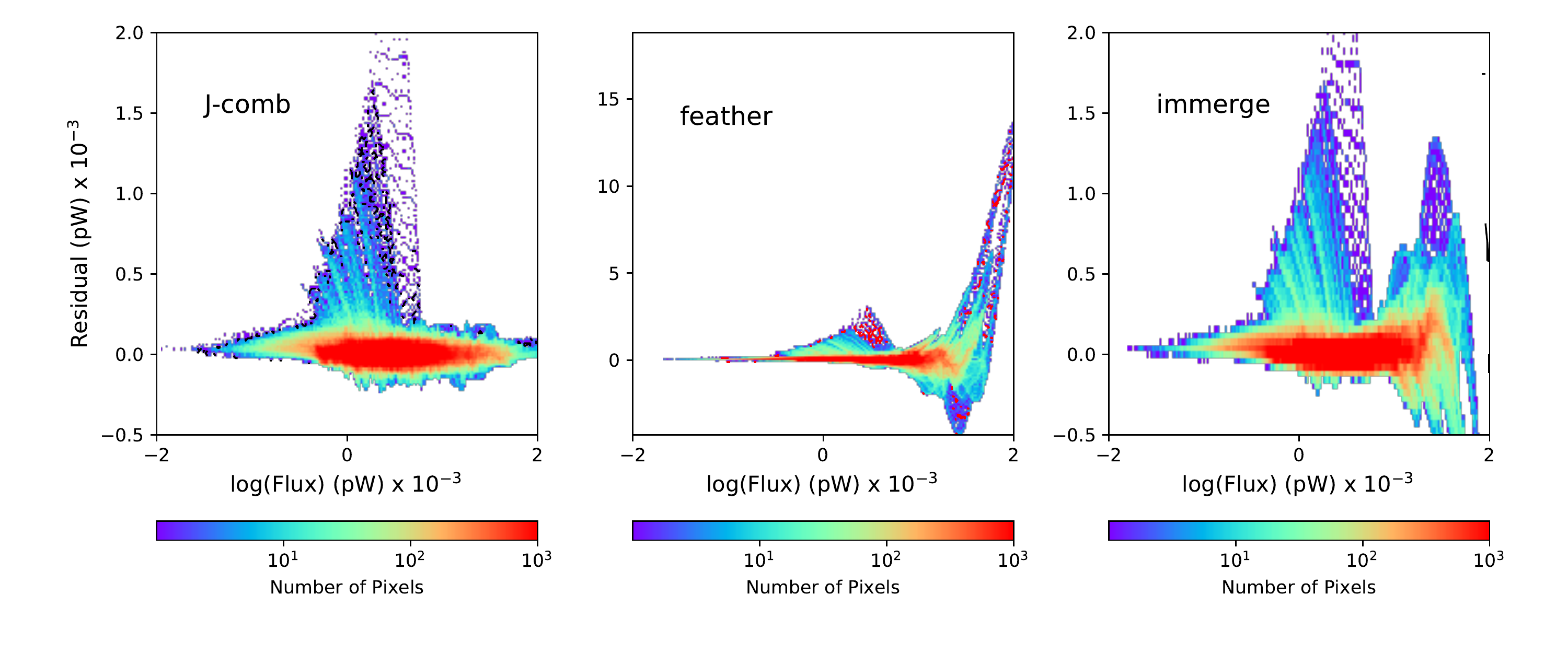}
\vspace{-0.6cm}
\caption{
Similar to Figure \ref{fig:pixel} but for 450 $\mu$m band.
}
\label{fig:pixel_450}
\end{figure*}

\begin{figure*}
\hspace{2.7cm}
\vspace{-0.1 cm}
\begin{tabular}{ p{0.26\linewidth} p{0.26\linewidth} p{0.26\linewidth} }
\hspace{-2.9cm} \includegraphics[scale=0.29]{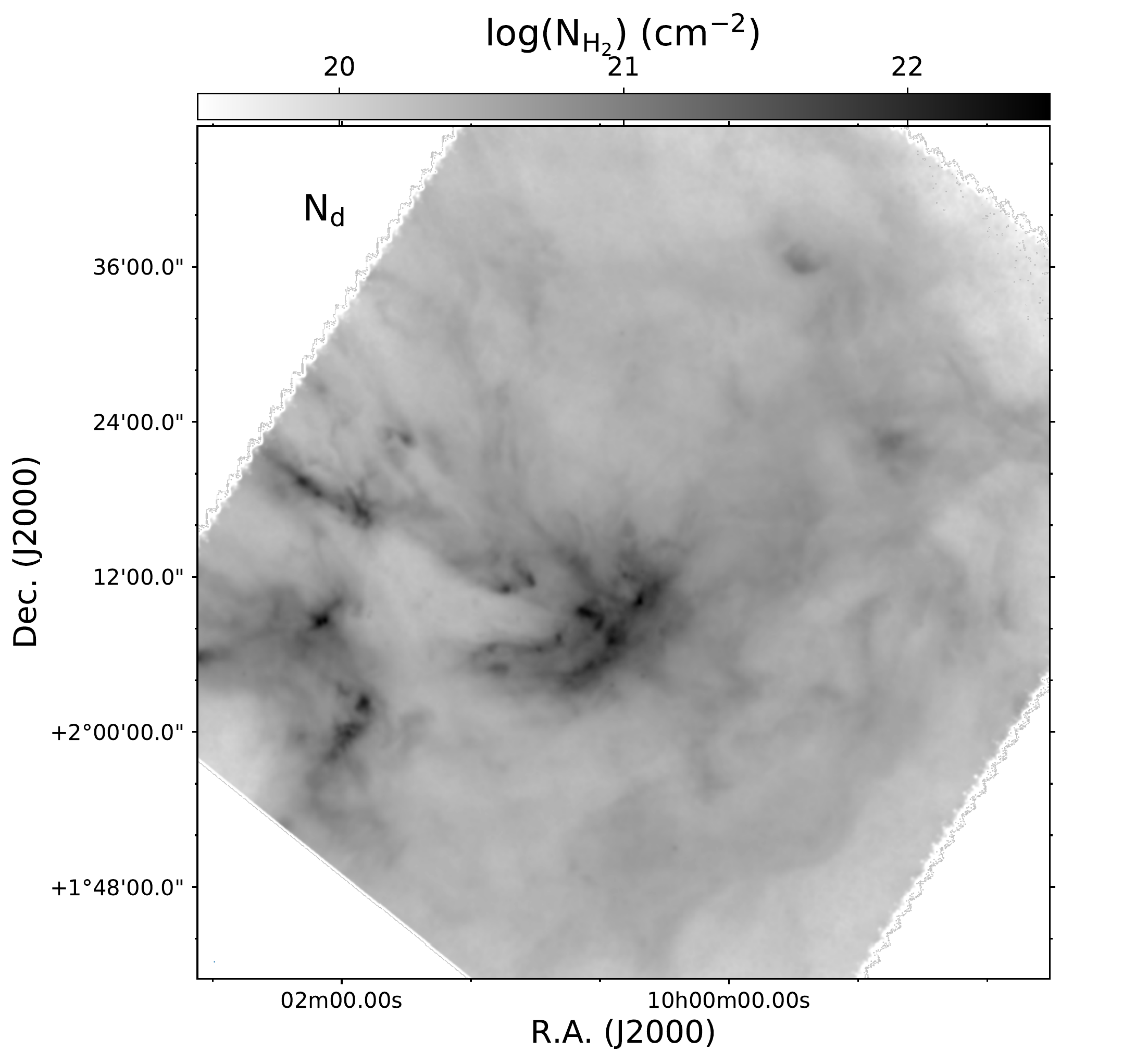} & \vspace{-6.cm} 
\hspace{-1.7cm}
\includegraphics[scale=0.29]{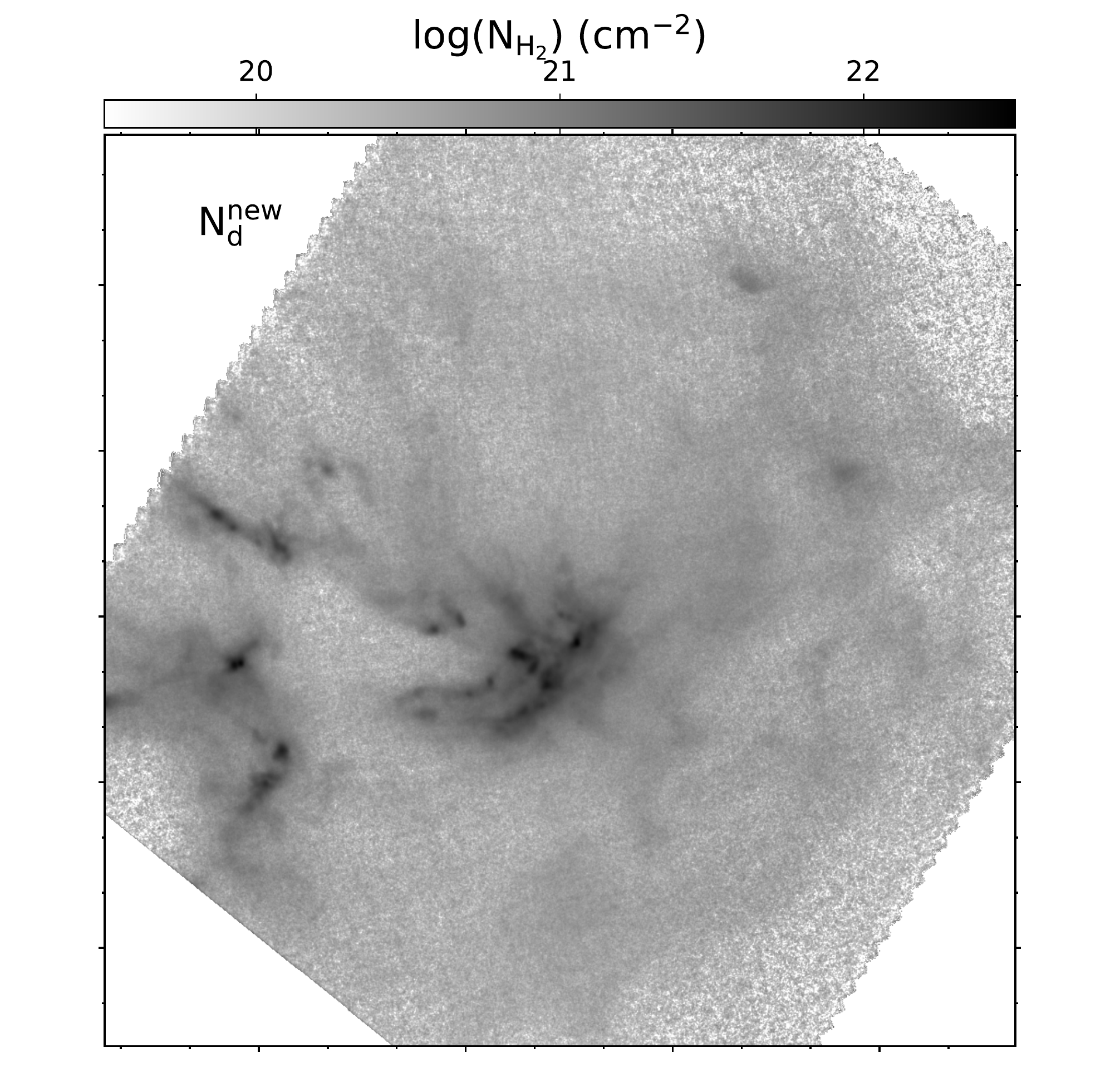} & 
\vspace{-5.7cm} 
\hspace{-1.1cm}
\includegraphics[scale=0.29]{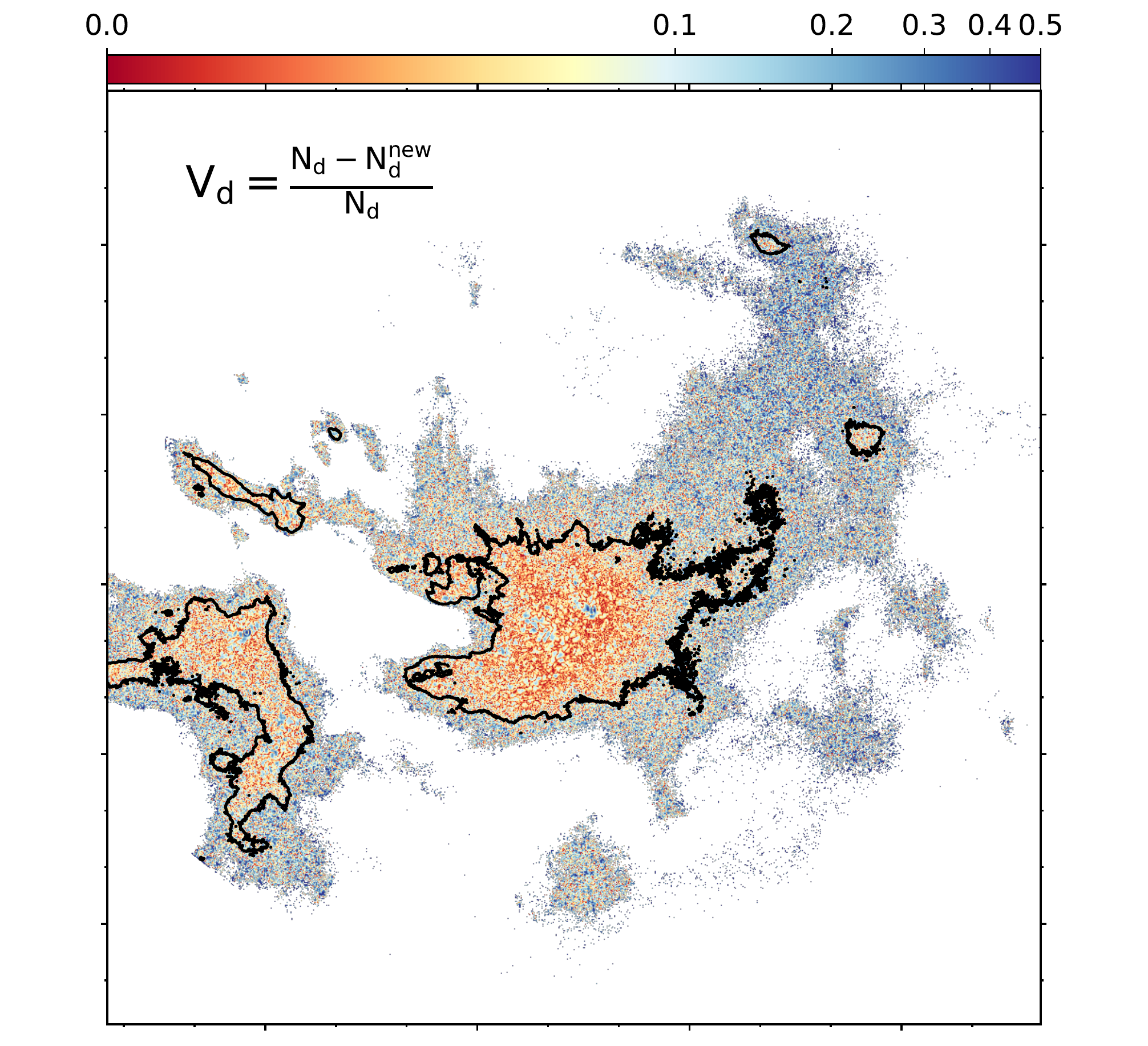}\\
\end{tabular}
\vspace{-0.3cm}
\caption{
Comparison between the model $H_{2}$ column density ($N_{\mbox{\scriptsize d}}$) map (left panel) and the column density (${N}_{\mbox{\scriptsize d}}^{{\mbox{\scriptsize new}}}$) map (middle panel) from the SED fitting based on 450 and 850 $\mu$m combined image.
The right panel shows $V_{\mbox{\scriptsize N}}$ image, which is calculated with Equation \ref{eq:residual}.
The black contour indicate the 5$\sigma$ detection in the combined image.
}
\label{fig:N}
\end{figure*}

We benchmark the performance of the J-comb, \textsc{CASA}-feather, and \textsc{MIRIAD}-immerge algorithms based on the simulated JCMT-SCUBA2 450/850 $\mu$m observations.
The procedure for linear image combinations is summarized in the yellow blocks of Figure \ref{fig:flowchart_benchmarking}, while the reliability of different algorithms is summarized in the green blocks of Figure \ref{fig:flowchart_benchmarking}.

Before applying the linear image combination algorithms to the input simulated space and ground-based telescopes images, we need to check whether the input low and high-resolution images have overlapping spatial frequencies.
If the spatial frequencies of input images do not overlap, we need to deconvolve the low-resolution image.
At 450 $\mu$m band, the spatial frequency coverage of the JCMT-SCUBA2 observations overlaps with the {\it Herschel}-SPIRE observations.
However, at 850 $\mu$m wavelength, the spatial frequency coverage of the JCMT-SCUBA2 observations does not overlap well with the {\it Planck} images.

We use the Lucy-Richardson algorithm \cite{Lucy1974} to deconvolve the low-resolution space telescope image. 
This guarantees that every pixel of the deconvolved image has a positive value, and the total flux of the deconvolved image is conserved.  
We first simulated {\it Herschel} observations by smoothing $I_{\nu}$ to the same angular resolution of {\it Herschel} observations at each band.
Based on these simulated {\it Herschel} images, we extrapolate an 850 $\mu$m image of 40$''$ resolution and then use this extrapolated image as the model image to initialize the deconvolution procedure. 
Using a model image helps speed up the convergence of compact sources.
The deconvolved image is presented in the lower right panel in Figure \ref{fig:input}).

We combine the {\tt fakemap} image with the deconvolved image with different methods (J-comb, immerge, and feather).
The top panels in Figure \ref{fig:2dcombinedresidual} show the combined images.
By comparing the power spectra of the model image and output combined image, it appears that the J-comb algorithm works well quantitatively.
In Figure \ref{fig:2dpsd}, the left column shows the power spectrum for all spatial frequencies.
At both large ( \textgreater 800$''$) and small ( \textless 40$''$) angular scales, the power spectrum of all combined images well preserve the power spectrum of $I_{\nu}$$\ast$$B_{\mbox{\scriptsize ground}}$.
The right panel shows the zoom-in plot of angular scales from 50$''$ to 400$''$. 
It can be seen that the power spectrum of the combined image derived by the J-comb algorithm is more consistent with the power spectrum of $I_{\nu}$$\ast$$B_{\mbox{\scriptsize ground}}$ than the results of immerge and feather.
As a further check, we convolve the model image and the combined images to 50$''$ angular resolution.
We obtain the corresponding residuals generated by subtracting the 50$''$ angular resolution combined image from the 50$''$ angular resolution model image.
The residuals are shown in the bottom panels of Figure \ref{fig:2dcombinedresidual}. 
We can see that the residuals of the feather algorithm and immerge algorithm have noticeable artifacts, which follow the shape of the molecular clouds and will likely affect follow-up analysis on the cloud structures and physical properties.
On the contrary, the residual of the J-comb algorithm does not present a visually recognizable pattern of artifact and appears everywhere uniform.

As a quantitative comparison, we summarize the pixel values of the 50$''$ combined images and the residual with two-dimensional histograms in Figure \ref{fig:pixel}. 
The RMS noise level of the 50$''$ combined image is $0.005\times$10$^{-3}$ pW.
From the two-dimensional histograms, the residuals from the feather algorithm and immerge algorithm increase with the pixel values of the combined images.
In contrast, the residuals from J-comb are mostly in the range of $-0.005\times$10$^{-3}$ pW to 0.005$\times$10$^{-3}$ pW.

Figure \ref{fig:input_450}-\ref{fig:pixel_450} are similar to Figures \ref{fig:input}-\ref{fig:pixel} but for 450 $\mu$m band.
The selected parameters used in J-comb are listed in Table \ref{tab:parameters} and $\theta_{\mbox{\scriptsize threshold}}^{\mbox{\scriptsize low}}$ and $\theta_{\mbox{\scriptsize threshold}}^{\mbox{\scriptsize high}}$ are the corresponding angular scales of  $k_{\mbox{\scriptsize threshold}}^{\mbox{\scriptsize low}}$ and $k_{\mbox{\scriptsize threshold}}^{\mbox{\scriptsize high}}$, respectively.

\subsection{Simulation for SED fitting}\label{subsub:sed}

In this section, we test the influence of the combined image on the SED fitting analyses.
The procedure is summarized in the orange blocks of Figure \ref{fig:flowchart_benchmarking}.

We used the 450/850 $\mu$m combined images from the J-comb algorithm and the model $H_{2}$ column density ($N_{\mbox{\scriptsize d}}$) map to analyze the influence.
The combined images are introduced in Section \ref{subsub:2dcombine}.
For the 450/850 $\mu$m combined images, we smoothed them to have the same angular resolution of 15$''$, which is larger than the beam size of the model images (8$''$ at JCMT-SCUBA2 450 $\mu m$, and 14$''$ at JCMT-SCUBA2 850 $\mu m$).
We re-gridded them to have the same pixel size.
From these 15$''$ images, we obtain dust temperature (${T}_{\mbox{\scriptsize d}}^{{\mbox{\scriptsize new}}}$) and column density (${N}_{\mbox{\scriptsize d}}^{{\mbox{\scriptsize new}}}$) maps by performing one-component modified blackbody SED fitting.
The model $H_{2}$ column density ($N_{\mbox{\scriptsize d}}$) map is introduced in Section \ref{subsub:model}.
We generated the column density fractional residual ($V_{\mbox{\scriptsize N}}$) of ${N}_{\mbox{\scriptsize d}}^{{\mbox{\scriptsize new}}}$ and ${N}_{\mbox{\scriptsize d}}$ with the following equation:
\begin{equation}\label{eq:residual}
V_{\mbox{\scriptsize N}} = ({N}_{\mbox{\scriptsize d}}-{N}_{\mbox{\scriptsize d}}^{{\mbox{\scriptsize new}}})/{N}_{\mbox{\scriptsize d}}.
\end{equation}
The difference between the two column density maps is smaller at the higher column density region.
For the region with the detection higher than 5 times RMS in the combined image, $V_{\mbox{\scriptsize N}}$ is lower than 10\% (within the black contours in the right panel of Figure \ref{fig:N}).

\begin{figure}[H]
\hspace{-0.3cm}
\vspace{-0.1cm}
\includegraphics[width=9.cm]{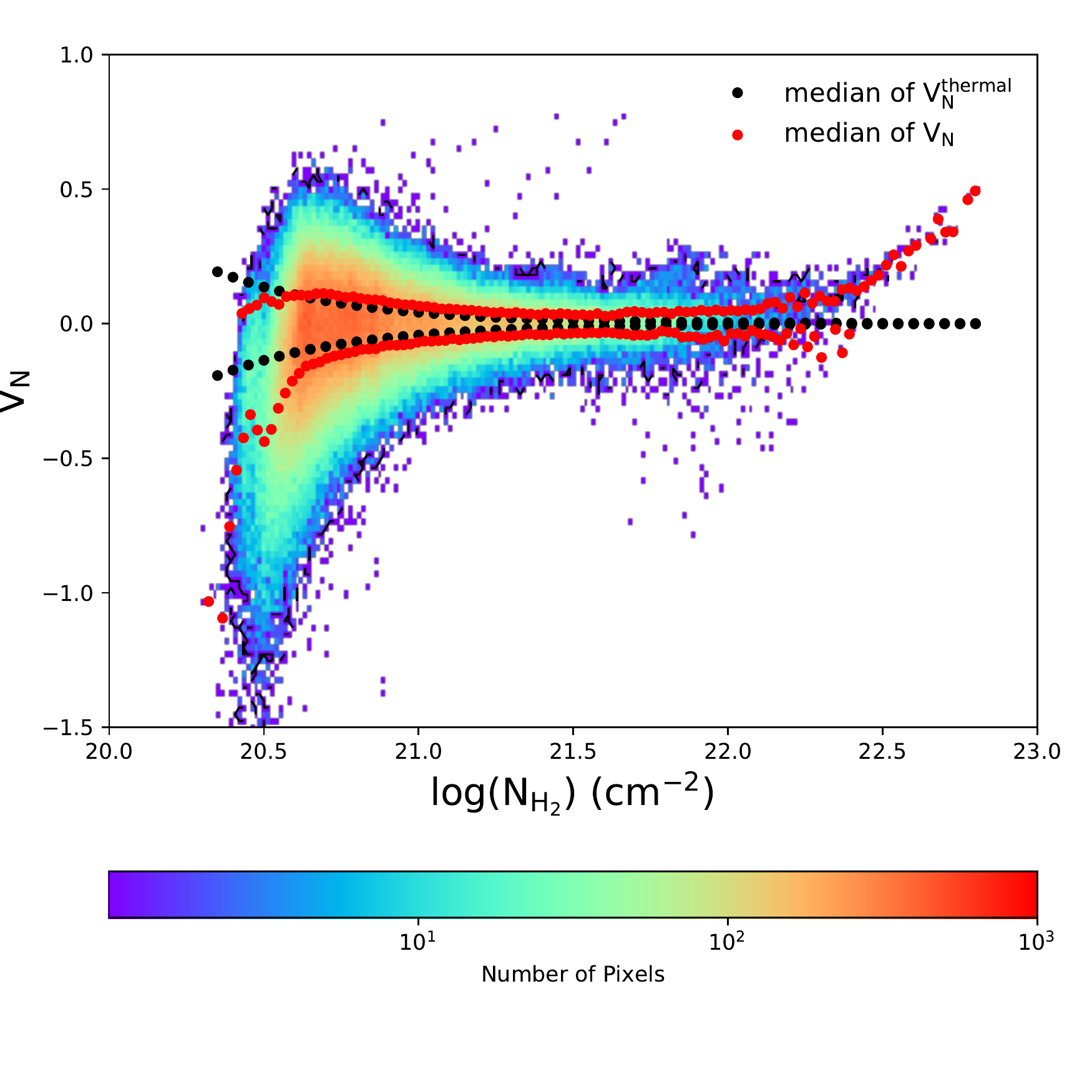}
\caption{
Distribution of ${N}_{\mbox{\scriptsize d}}$ and $V_{\mbox{\scriptsize N}}$ pixel-by-pixel. 
The black points and red points show the median values of $V_{\mbox{\scriptsize N}}^{{\mbox{\scriptsize thermal}}}$ and $V_{\mbox{\scriptsize N}}$ with different column densities respectively.
}
\label{fig:res_compare}
\end{figure}

The systemic errors from the combination and the thermal noise of the observations lead to the error in the derived column density map. 
We compared the column density errors induced by image combination and thermal noise to quantify this.
To estimate column density error induced by thermal noise, we used a Monte-Carlo technique.
Assuming a characteristic dust temperature of 20 K and dust opacity index of 1.8, we obtained the correspondent flux densities at 450 $\mu$m and 850 $\mu$m with different column densities.
We created the artificial 450/850 $\mu$m flux densities by adding normal-distributed thermal noise.
We then calculated the standard deviation within a map region without obvious emission as the real thermal noise.
We then estimated the median of fractional residual induced by thermal noise ($V_{\mbox{\scriptsize N}}^{{\mbox{\scriptsize thermal}}}$), following Equation \ref{eq:residual} for each flux density with 1000 fittings based on the artificial 450/850 $\mu$m flux densities.
The two-dimensional histogram of Figure \ref{fig:res_compare} shows the distribution of ${N}_{\mbox{\scriptsize d}}$ and $V_{\mbox{\scriptsize N}}$ pixel-by-pixel.
The black points show the median values of $V_{\mbox{\scriptsize N}}^{{\mbox{\scriptsize thermal}}}$ with different column densities, while the red points show the median values of $V_{\mbox{\scriptsize N}}$.
The difference between the median value of $V_{\mbox{\scriptsize N}}$ and the median value of $V_{\mbox{\scriptsize N}}^{{\mbox{\scriptsize thermal}}}$ is lower than 5 \% within the column density range between 4.0 $\times$$10^{20}$ cm$^{-2}$ and 1.6 $\times$$10^{22}$ cm$^{-2}$.
The difference grows at the low and high column density regimes and is likely due to different angular resolutions.
We assume that the model $T_{\mbox{\scriptsize d}}$, $N_{\mbox{\scriptsize d}}$ images represent the properties of a molecular cloud precisely, without a limit of angular resolution.
The dust temperature (${T}_{\mbox{\scriptsize d}}^{{\mbox{\scriptsize new}}}$) and column density (${N}_{\mbox{\scriptsize d}}^{{\mbox{\scriptsize new}}}$) maps derived from SED fitting have an angular resolution of 15$''$. 
The value of $\lvert V_{\mbox{\scriptsize N}} \rvert$ grows at the low and high column density regimes and is likely due to different angular resolutions.
The precision of the derived dust column density map is therefore mainly limited by the thermal noise of the JCMT-SCUBA2 images.

\section{Application to real observations}\label{subsub:orion}

\begin{figure*}
\centering
\includegraphics[width=18cm]{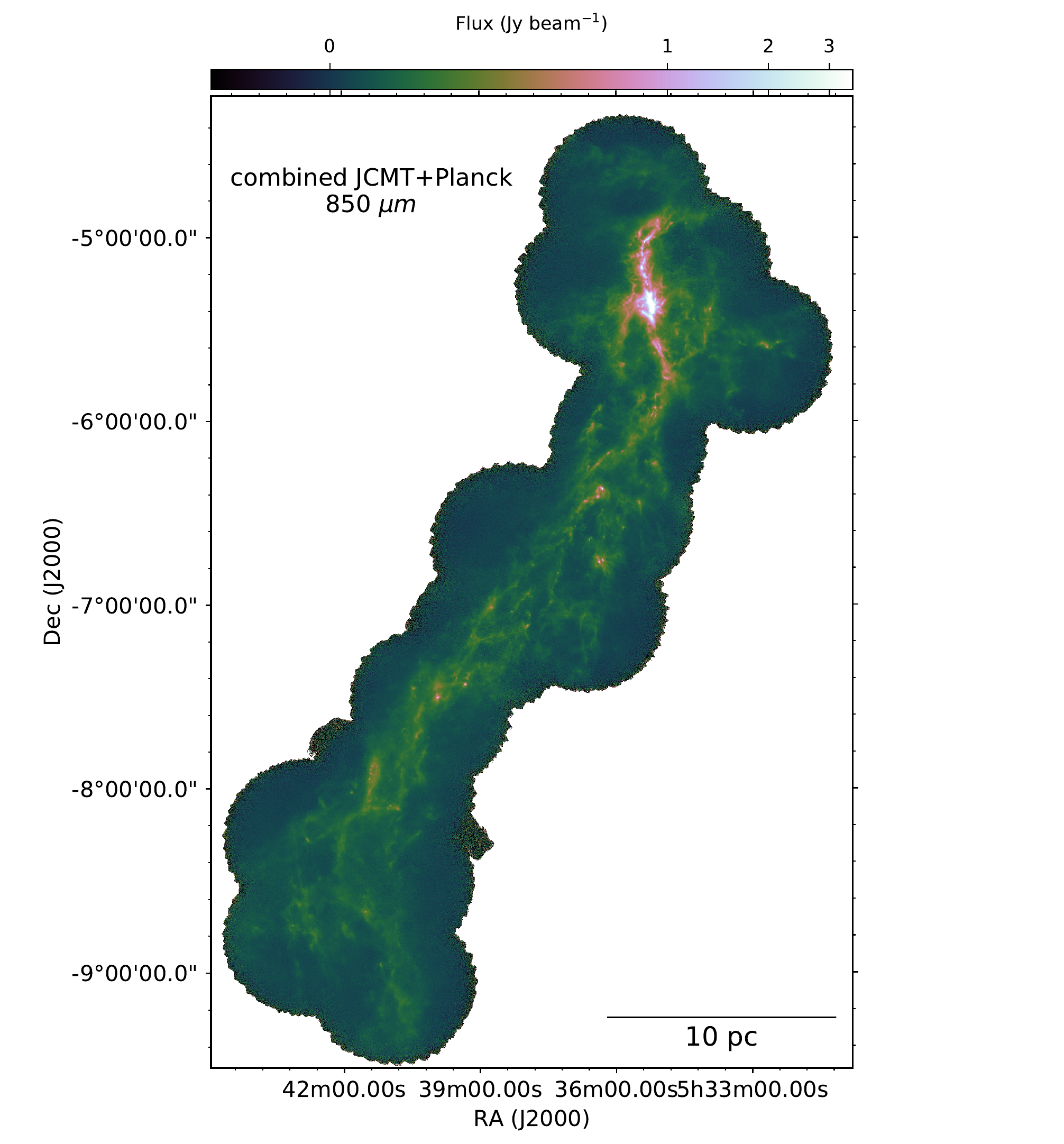}
\caption{The 14$''$ resolution, 850 $\mu$m continuum image of the Orion A region produced by combining {\it Planck}, JCMT-SCUBA2.
}
\label{fig:ori_850}
\end{figure*}

\begin{figure*}
\vspace{-1.1cm}
\includegraphics[width=18cm]{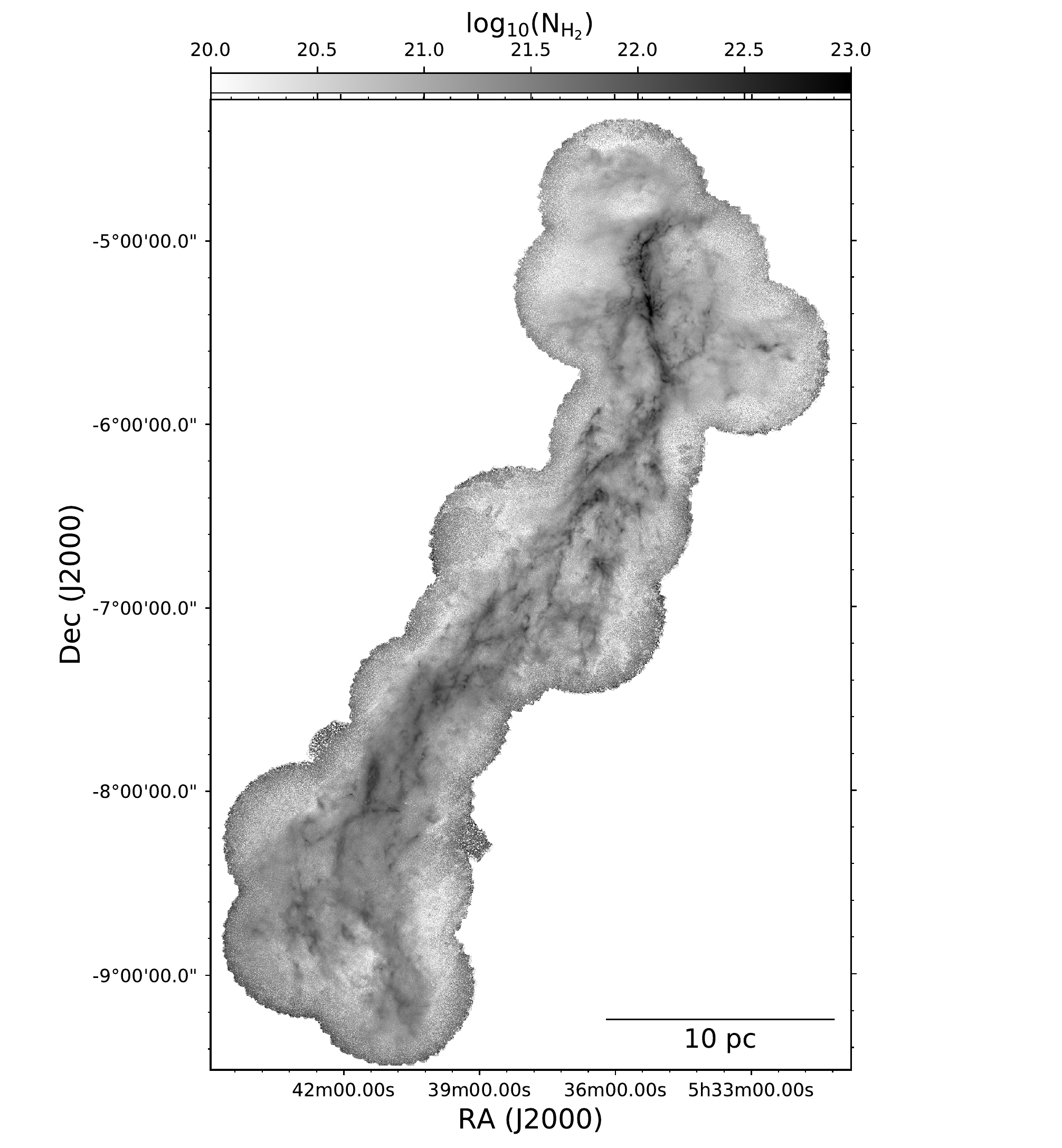}
\caption{The 10$''$ column density map of Orion A, derived based on modified black body SED fitting iteratively to the {\it Herschel} 70/160/250/350/500 $\mu$m, and the combined 450/850 $\mu$m images.
}
\label{fig:ori_sed_high_nh2}
\end{figure*}

\begin{figure*}
\vspace{-1.1cm}
\includegraphics[width=18cm]{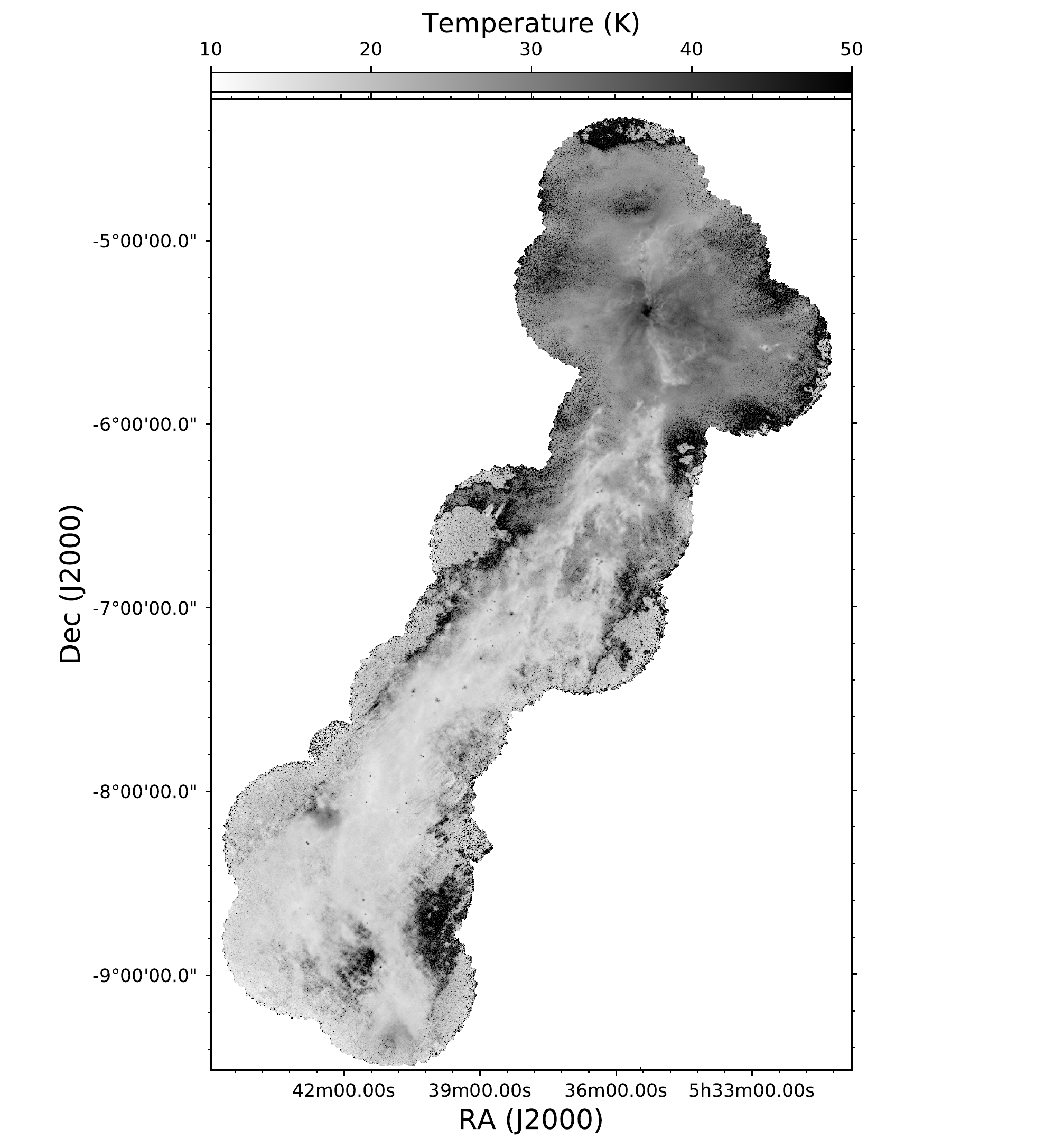}
\caption{The 10$''$ dust temperature map of Orion A, derived based on modified black body SED fitting iteratively to the {\it Herschel} 70/160/250/350/500 $\mu$m, and the combined 450/850 $\mu$m images.
}
\label{fig:ori_sed_high_td}
\end{figure*}


\begin{figure*}
\begin{tabular}{ p{0.5\linewidth}p{0.5\linewidth} }
\hspace{-0.5cm}\includegraphics[scale=0.65]{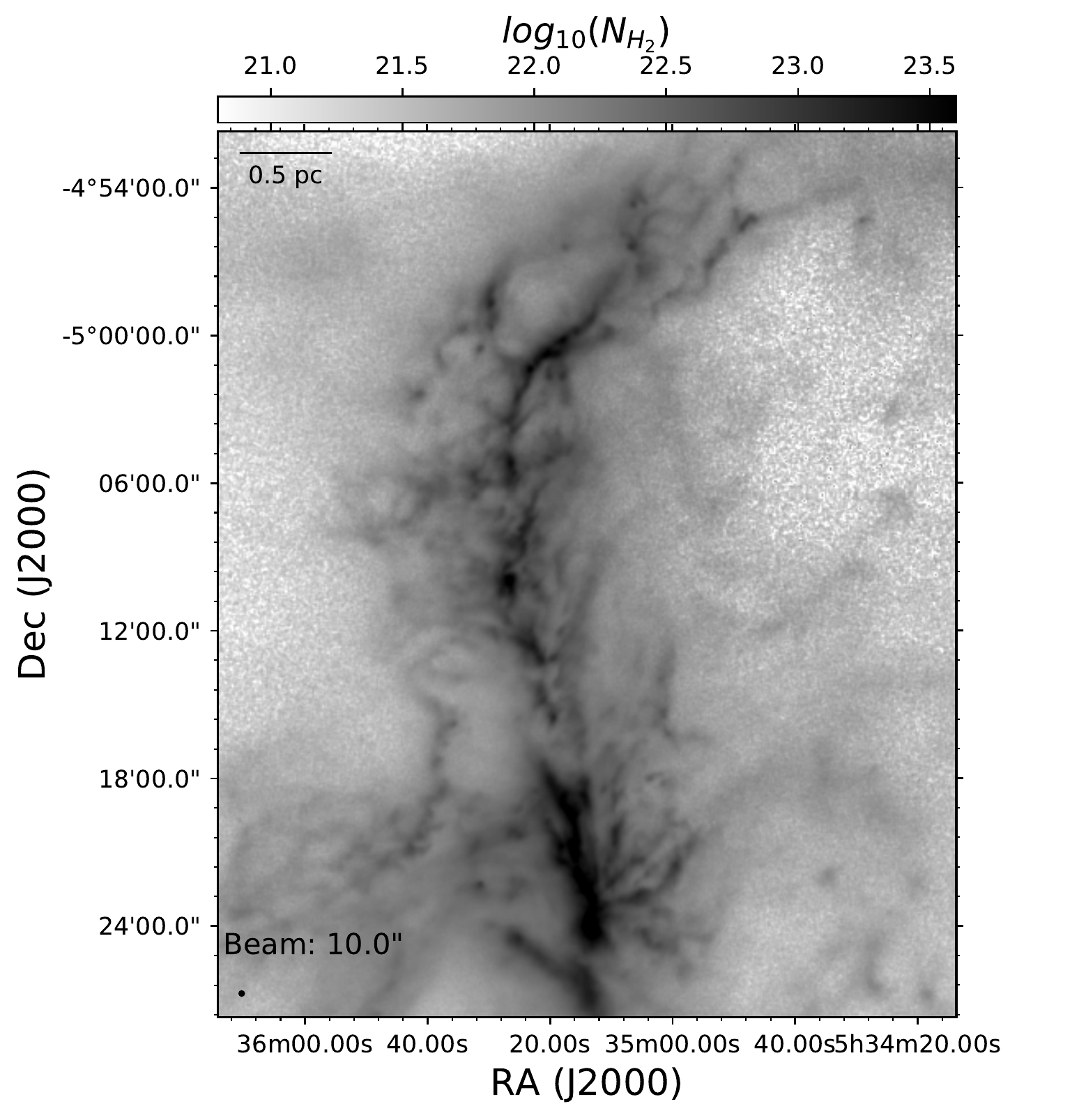} & \vspace{-10.6cm}\hspace{-0.37cm}\includegraphics[scale=0.65]{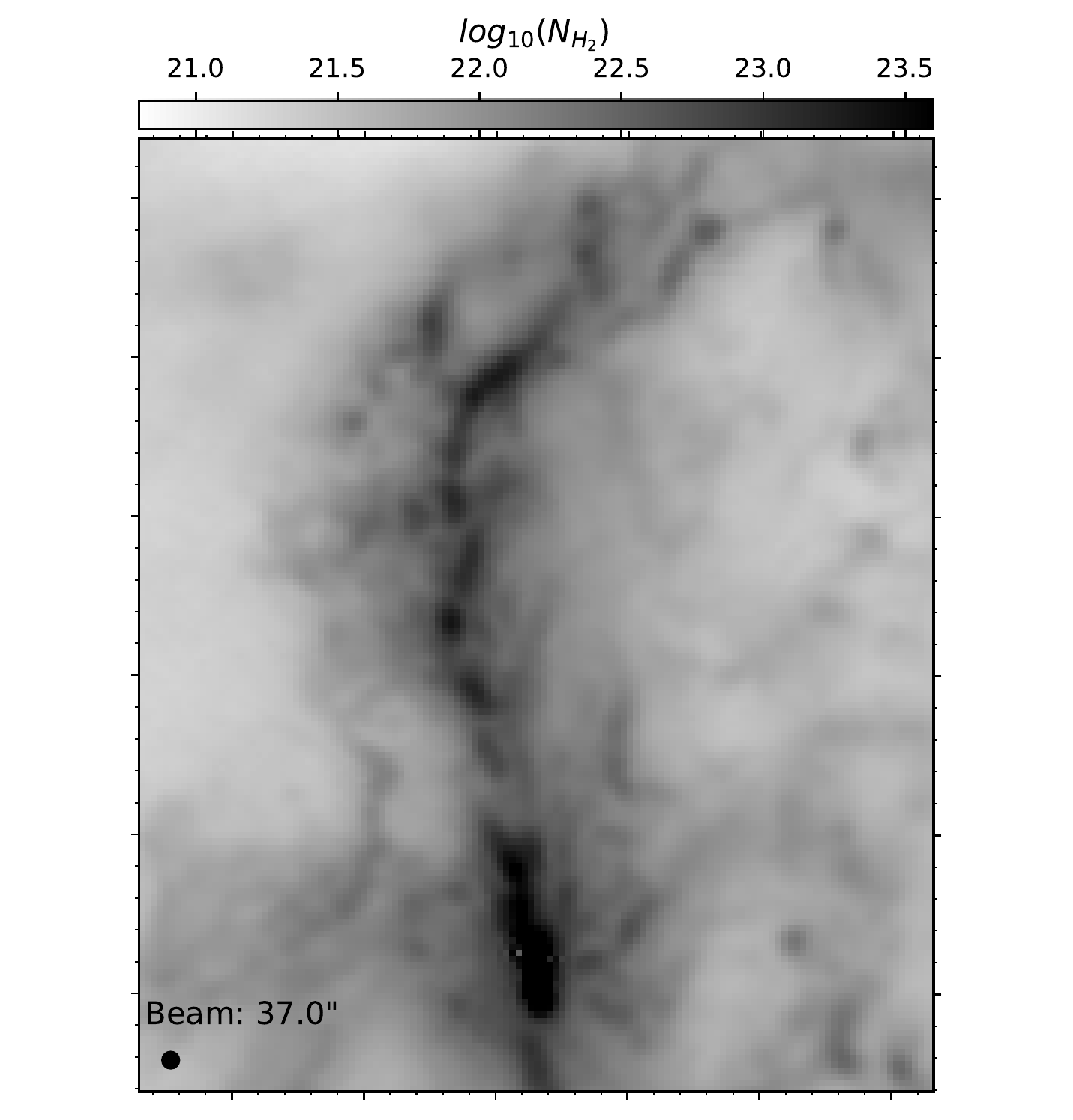}  \\
\end{tabular}
\vspace{-0.3cm}
\caption{
Comparing the 10$''$ column density map of Orion 1, 2, 3 (left panel) derived based on modified black body SED fitting iteratively with the 37$''$ column density map derived from {\it Herschel} data (right panel). 
}
\label{fig:orion_compare}
\end{figure*}


To further test the performance of the J-comb algorithm, we applied it to real observations of the Orion A region.
We include the implementation of the J-comb algorithm on OMC3 as an example at \href{https://github.com/SihanJiao/J-comb}{https://github.com/SihanJiao/J-comb/blob/main/example\_omc3.ipynb}.
The Orion A molecular cloud is the largest nearby molecular cloud hosting massive cluster formation.
Thanks to the proximity, the Orion A region can be very well resolved in (sub)millimeter bands by {\it Herschel}, {\it Planck} Space Observatory, and JCMT.
We retrieved the available night observations of JCMT Submillimetre Common-user Bolometer Array 2 \cite{Chapin2013} at 450 and 850 $\mu$m from the online data archive. (Program ID is MJLSG22 and MJLSG31).
We retrieved the level 2.5 processed, archival {\it Herschel} images that were taken at 70/160 $\mu$m using the PACS instrument \cite{Poglitsch2010} and at 250/350/500 $\mu$m using the SPIRE instrument \cite{Griffin2010}, (obsID: 1342218967, 1342218968).
{\it Planck}/High Frequency Instrument (HFI) 353 GHz images are also used.\footnote{Based on observations obtained with {\it Planck} (http://www.esa.int/Planck), an ESA science mission with instruments and contributions directly funded by ESA Member States, NASA, and Canada.}
For our combination purpose, we used the {\it Herschel} SPIRE extended emission products from the data base.


We have performed single-component, modified black body SED fitting to each pixel of the input images.
Before performing any SED fitting, we smoothed all images to a common angular resolution of the largest telescope beam and all images were re-gridded to have the same pixel size.
In the least-squares fits, we weighted the data points by the noise level.
We adopted the dust opacity per unit mass at 230 GHz of 0.09 $cm^{2}$ $g^{-2}$ \cite{Ossenkopf1994}, and we assumed a gas-to-dust mass ratio of 100.
As modified black-body assumption, the flux density $S_{\nu}$ at a certain observing frequency $\nu$ is given by
\begin{equation}
S_{\nu} = \Omega_{m}B_{\nu}(T_{\mbox{\scriptsize d}})(1-e^{-\tau_{\nu}}),
\end{equation}
where the gas column density $N$ can be approximated by
\begin{equation}
N=g(\tau_{\nu}/\kappa_{\nu}\mu m_{H}),
\end{equation}
$B_{\nu}(T_{\mbox{\scriptsize d}})$ is the Planck function at temperature $T_{\mbox{\scriptsize d}}$, the dust opacity $\kappa_{\nu}=\kappa_{\mbox{\tiny{230 GHz}}}(\nu/230\,GHz)^{\beta}$, $\Omega_{m}$ is the solid angle.
The effect of scattering opacity \cite{Liu2019ApJ} can be ignored in our case given that we are focusing on $>$10$^{3}$ AU scale cloud structures, where the averaged maximum grain size is expected to be well below 100 $\mu$m \cite{Wong2016PASJ}.

450 $\mu$m Band --- We made interpolations to the 450 $\mu$m images, based on the modified blackbody SED fitting to the PACS 70/160 $\mu$m and the SPIRE 250/350/500 $\mu$m images. 
We pre-smoothed all images for the SED fitting to the same angular resolution of SPIRE 500 $\mu$m image ($\sim$37$''$) and re-gridded these images to the same pixel size of PACS 70 $\mu$m image (3.2$''$).
We linearly combined the SCUBA2 450 $\mu$m and the interpolated 450 $\mu$m image in the Fourier domain.
In this way, the combined 450 $\mu$m map has the same angular resolution as the original SCUBA2 image (8.0$''$) but recovers the extended emission structures.

850 $\mu$m Band --- Following a similar process as the 450 $\mu$m image, we made interpolations to the 850 $\mu$m image, based on the modified blackbody SED fitting to the PACS 70/160 $\mu$m and the SPIRE 250/350/500 $\mu$m images.
We used this extrapolated {\it Herschel} 850 $\mu$m image as the model image in our deconvolution procedure applied to {\it Planck} 353 GHz image.
We convolved the deconvolved {\it Planck} image to a resolution of 60.0$''$ and combined it with SCUBA2 850 $\mu$m image.
The combined SCUBA2 and {\it Planck} image was used as the model image to deconvolve the original {\it Planck} image.
We obtained a further deconvolved {\it Planck} image with a resolution of 60.0$''$ after the iteration.
After these, we combined the final deconvolved {\it Planck} image with the SCUBA2 image to obtain the combined map, which has an angular resolution of 14.0$''$.
The combined 850 $\mu$m image is shown in Figure \ref{fig:ori_850}.

Our procedure to iteratively derive the high angular resolution ($\sim$10$''$) dust temperature and dust/gas column density images is similar to what was introduced in \cite{Lin2016,Lin2017}.
Figure \ref{fig:ori_sed_high_nh2} and Figure \ref{fig:ori_sed_high_td} show the derived gas column density ($N_{H_{2}}$) and dust temperature ($T_{\mbox{\scriptsize d}}$) maps.
The Orion A molecular cloud appears as an integrally shaped filament on $\sim$30 pc scale. 
The high angular resolution column density map reveals that the large-scale filament consists of numerous bundles of small filamentary structures (or fibers).
The dominant dense gas structure in the north of Orion A is OMC-1, which has a higher dust temperature ( \textgreater 50 K) than other subregions.
The high temperatures are likely caused by the energetic explosion occurring in the Orion KL nebula about $\sim$500 yr ago \cite{Bally2011,Bally2017}.
The rest of the dense filamentary structures have on average $\sim$15-20 K dust temperature, along with some localized heated sources.

\begin{figure}[H]
\hspace{-0.3cm}
\vspace{-0.1cm}
\includegraphics[width=8.2cm]{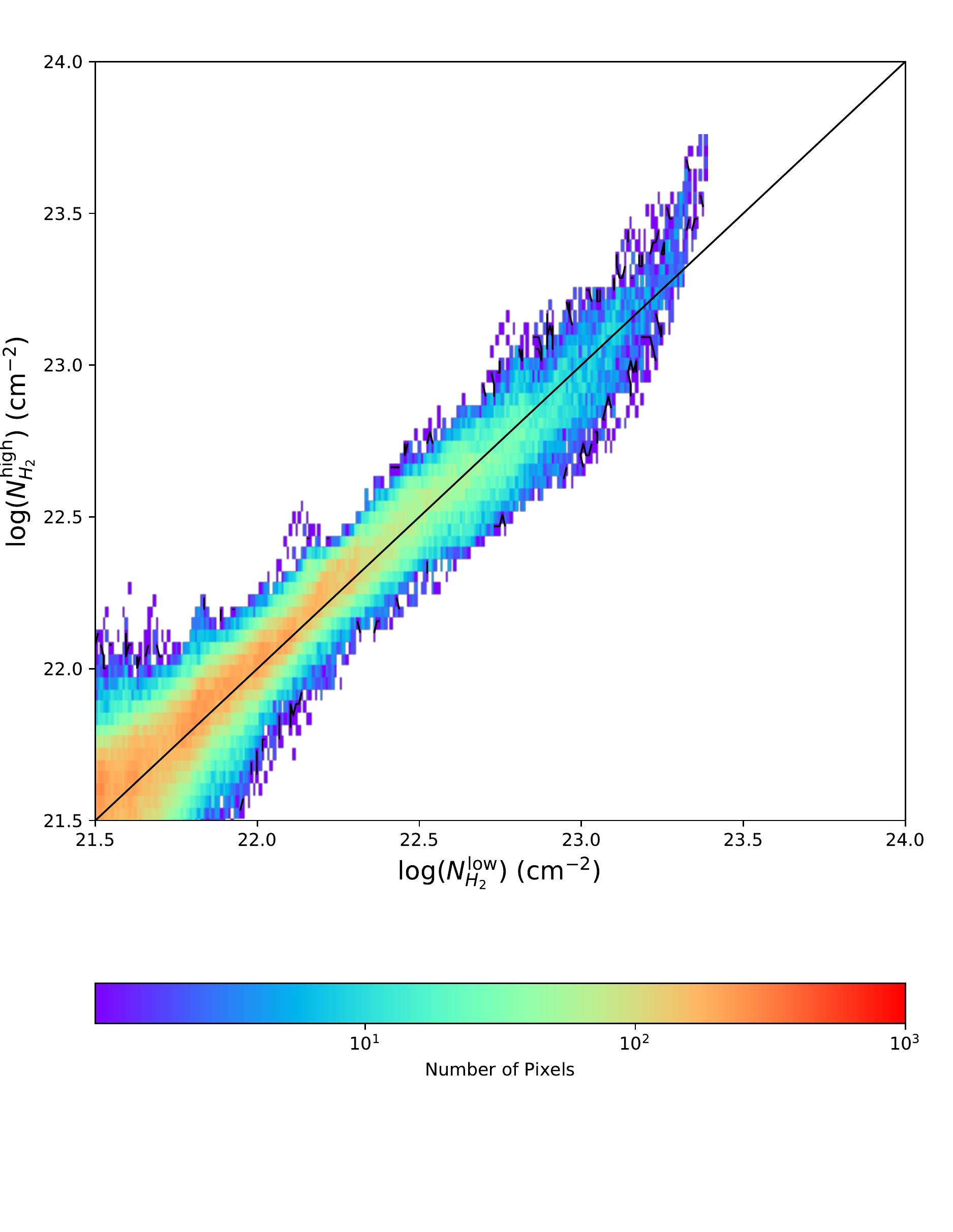}
\vspace{-0.5cm}
\caption{
Distribution of the pixel-by-pixel column density values from the 10$''$ the 37$''$ column density maps. 
The solid black line shows the line of slope=1.
}
\label{fig:N_compare}
\end{figure}

As a consistency check of our method with the existing studies based on {\it Herschel} observations \cite{Stutz2015}, we presented gas column density maps of OMC-1, 2, 3 regions, obtained from our iterative procedure and that from {\it Herschel} data only ($\beta$ fixed to a constant 1.8) respectively in Figure \ref{fig:orion_compare}.
Figure \ref{fig:orion_compare} shows consistent ${N}_{H_{2}}$ distributions in general.
However, the high-resolution ${N}_{H_{2}}$ image separates the localized dense clumps/cores from the fluffy or filamentary cloud structures in greater detail, which can be crucial for structure identification and analyses of the hierarchical cloud morphology.
The difference between the 10$''$ column density map and the 37$''$ column density map is shown in the two-dimensional histogram (see Figure \ref{fig:N_compare}).
For the dense region (${N}_{H_{2}}$>$10^{23}$ cm$^{-2}$), the values of the 37$''$ column density map are smaller than that of the 10$''$ column density map.
This will facilitate quantitative study of the physical (e.g.~\cite{li2013}) and chemical (e.g.~\cite{xie2021}) status of such star forming regions.

\section{Summary}\label{section:summary}

Linear methods for image combination in the Fourier domain have been widely applied to fix the missing flux problem of the ground-based bolometric observations.
Mathematically, the beams of the combined images derived from the previous linear data combination algorithms are not Gaussian and different from that of the high-resolution images.
The non-Gaussianity of the resultant beam makes the angular resolution of the combined image ill-defined.
To retain the same Gaussian beam as the high-resolution data and improve the accuracy of the combination procedure, we developed a new algorithm, J-comb, to combine high-resolution data that have large-scale missing information with low-resolution data containing the short spacing.
By applying a taper function to the low-pass filtered image and combining it with a high-pass filtered one using proper weights, the beam response functions of our combined images are guaranteed to have near-Gaussian shapes.
J-comb was tested on the realistic model and benchmarked with two widely used image combination algorithms, \textsc{CASA}-feather and \textsc{MIRIAD}-immerge.
We demonstrate that the performance of the J-comb algorithm is superior to the other two algorithms for our specific purpose of combining bolometric observations from ground-based and space telescopes.

The J-comb algorithm is applied to bolometric observations from ground-based and space telescopes for the Orion A region.
We obtained combined images that simultaneously achieve high angular resolutions comparable to the best angular resolution ground-based observations and suffer from minimum loss of extended structures. 
We obtained the $\sim$ 10 $''$ angular resolution dust temperature and column density maps using the iterative SED fitting procedure. 
The maps show more details than those obtained with standard analysis techniques.

While developed specifically for (sub)millimeter bolometers, the J-comb algorithm is based on general principles and should be applicable to data combinations in ($u$, $v$) space in general. 
In particular, the upcoming FAST HI surveys \cite{li2018, li2019} will benefit from combining with high-resolution interferometric data \cite{Stil2006} to provide the best possible HI sky survey.

\vspace*{2mm} \Acknowledgements{\bahao This work was supported by National Natural Science Foundation of China (NSFC) (Grant Nos. 11988101, 11725313, 11911530226, and 11403041), and the Chinese Academy of Sciences (CAS) International Partnership Program (Grant No. 114A11KYSB20160008).
The authors thank the referees for the constructive comments on improving this work.
S.J. thanks Hauyu Baobab Liu for helpful discussions on the method design and benchmark, and Chao-Wei Tsai for helpful discussions on the missing flux problem.
The James Clerk Maxwell Telescope is operated by the East Asian Observatory on behalf of The National Astronomical Observatory of Japan; Academia Sinica Institute of Astronomy and Astrophysics; the Korea Astronomy and Space Science Institute; the National Astronomical Research Institute of Thailand; Center for Astronomical Mega-Science (as well as the National Key R\&D Program of China with No. 2017YFA0402700). Additional funding support is provided by the Science and Technology Facilities Council of the United Kingdom and participating universities and organizations in the United Kingdom and Canada.
}

\end{multicols}

\end{document}